\documentclass[12pt]{article}
\usepackage{amsmath}
\usepackage{graphicx}
\usepackage{enumerate}
\usepackage{url} 
\usepackage[utf8]{inputenc}
\usepackage{booktabs}
\usepackage{amsthm}
\usepackage{amssymb}
\usepackage{algorithmic}
\usepackage[colorinlistoftodos]{todonotes}
\usepackage{caption}
\usepackage{subcaption}
\usepackage{lscape}
\usepackage{verbatim}
\usepackage{csvsimple}
\usepackage{multirow}
\usepackage{textcomp}
\usepackage{natbib}
\usepackage[linesnumbered,ruled,vlined]{algorithm2e}
\usepackage{bm}
\usepackage{wrapfig}
\usepackage{pdfpages,setspace}
\usepackage[normalem]{ulem}
\usepackage{mathabx}

\usepackage{enumitem}

\newcommand{\blind}{1}

\usepackage{tablefootnote}

\begin{document}

\def\spacingset#1{\renewcommand{\baselinestretch}%
{#1}\small\normalsize} \spacingset{1}


\if1\blind
{
  \title{\bf A Hierarchical Random Effects State-space Model for Modeling Brain Activities from Electroencephalogram Data}
  \author{ \bf Xingche Guo$^{1}$, Bin Yang$^{1}$, Ji Meng Loh$^{2}$, Qinxia Wang$^{3}$, \vspace{0.6em}\\
  \bf and Yuanjia Wang$^{1,4}$ \vspace{1em} \\
  \small{$^{1}$Department of Biostatistics, Columbia University} \\
  \small{$^{2}$Department of Mathematical Sciences, New Jersey Institute of Technology} \\
  \small{$^{3}$Novartis Pharmaceuticals Corporation} \\
  \small{$^{4}$Department of Psychiatry, Columbia University}
     }
     \date{}
  \maketitle
} \fi

\if0\blind
{
 \title{\bf A Hierarchical Random Effects State-space Model for Modeling Brain Activities from Electroencephalogram Data}
      \date{}
  \maketitle
} \fi

\bigskip
\begin{abstract}
Mental disorders present challenges in diagnosis and treatment due to their complex and heterogeneous nature. Electroencephalogram (EEG) has shown promise as a potential biomarker for these disorders. However, existing methods for analyzing EEG signals have limitations in addressing heterogeneity and capturing complex brain activity patterns between regions.
This paper proposes a novel random effects state-space model (RESSM) for analyzing large-scale multi-channel resting-state EEG signals, accounting for the heterogeneity of brain connectivities between groups and individual subjects.
We incorporate multi-level random effects for temporal dynamical and spatial mapping matrices and address nonstationarity 
so that the brain connectivity patterns can vary over time. 
The model is fitted under a Bayesian hierarchical model framework coupled with a Gibbs sampler. 
Compared to previous mixed-effects state-space models, we directly model high-dimensional random effects matrices without structural constraints and tackle the challenge of identifiability. 
Through extensive simulation studies, we demonstrate that our approach yields valid estimation and inference. 
We apply RESSM to a multi-site clinical trial of Major Depressive Disorder (MDD). Our analysis uncovers significant differences in resting-state brain temporal dynamics among MDD patients compared to healthy individuals. In addition, we show the subject-level EEG features derived from RESSM exhibit a superior predictive value for the heterogeneous treatment effect compared to the EEG frequency band power, suggesting the potential of EEG as a valuable biomarker for MDD.
\end{abstract}

\noindent%
\vfill

\newpage
\spacingset{1.9} 


\maketitle


%

\section{Introduction}
\label{s:intro}
Mental disorders, such as Major Depressive Disorder (MDD), pose significant challenges for diagnosis and treatment due to their complex and heterogeneous nature. Unlike many physical illnesses, mental disorders lack objective biomarkers, resulting in their assessment primarily relying on subjective measures and clinical observations. This limitation makes it challenging to develop effective therapy targeting biology. Thus, it is desirable to identify reliable biomarkers for mental disorders. 

Emerging evidence suggests that electroencephalogram (EEG), a non-invasive neuroimaging technique, holds promise as a potential objective biomarker for mental disorders \citep{niedermeyer2005electroencephalography}. One of the key advantages of EEG is its ability to capture dynamic changes in brain activity with high temporal resolution, typically in milliseconds. By placing electrodes on the scalp, EEG records electrical signals generated by the underlying cortical networks, offering insights into the temporal dynamics of neural processing.  Unlike other imaging techniques, such as functional magnetic resonance imaging (fMRI), EEG provides a direct measure of neural electrical activity, making it particularly well-suited for studying real-time cognitive processes and event-related responses. In recent years, there has been a growing interest in exploring the intrinsic patterns of brain activity during the absence of explicit tasks (commonly referred to as the resting-state) due to its ease of implementation. In a resting-state, the brain is not idle but rather exhibits spontaneous and self-generated patterns of neural activity \citep{pizoli2011resting}. 

The main goal of our paper is to introduce statistical methodologies that can effectively analyze large-scale multichannel resting-state EEG signals while considering the heterogeneity observed among various clinical groups, different subjects, and different recording periods of EEG within subjects. 
For resting-state EEG data, a widely-used approach uses parametric modeling techniques, such as the autoregressive moving-average (ARMA) model, to directly analyze each EEG channel in the time domain \citep{pardey1996review}.
An alternative method utilizes the Fourier transform to convert the original signals into the frequency domain. Subsequently, the power spectral densities (PSD) are computed and condensed into summary measures specific to frequency bands (e.g., alpha or theta band powers) for individual channels \citep{sanei2013eeg}. While the power spectrum is completely unlocalized in time, the time-frequency methods such as short-time Fourier transform \citep{allen1977short}, wavelet transform \citep{adeli2003analysis}, and Wigner-Ville distribution \citep{cohen1989time} overcome this limitation by capturing local transient features in both time and frequency domains. 
The Fourier and wavelet cross-spectrum were used to model the interaction between different EEG channels \citep{khan2022development, ombao2022spectral}. 

Extensive research has been conducted to model EEG signals focusing on brain connectivity. Recognizing that EEG signals on the scalp is often a complex mixture of unknown underlying brain sources,
Independent Component Analysis (ICA) offers a data-driven approach to decompose observed EEG signals into independent non-Gaussian sources to study spatial patterns of brain signals \citep{jung2001imaging, james2004independent}. However, ICA does not model spatial patterns and temporal dynamics simultaneously.
To capture temporal patterns, multichannel EEG signals can be modeled as a system connected by additive differential equations \citep{chen2017network}. 
\cite{sun2021modelling} proposed semiparametric nonlinear dynamical systems under a single-index model framework to relax the additive assumption.
However, dynamical system-based methods impose assumptions on signal derivatives and require solving a system of differential equations, which can be sensitive to initial conditions. On the other hand, hidden Markov models have been utilized to capture brain activity by modeling transitions among a finite number of hidden states  \citep{williams2018markov}. A related approach, the state-space model, was proposed to analyze EEG data by representing observed noisy signals originating from low-dimensional latent underlying sources \citep{cheung2010estimation, yang2016state}. \cite{li2021mapping} introduce a modular state-space multivariate autoregressive model by dividing the brain network into multiple clusters. However, a common limitation of the above methods is that they apply to single-subject EEG signals, making it challenging to conduct inference in multi-group, multi-subject EEG studies.

When analyzing multi-subject EEG data,  it is critical to address the considerable heterogeneity among patients  in various clinical groups while also enabling comparisons within and between groups. To this end, 
\cite{wang2022latent} introduced a latent state-space model incorporating subject-level covariates to analyze multichannel, multi-subject EEG data. They accounted for between-subject heterogeneity by modeling covariates with fixed effects on the scale of the temporal dynamical matrix in a multivariate vector autoregressive (MVAR) model. They assumed a common spatial mapping matrix in the sensor model for all subjects (temporal dynamical and spatial mapping matrices are defined in Section \ref{sec:methods}). However, there is evidence that spatial maps may vary across subjects \citep{davison2016individual}, and this approach does not account for the unobserved heterogeneity beyond a few given covariates.

In this work, we propose a hierarchical random effects state-space model (RESSM) that can effectively analyze large-scale multichannel resting-state EEG signals while considering the heterogeneity observed among various clinical groups and subjects in both brain temporal dynamics and spatial mapping patterns. Specifically, we introduce subject-specific random effects for the temporal dynamical and spatial mapping matrices under a latent state-space model framework. These matrices can be used to derive directional brain connectivity measures \citep{li2021mapping}. Furthermore, we divide the signals, typically recorded over an extended duration, into multiple time segments to address nonstationarity observed in the resting state EEG signals. We introduce segment-specific random effects to capture the local stationary characteristics within each time segment and allow the connectivity patterns to change over time. 
The advantage of incorporating random effects lies in its ability to capture the inherent heterogeneity found in EEG data, thereby enabling accurate inference. In contrast, fixed-effect models often underestimate the uncertainty, leading to inaccurate inference.
Other methods that characterize time-varying brain connectivity patterns include slowly evolving locally stationary process \citep{fiecas2016modeling} and state-switching state-space model \citep{ombao2018statistical}.

Our method falls under the mixed-effects state-space model framework \citep{liu2011mixed}. However,  our model entails a multi-level structure and a significantly higher computational challenge due to the higher dimension of the random effects in the temporal dynamical and spatial mapping matrices (e.g., the random effects to model the spatial maps has dimension $54\times 5=270$ in our motivating study), in contrast to the dimension of random effects utilized in the usual latent state space models. We introduce novel approaches to tackle this dimensionality challenge. 

Fitting a latent state-space model often involves using the Expectation-maximization (EM) algorithm in conjunction with the Kalman filter and Kalman smoother \citep{shumway2000time, wang2022latent}. However, this approach has difficulties when dealing with random effects. Instead, we use a Bayesian hierarchical framework coupled with a Gibbs sampler to fit the model \citep{carlin1992monte}, which is more effective for accommodating random effects. 
We also impose fewer constraints on the heterogeneity of the temporal dynamical and spatial mapping matrices compared to representing them using a small set of parameters \citep{liu2011mixed}.
Additionally, we introduce a novel two-stage method that tackles the issue of identifiability.
We apply our approach to the EMBARC study, a large-scale randomized clinical trial that includes subjects diagnosed with major depressive disorder (MDD) and control groups comprising healthy subjects at four study centers. Our analysis reveals a significant difference in the temporal dynamics of resting-state brain activity between individuals with MDD and healthy individuals. Moreover, we show that the subject-level EEG features extracted from our approach demonstrate a higher predictive accuracy for the conditional average treatment effect (CATE) compared to alternative measures (e.g., EEG band powers), suggesting potential utilities of EEG as a biomarker for MDD treatment response.



The remainder of the article is organized as follows. Section \ref{sec:methods} details our random effects state-space model, addressing the identifiability issue and presenting the hierarchical structure and prior distributions. The MCMC computations and model selection are described in Section \ref{sec:computation}. In Section \ref{sec:simu_study}, we show results from our simulation studies and sensitivity analysis, while Section \ref{sec:real} focuses on the analysis of EEG signals from the EMBARC study. In Section \ref{sec:discussion}, we summarize our approach, highlight key findings, and discuss potential extensions for future research.

\section{Methods}
\label{sec:methods}
\subsection{Random effects state-space models (RESSM)}
Let $\mathbf{Y}_{rij}(t_k) = \left(Y_{rij1}(t_k), \dots, Y_{rijP}(t_k) \right)^{\top}$ denote the resting-state EEG data collected from $P$ scalp electrodes for the $r$-th subgroup ($r=1,\dots, R$),
$i$-th subject ($i=1,\dots,N_r$), and $j$-th time segment ($j=1,\dots, J_{ri}$) at time $t_k$ ($k = 1, \dots, K$). 
A time segment (or segment) is defined by dividing a single EEG scan into multiple time intervals.
The $P$-dimensional EEG signals (observed states) are often a mixture of unknown
underlying brain source  (latent states) with lower dimensions.
A commonly used sensor model to describe the observed EEG signals is 
\vspace{-1em}
\begin{align}
&\mathbf{Y}_{rij}\left(t_{k}\right) = \mathbf{\Theta}_{rij} \mathbf{M}_{rij}\left(t_{k}\right)+\boldsymbol{\epsilon}_{rij}\left(t_{k}\right), \quad \boldsymbol{\epsilon}_{rij}\left(t_{k}\right) \sim \mathcal{N}\left(\mathbf{0}, \boldsymbol{\Sigma}_{rij}\right) \label{model1_eq1},
\end{align}
where $\mathbf{M}_{rij}(t_k) = \left(M_{rij1}(t_k), \dots, M_{rijQ}(t_k) \right)^{\top}$ is a $Q$-dimensional $(Q < P)$ vector representing the latent states, $\boldsymbol{\epsilon}_{rij}\left(t_{k}\right)$ is independent background white noise with its variance depending on each time segment. The spatial mapping matrix $\mathbf{\Theta}_{rij}$ with dimension $P \times Q$ is a subject- and segment-specific random effect matrix that represents how the lower-dimensional latent state $\mathbf{M}_{rij}(t_k)$ maps to the higher-dimensional EEG signals $\mathbf{Y}_{rij}(t_k)$.
To model the temporal dynamics of the latent states $\mathbf{M}_{rij}(t_k)$, we assume that they follow a multivariate vector autoregressive (MVAR) model with order $m$, 
\begin{align}
&\mathbf{M}_{rij}\left(t_{k}\right) = \sum_{h=1}^m \mathbf{A}_{rijh}   \mathbf{M}_{rij}\left(t_{k-h}\right) + \mathbf{W}_{rij}\left(t_{k}\right), \quad 
\mathbf{W}_{rij}\left(t_{k}\right) \sim \mathcal{N}\left(\mathbf{0}, \boldsymbol{\Sigma}_{w}\right), \label{model1_eq2} 
\end{align}
where $\mathbf{W}_{rij}\left(t_{k}\right)$ is i.i.d Gaussian noise for all time segments. The temporal dynamical matrix $\mathbf{A}_{rijh}$ with dimension $Q \times Q$ is a subject- and segment-specific random effect matrix that represents how latent state $\mathbf{M}_{rij}(t_k)$ depends on its own $h$-step lagged state $\mathbf{M}_{rij}(t_{k-h})$.  
Equation \eqref{model1_eq2} can be easily reduced to a simplified first-order model (see Web Appendix A.1). The sensor model \eqref{model1_eq1} and state model \eqref{model1_eq2} form the RESSM for the observed EEG signals.
We incorporate flexibility in the temporal dynamical matrices and spatial mapping matrices by allowing them to vary from segment to segment, thus accommodating non-stationarity.
In contrast, modeling the entire EEG scan using Equations (1) and (2) as in \cite{wang2022latent} would not capture non-stationarity effectively. 
If we define $\widetilde{\mathbf{Y}}_{rij}\left(t_{k}\right)$ as the corresponding EEG signal without any random measurement error, i.e., $\widetilde{\mathbf{Y}}_{rij}\left(t_{k}\right) = \mathbf{\Theta}_{rij} \mathbf{M}_{rij}\left(t_{k}\right)$, we obtain the following equation:
\begin{align*}
    \widetilde{\mathbf{Y}}_{rij}\left(t_{k}\right) = \sum_{h=1}^m \mathbf{B}_{rijh}   
    \widetilde{\mathbf{Y}}_{rij}\left(t_{k-h}\right) + \mathbf{U}_{rij}\left(t_{k}\right), \quad 
\mathbf{U}_{rij}\left(t_{k}\right) = \mathbf{\Theta}_{rij} \mathbf{W}_{rij}\left(t_{k}\right),
\end{align*}
where 
\vspace{-2em}
\begin{align}
    \mathbf{B}_{rijh} = \mathbf{\Theta}_{rij} \mathbf{A}_{rijh} \left( \mathbf{\Theta}_{rij}^{\top} \mathbf{\Theta}_{rij} \right)^{-1} \mathbf{\Theta}_{rij}^{\top}, \quad h = 1, \dots, m 
    \label{equ:directional_connectivity_matrix}
\end{align}
is referred to as the subject- and segment-specific directional connectivity matrix \citep{li2021mapping}, and captures the complex interaction between the spatial mapping matrix and the temporal dynamical matrix. As these connectivities can vary across different time segments, they provide  dynamic measures of connectivity.

\subsection{Identifiability}

To begin with, note that there exist infinitely many equivalent RESSM configurations with different values of $\boldsymbol{\Theta}_{rij}$ and $\mathbf{A}_{rij}$ when no constraints are imposed on $\boldsymbol{\Sigma}_{w}$ and $\boldsymbol{\Sigma}_{rij}$.
In light of this, we take the initial step by setting $\boldsymbol{\Sigma}_{w} = \mathbf{I}$ and constraining $\boldsymbol{\Sigma}_{rij}$ as diagonal matrices to ensure model identifiability (see Web Appendix A.2 for a detailed derivation).
Similar to other factor models, $\mathbf{M}_{rij}(t_k)$ in the RESSM \eqref{model1_eq1}-\eqref{model1_eq2} is only identifiable up to a rotation. To see this, note that for any orthogonal matrix $\mathbf{R}$, we can define
\begin{align}
\mathbf{M}_{rij}^{*}\left(t_{k}\right)=\mathbf{R M}_{rij}\left(t_{k}\right), \quad \mathbf{\Theta}^{*}_{rij} = \mathbf{\Theta}_{rij} \mathbf{R}^{\top}, \quad \mathbf{A}^{*}_{rijh} =\mathbf{R A}_{rijh} \mathbf{R}^{\top}. \label{equ:reparameterization}
\end{align}
Replacing $\mathbf{\Theta}_{rij}$, $\mathbf{M}_{rij}\left(t_{k}\right)$, and $\mathbf{A}_{rijh}$ in \eqref{model1_eq1}-\eqref{model1_eq2} with \eqref{equ:reparameterization} does not change the latent state-space model. Using QR decomposition, we can find an orthogonal matrix $\widecheck{\mathbf{R}}$ such that $\widecheck{\mathbf{\Theta}}_{rij} = \mathbf{\Theta}_{rij} \widecheck{\mathbf{R}}^{\top}$. Here, $\widecheck{\mathbf{\Theta}}_{rij}$ is a lower-triangular matrix. The QR factorization is unique up to the sign of each column of $ \mathbf{\Theta}_{rij}$ if $\mathbf{\Theta}_{rij}$ is full-rank \citep{trefethen2022numerical}. 
Therefore, we force the upper-triangular entries of all mapping matrices $\mathbf{\Theta}_{rij}$ in \eqref{model1_eq1} to be zero to ensure that each latent state $M_{rijq}(t_k)$ ($q=1,\dots, Q$) is identifiable up to a sign. 

We implement a two-stage procedure to achieve sign identifiability: in the first stage (referred to as the initialization stage), a Markov Chain Monte Carlo (MCMC) algorithm is  applied to a simplified RESSM, where we set $\mathbf{\Theta}_{rij} \equiv \mathbf{\Theta}_0$ in \eqref{model1_eq1}. 
The posterior mean of $\mathbf{\Theta}_0$ obtained from the first step is utilized as the initial values for the spatial mapping matrices at each level in the proposed method.
The initialization stage ensures an identical initial value for spatial mapping matrices at all levels that is close to the truth, thereby mitigating the concentration of random effects around two means with opposite signs. 
In the second stage (referred to as the sign-tracking stage), we monitor the signs for each column of the spatial mapping matrices at each MCMC iteration and adjust the signs if potential errors are identified. A cosine correlation (i.e., $\cos(\mathbf{x}, \mathbf{y}) = \mathbf{x}^{\top}\mathbf{y}/\sqrt{(\mathbf{x}^{\top} \mathbf{x}) (\mathbf{y}^{\top} \mathbf{y})}$) is used to determine whether a segment-, subject-, or group-level spatial mapping matrices (i.e., each column of $\boldsymbol\Theta$)  share the same sign as their parent level. Note that the cosine correlation always falls within the range of $-1$ and $1$. 
A substantially negative value indicates an incorrect sign identification, and in such cases, we switch the sign of the latent signals and the corresponding column of spatial mapping matrices. Additional computational details and a simulation study demonstrating the effectiveness of both stages of the proposed two-stage algorithm are presented in Web Appendix A.2 and C.3, respectively. 

\vspace{-1em}

\subsection{Hierarchical structure and prior distributions}

The RESSM establishes a multi-level framework wherein time segments are nested in subjects and groups.
The hierarchical structure of segment-level, subject-level, and group-level temporal dynamical matrices is modeled as follows:
  \begin{gather}
 \left[ \mbox{vec}\left( \mathbf{A}_{rij} \right) \mid 
         \mbox{vec}\left( \mathbf{A}_{ri} \right) \right] \sim 
\mathcal{N}\Big(\mbox{vec}\left( \mathbf{A}_{ri} \right), \boldsymbol{\Sigma}_{v,r} \Big), \label{equ:Arij} \\
 \left[ \mbox{vec}\left( \mathbf{A}_{ri} \right) \mid 
         \mbox{vec}\left( \mathbf{A}_{r} \right) \right] \sim 
\mathcal{N}\Big(\mbox{vec}\left( \mathbf{A}_{r} \right), \boldsymbol{\Sigma}_{\gamma,r} \Big), \label{equ:Ari} \\
 \left[ \mbox{vec}\left( \mathbf{A}_{r} \right) \mid 
         \mbox{vec}\left( \mathbf{A} \right) \right] \sim 
\mathcal{N}\Big(\mbox{vec}\left( \mathbf{A} \right), \boldsymbol{\Sigma}_{a} \Big), \nonumber
\end{gather}
where $\mathbf{A}_{rij} = \left[ \mathbf{A}_{rij 1}, \dots, \mathbf{A}_{rij m}  \right]$, and similarly for $\mathbf{A}_{ri}$ and $\mathbf{A}_{r}$. 
For any $m \times n$ matrix $\mathbf{X}$, $\mbox{vec}(\mathbf{X})$ is defined to be the vector of length $mn$ obtained by stacking the columns of $\mathbf{X}$ on top of one another. 
The matrix $\mathbf{A}$ represents the population-level \textit{prior} for the temporal dynamics, and $\boldsymbol{\Sigma}_{v,r}$, $\boldsymbol{\Sigma}_{\gamma,r}$, and $\boldsymbol{\Sigma}_{a}$ are the covariance matrices of the (vectorized) temporal dynamical matrices for the three levels.
For the spatial mapping matrices, we  model the hierarchical structure of the lower-triangular entries as
\begin{gather}
 \left[ \mbox{low}\left( \mathbf{\Theta}_{rij} \right) \mid 
         \mbox{low}\left( \mathbf{\Theta}_{ri} \right) \right] \sim 
\mathcal{N}\Big(\mbox{low}\left( \mathbf{\Theta}_{ri} \right), \boldsymbol{\Sigma}_{u,r} \Big), \label{equ:Theta_rij} \\
 \left[ \mbox{low}\left( \mathbf{\Theta}_{ri} \right) \mid 
         \mbox{low}\left( \mathbf{\Theta}_{r} \right) \right] \sim 
\mathcal{N}\Big(\mbox{low}\left( \mathbf{\Theta}_{r} \right), \boldsymbol{\Sigma}_{\psi,r} \Big), \label{equ:Theta_ri} \\
 \left[ \mbox{low}\left( \mathbf{\Theta}_{r} \right) \mid 
         \mbox{low}\left( \mathbf{\Theta} \right) \right] \sim 
\mathcal{N}\Big(\mbox{low}\left( \mathbf{\Theta} \right), \boldsymbol{\Sigma}_{\theta} \Big), \nonumber 
\end{gather}
\noindent where $\mbox{low}(\mathbf{X})$ for an $m \times n$ matrix $\mathbf{X}$ is defined to be the vector of length $(2m-n+1)n/2$ obtained by deleting the upper-triangular elements of $\mathbf{X}$ in $\mbox{vec}(\mathbf{X})$. 
In this case, the spatial mapping matrices and the latent states are identifiable up to the sign. The group- and subject-level directional connectivity matrices (i.e., $\mathbf{B}_{rh}$ and $\mathbf{B}_{rih}$) can be derived by replacing $\mathbf{\Theta}_{rij}$ and $\mathbf{A}_{rijh}$ in \eqref{equ:directional_connectivity_matrix} with their corresponding group- and subject-level means. 

We choose conjugate inverse Wishart priors $\mathcal{IW}(\nu, \mathbf{H})$ for the variance components. Here, $\nu$ represents the degrees of freedom, and $\mathbf{H}$ is the scale matrix.
To reduce overfitting, we assume that $\mathbf{A}_{rij}$ and $\mathbf{\Theta}_{rij}$ share the same conditional covariance for each subgroup in \eqref{equ:Arij} and \eqref{equ:Theta_rij}, respectively. This assumption can be relaxed by adding one more hierarchy on $\boldsymbol{\Sigma}_{u,r}$ and $\boldsymbol{\Sigma}_{v,r}$, for instance, letting $\boldsymbol{\Sigma}_{u,ri} \sim \mathcal{IW}(\nu_u, \boldsymbol{\Sigma}_{u,r})$ and $\boldsymbol{\Sigma}_{v,ri} \sim \mathcal{IW}(\nu_v, \boldsymbol{\Sigma}_{v,r})$.
To avoid matrix inversions in the posterior computations,
we employ direct sampling of the precision matrices (i.e., the inverse of the covariance matrices, $\boldsymbol{\Sigma}^{-1}$), which has a prior following a conjugate Wishart distribution $\mathcal{W}(\nu, \mathbf{H}^{-1})$. 
Web Appendix B.2 provides detailed information on the prior distributions and posterior computations for the variance components, as well as a discussion on the selection of prior hyperparameters. A simulation study demonstrating the robustness of our model to hyperparameter choices can be found in Web Appendix C.4.
Furthermore, we let $\boldsymbol{\Sigma}_{rij}=\sigma_{rij}^2 \mathbf{I}$, and a conjugate inverse gamma prior $\mathcal{IG}(a_0, b_0)$ is assigned to each $\sigma_{rij}^2$. We assign improper priors to $\mathbf{\Theta}$ and $\mathbf{A}$, i.e. $p \left( \mbox{low}\left( \mathbf{\Theta} \right) \right) \propto 1, \ p \left( \mbox{vec}\left( \mathbf{A} \right) \right) \propto 1 $.

\section{MCMC computations}
\label{sec:computation}

\vspace{-0.8em}

\subsection{Sampling from full conditional distributions}

Given the structure of our model, we can use a Gibbs sampler for posterior computation. To simplify the presentation of the posterior, we use the canonical form of the Gaussian distribution
$\mathcal{N}_C(\boldsymbol{b}, \boldsymbol{Q})$ to represent a Gaussian distribution in the form $\mathcal{N}(\boldsymbol{Q}^{-1} \boldsymbol{b}, \boldsymbol{Q}^{-1})$. 
The full conditional distributions of $\mathbf{M}_{rij}\left(t_{k}\right)$, $\mathbf{A}_{rij}$, and $\mathbf{\Theta}_{rij}$ are given below. 

  \noindent \textbf{Sample $\mathbf{M}_{rij}\left(t_{k}\right)$}: For the $r$-th subgroup, $i$-th subject in the $j$-th segment, at each time $t_k$, we sample the latent EEG signal $\mathbf{M}_{rij}\left(t_{k}\right)$ from its full conditional posterior:
  \begin{align*}
\left[ \mathbf{M}_{rij}\left(t_{k}\right) \mid \cdots \right] &\propto p\Big( \mathbf{Y}_{rij}\left(t_{k}\right) \Big| \mathbf{M}_{rij}\left(t_{k}\right) \Big)  \prod_{s=k}^{k+m} p\Big( \mathbf{M}_{rij}\left(t_{s}\right) \Big| \left\{ \mathbf{M}_{rij}\left(t_{s-h}\right) \right\}_{h=1}^m \Big) 
\sim \mathcal{N}_C \left( \boldsymbol{b}_{rijk}^{(M)}, \boldsymbol{Q}_{rijk}^{(M)}  \right),
\end{align*}
where
\vspace{-2em}
\begin{gather*}
\boldsymbol{Q}_{rijk}^{(M)} = \sigma_{rij}^{-2} \left(\mathbf{\Theta}_{rij} \right)^{\top}  \mathbf{\Theta}_{rij} + \sum_{h=0}^{m} \mathbf{A}_{rijh}^{\top} \mathbf{A}_{rijh}, \\
\boldsymbol{b}_{rijk}^{(M)} = \sigma_{rij}^{-2} \left(\mathbf{\Theta}_{rij} \right)^{\top} \mathbf{Y}_{rij}\left(t_{k}\right) -  \sum_{h_1=0}^m \mathbf{A}_{rijh_1}^{\top} \sum_{h_2=0, h_2 \neq h_1}^m \mathbf{A}_{rijh_2}
\mathbf{M}_{rij}\left(t_{k+h_1-h_2}\right),
\end{gather*}
with the $Q \times Q$ matrix $\boldsymbol{A}_{rij0}:= - \mathbf{I}$. Note that since $\boldsymbol{Q}_{rijk}^{(M)}$ is identical at each time point, we only need to compute it once.

  \noindent \textbf{Sample $\mathbf{A}_{rij}$}:
Define 
$\mathbf{M}_{rijk}^{*} = \left( \mathbf{M}_{rij}^{\top}\left(t_{k-1}\right), \dots,  \mathbf{M}_{rij}^{\top}\left(t_{k-m}\right) \right)^{\top}.$
We sample the temporal dynamical matrices $\mathbf{A}_{rij}$ from its full conditional posterior:
  \begin{align*}
\left[ \mbox{vec} \left( \mathbf{A}_{rij} \right) \Big| \cdots \right] &\propto
p\left( \mbox{vec} \left( \mathbf{A}_{rij} \right)  \Big| \mbox{vec} \left( \mathbf{A}_{ri} \right) \right)
\prod_{k=m+1}^K p\Big( \mathbf{M}_{rij}\left(t_{k}\right) \Big| \mathbf{M}_{rijk}^{*}, \mathbf{A}_{rij} \Big)  
\sim \mathcal{N}_C \left( \boldsymbol{b}_{rij}^{(A)}, \boldsymbol{Q}_{rij}^{(A)} \right),
\end{align*}
\vspace{-2em}
where
\begin{gather*}
\boldsymbol{Q}_{rij}^{(A)} = \boldsymbol{\Sigma}_{v,r}^{-1} + 
\left\{\sum_{k=m+1}^K \mathbf{M}_{rijk}^* \left(\mathbf{M}_{rijk}^*\right)^{\top} \right\} \otimes \mathbf{I}_Q, \\
\boldsymbol{b}_{rij}^{(A)} = \boldsymbol{\Sigma}_{v,r}^{-1} \ \mbox{vec}\left( \mathbf{A}_{ri} \right) + \sum_{k=m+1}^K  \mathbf{M}_{rijk}^* \otimes  \mathbf{M}_{rij}\left(t_{k}\right).
\end{gather*}

  \noindent \textbf{Sample $\mathbf{\Theta}_{rij}$}: We sample spatial mapping matrices $\mathbf{\Theta}_{rij}$ from its full conditional posterior:
  \begin{align*}
\left[ \mbox{low} \left( \mathbf{\Theta}_{rij} \right) \Big| \cdots \right] &\propto
p\left( \mbox{low} \left( \mathbf{\Theta}_{rij} \right)  \Big| \mbox{low} \left( \mathbf{\Theta}_{ri} \right) \right) 
\prod_{k=1}^K p\Big( \mathbf{Y}_{rij}\left(t_{k}\right) \Big| \mathbf{M}_{rij}\left(t_{k}\right), \mathbf{\Theta}_{rij} \Big)   
\sim \mathcal{N}_C \left( \boldsymbol{b}_{rij}^{(\Theta)}, \boldsymbol{Q}_{rij}^{(\Theta)} \right),
\end{align*}
\vspace{-2em}
where
\begin{gather*}
\boldsymbol{Q}_{rij}^{(\Theta)} = \left( \boldsymbol{\Sigma}_{u,r} \right)^{-1} + \sigma_{rij}^{-2} \left[ \left\{ \sum_{k=1}^K  \mathbf{M}_{rij}\left(t_{k}\right) \mathbf{M}_{rij}^{\top}\left(t_{k}\right) \right\} \otimes \mathbf{I}_P \right]_{\mathcal{F}, \mathcal{F}}, \\
\boldsymbol{b}_{rij}^{(\Theta)} = \left( \boldsymbol{\Sigma}_{u,r} \right)^{-1} \mbox{low}\left( \mathbf{\Theta}_{ri} \right) + \sigma_{rij}^{-2} \left[ \sum_{k=1}^K  \mathbf{M}_{rij}\left(t_{k}\right) \otimes  \mathbf{Y}_{rij}\left(t_{k}\right) \right]_{\mathcal{F}}.
\end{gather*}
Here, $\mathcal{F}$ is the index mapping from the elements of $\mbox{vec} \left( \mathbf{\Theta}_{rij} \right)$ to the elements of $\mbox{low} \left( \mathbf{\Theta}_{rij} \right)$ in $\mbox{vec} \left( \mathbf{\Theta}_{rij} \right)$, i.e. $\left[ \mbox{vec} \left( \mathbf{\Theta}_{rij} \right) \right]_{\mathcal{F}} = \mbox{low} \left( \mathbf{\Theta}_{rij} \right)$.

We leverage the block structure of $\boldsymbol{Q}_{rij}^{(A)}$ and $\boldsymbol{Q}_{rij}^{(\Theta)}$ to ensure efficient computation when sampling from high-dimensional temporal dynamical and spatial mapping matrices. Detailed computation procedures, full conditional distributions of the parent random effects of $\mathbf{A}_{rij}$ and $\mathbf{\Theta}_{rij}$, as well as a discussion on the computational complexity of MCMC sampling are provided in Web Appendix B.

\subsection{Selecting the number of latent states and MVAR order}

The challenge of comparing Bayesian hierarchical models arises from the unclear definition of the number of parameters, hindering the use of classical model comparison criteria, such as AIC and BIC, both of which impose penalties based on the number of parameters. Instead, the deviance information criterion \citep[DIC;][]{spiegelhalter2002bayesian}, which offers a method to compute the effective number of parameters directly from the MCMC samples, emerges as an alternative for comparing models within the Bayesian hierarchical framework.
To address cases with latent variables (the latent EEG signals in our case), \cite{celeux2006deviance} proposed Complete DIC (cDIC) by replacing the computation of the observed likelihood in DIC with the complete likelihood.
Specifically, denote $\mathbf{Y}$ and $\mathbf{M}$ as the collection of observed EEG signals $\left\{\mathbf{Y}_{rij}(t_k)\right\}_{r,i,j,k}$ and latent EEG signals $\left\{\mathbf{M}_{rij}(t_k)\right\}_{r,i,j,k}$, respectively. Let $\log p(\mathbf{Y} , \mathbf{M} \mid \mathbf{\Lambda})$ be the complete-data log-likelihood function of $\mathbf{Y}$ and $\mathbf{M}$, where $\mathbf{\Lambda}$ is the collection of the parameters in the log-likelihood.
Then, the cDIC is defined as follows,
\begin{align}
    \mathrm{cDIC} & =-4 \mathbb{E}_{\mathbf{\Lambda}, \mathbf{M} \mid \mathbf{Y}}\big[\log p(\mathbf{Y}, \mathbf{M} \mid \mathbf{\Lambda}) \big] + 2 \mathbb{E}_{\mathbf{M} \mid \mathbf{Y}}\Big[\log p\Big( \mathbf{Y}, \mathbf{M} \ \Big\vert \  \mathbb{E}_{\mathbf{\Lambda} \mid  \mathbf{Y}, \mathbf{M}}[\mathbf{\Lambda}] \Big) \Big].
    \label{equ:DIC}
\end{align}
We can approximate \eqref{equ:DIC} with
\begin{align*}
    \widetilde{\mathrm{cDIC}} & =-\frac{4}{L} \sum_{l=1}^L \log p(\mathbf{Y}, \mathbf{M}^{(l)} \mid \mathbf{\Lambda}^{(l)})+ \frac{2}{L} \sum_{l=1}^L \log p\left(\mathbf{Y}, \mathbf{M}^{(l)} \mid \widehat{\mathbf{\Lambda}}^{(l)} \right),
\end{align*}
where $L$ is the total number of MCMC iterations (after burn-in and thinning), $\mathbf{M}^{(l)}$ and $\mathbf{\Lambda}^{(l)}$ are the MCMC samples at the $l$-th iteration, and $\widehat{\mathbf{\Lambda}}^{(l)}$ is the posterior mean of the distribution $p( \mathbf{\Lambda} \mid \mathbf{Y}, \mathbf{M}^{(l)})$. 
A simulation study in Section \ref{sec:sensitivity} demonstrated that the cDIC is effective for selecting the number of latent states $Q$. 
Several alternative model selection methods based on DIC, along with a simulation comparing cDIC with these methods, are presented in Web Appendix C.5. The comparison results show the superiority of cDIC, demonstrating its effectiveness in preventing overfitting. 
\section{Simulation Studies}
\label{sec:simu_study}

\subsection{Simulation design}
We conducted extensive simulation studies to assess the finite sample performance of our proposed method. To mimic the real study, we simulated data using parameters close to the parameters obtained from the real data analysis described in Section \ref{sec:real}.
We used 54 channels ($P=54$) of resting-state EEG signals measured at $125$ Hz for each subject, generated from two underlying latent neuronal states ($Q=2$), with observed EEG signals divided into 30 equal-length time segments (2 seconds per segment). We considered two subgroups, corresponding to Cases and Controls, with $n_1$ and $n_2$ subjects, respectively. 

We considered two scenarios with balanced and imbalanced sample sizes: Scenario I, with $n_1 = n_2 = 75$, and Scenario II, with $n_1 = 75, n_2 = 20$. For both scenarios, we set $\mathbf{\Theta}_{1} = \mathbf{\Theta}_{2} = \mathbf{\Theta}^{0}$, where $\mathbf{\Theta}^{0}$ is the first two columns of the posterior mean of $\mathbf{\Theta}$ (population level) from the real data analysis. 
We normalize $\mathbf{\Theta}^{0}$ by rescaling, setting the first element in the first row to $0.5$.
Then $\mathbf{\Theta}_{rij}$ and $\mathbf{\Theta}_{ri}$ are generated using \eqref{equ:Theta_rij} and \eqref{equ:Theta_ri}. 
In the MVAR model \eqref{model1_eq2}, the order $m$ can be any integer suitable for the real data, and we use a second-order MVAR ($m=2$) here. For both scenarios, we set $\mathbf{A}_{1} = (\mathbf{A}_{21}, \mathbf{A}_{22}) = (\mbox{diag}(0.95, 0.9), \mbox{diag}(-0.55, -0.5))$ and $\mathbf{A}_{2} = (\mathbf{A}_{11}, \mathbf{A}_{12}) = (\mathbf{I}, -0.6 \cdot \mathbf{I})$, and $\mathbf{A}_{rij}$ and $\mathbf{A}_{ri}$ are generated using \eqref{equ:Arij} and \eqref{equ:Ari}. 
Detailed information on the assignments of variance components for the spatial mapping matrices and temporal dynamical matrices is provided in Web Appendix C.1. Finally, we assume $\sigma_{rij} = \sigma_{r}$ and set $\sigma_{1}^2 = \sigma_{2}^2 = 0.16$.

\vspace{-0.5em}

\subsection{Simulation results}
We generated 100 replicates for each sample size scenario. The MCMC computation consisted of 7,500 iterations, with the first 2,500 iterations treated as burn-in and a thinning rate of 10. 
For parameter estimation, Web Figure S.1 presents the biases of the group-, subject-, and segment-level temporal dynamical matrices $\mathbf{A}$ for each element at each time lag (i.e., $h=1,2$) for both the case and control groups ($r=1,2$) under the two sample size scenarios.
The biases of the group/subject-level $\mathbf{A}$ are smaller than the segment-level $\mathbf{A}$. Meanwhile, biases of $\mathbf{A}$ at all levels are small, indicating that the model can accurately recover the temporal dynamics of the latent sources.
Web Figure S.2 shows the relative estimation errors (REEs) of the group-, subject- and segment-level spatial mapping matrices $\mathbf{\Theta}$. The detailed definition of the REE is given in the Web Appendix C.1. 
The REEs in segment-level $\mathbf{\Theta}$ are larger than the REEs in group- and subject-level $\mathbf{\Theta}$. The REEs at all levels are small, indicating that the model can accurately recover the spatial mapping relationships between the latent and observed states.
Furthermore, biases and REEs for group-level $\mathbf{A}$ and $\mathbf{\Theta}$ matrices decrease with larger sample sizes, indicating that our model can accurately estimate the true parameter values for large samples.

\begin{figure}[!ht]
    \begin{subfigure}{0.52\textwidth}
        \begin{flushright}
        \includegraphics[width=0.85\linewidth]{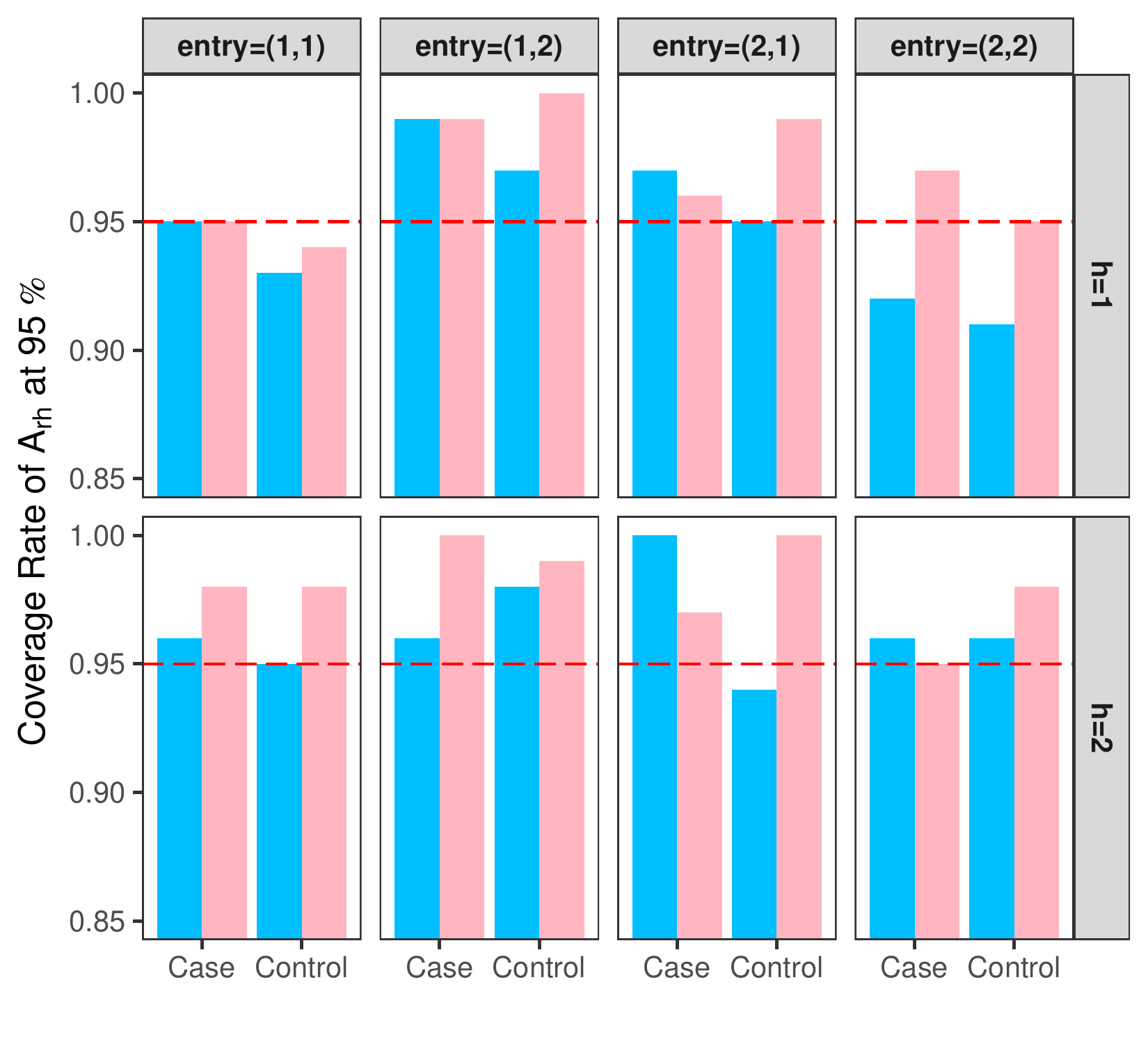}
        \end{flushright}
        \vspace{-1.5em}
        \caption{Coverage rate for $\mathbf{A}_{rh}(q_1, q_2)$}
    \end{subfigure}
     \vspace{1em}
    \begin{subfigure}{0.4\textwidth}
        \begin{flushright}
        \includegraphics[width=0.9\linewidth]{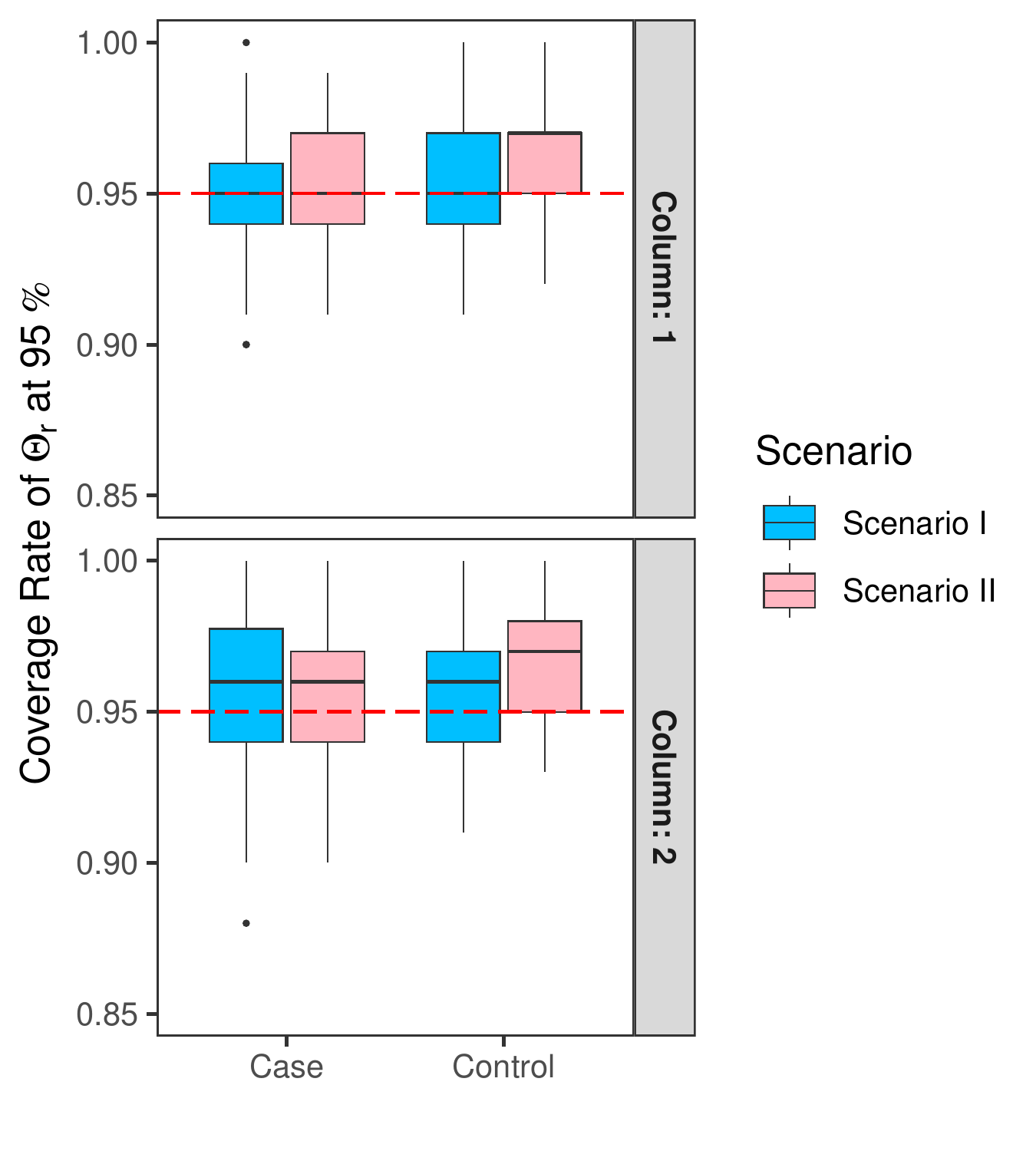}
        \end{flushright}
        \vspace{-1.5em}
        \caption{Coverage rate for $\mathbf{\Theta}_{r}(\cdot, q)$}
    \end{subfigure}
    \begin{subfigure}{0.9\textwidth}
        \begin{flushright}
        \includegraphics[width=0.9\linewidth]{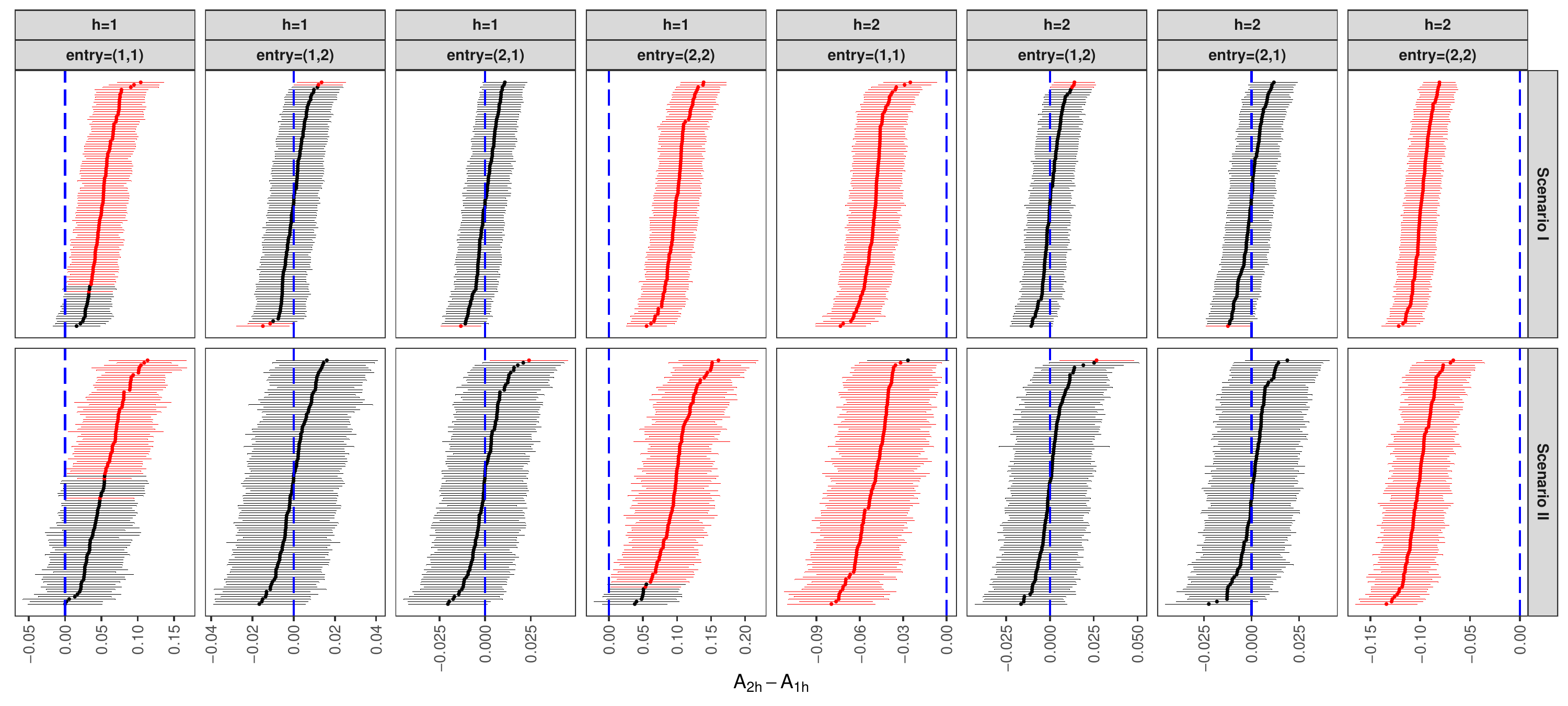}
        \end{flushright}
        \vspace{-1.5em}
        \caption{Inference on group difference $\mathbf{A}_{2h}(q_1, q_2) - \mathbf{A}_{1h}(q_1, q_2)$}
    \end{subfigure}
    \caption{(Panel a): Coverage rate for entries $(q_1, q_2) \in \{1,2\}\times\{1,2\}$ in the group-level temporal dynamical matrices $\mathbf{A}_{rh}$ $(r=\text{1:case}, \text{2:control}, h=1,2)$ for $100$ simulation replicates. 
    (Panel b): Coverage rate for entries in the group-level spatial mapping matrices $\mathbf{\Theta}_{r}$ for $100$ simulation replicates. Entries in each column are summarized by a box.  
    (Panel c): The posterior means (dots) and $95\%$ credible intervals (bars) of the group-level differences $\mathbf{A}_{2 h}(q_1, q_2) - \mathbf{A}_{1 h}(q_1, q_2)$. 
    The blue dashed lines correspond to $x = 0$ and the red error bars are $95\%$ credible intervals that fail to cover $0$. }
  	\label{img:simu_results}
\end{figure}

For inference, we calculate $95\%$ credible intervals using a normal approximation (i.e., $\text{posterior mean} \pm 1.96 \times \text{posterior s.d.}$) for group/subject-levels and each element of the temporal dynamical and spatial mapping matrices. Figure \ref{img:simu_results} (a) and (b) depict the coverage rate (CR) of the group-level $\mathbf{A}_{rh}$ and $\mathbf{\Theta}_r$ ($h=1,2, r=1,2$), respectively. CR is defined as the total number of credible intervals that cover the true values for the 100 replicates. 
From the figure, we observe the group-level $\mathbf{A}_{rh}$ and $\mathbf{\Theta}_{r}$ exhibit coverage rates of approximately $95\%$, aligning closely with the nominal level.
Figure \ref{img:simu_results} (c) shows the posterior mean (dots) and $95\%$ credible intervals (bars) for the group differences $\mathbf{A}_{2h} - \mathbf{A}_{1h}$ for $h=1,2$. The blue dashed lines indicate $x = 0$ and the red error bars are $95\%$ credible intervals that fail to cover $0$.
The figure reveals that in scenario I, most between-group differences (at the four diagonal values of $\mathbf{A}_{r1}$ and $\mathbf{A}_{r2}$) are detected. In scenario II (with a smaller sample size), one of the between-group differences with a smaller effect size is not well-identified. This suggests that either a sufficiently large sample size or a substantial group difference is necessary for detecting significant group differences.
For both scenarios, only a few false positives are observed at the four off-diagonal values of $\mathbf{A}_{r1}$ and $\mathbf{A}_{r2}$, where no group difference is present.
Moreover, Web Figure S.3 demonstrates that the CR for the subject-level temporal dynamical matrices are concentrated around $95\%$, the CR for the subject-level spatial mapping matrices are concentrated around $93\% \sim 94\%$. Combined with the results presented in Figure \ref{img:simu_results}, we conclude that our method provides accurate inference results at both the group- and subject-levels. Notably, the inference of segment-level parameters is not of interest, and only the posterior means of the segment-level parameters are retained in the model output to save space.
Additional details regarding the MCMC diagnostics are presented in Web Appendix C.2, demonstrating the good mixing and convergence of the MCMC posteriors using the proposed methods.
In Web Appendix C.6, we also conducted a comparison between our proposed RESSM and the latent state-space models (SSM) by \cite{wang2022latent}. In their model, subject-level heterogeneity is addressed by fixed effects only and no segment-level heterogeneity is considered. 
The results demonstrate that when heterogeneity is present, our methods yield significantly improved coverage rates in model inference and reduce the probabilities of false positives for group comparisons.

\subsection{Model selection and sensitivity analysis}
\label{sec:sensitivity}
We conducted simulations to investigate the performance of using cDIC for selecting the number of latent states $Q$ and its robustness under misspecification of the MVAR order $m$. We used the same simulation settings except that we set $Q=3$ and simulated a single group (the Control group with 75 subjects). In this scenario, our true model is $Q=3$ and $m=2$. We varied the number of latent states $Q$ from $2$ to $4$ and varied $m$ from $1$ to $3$.

We conducted MCMC analysis for the 9 models (1 true, 8 misspecified) and computed the cDIC values for each model. The results can be found in Web Figure S.4. We observed that cDIC can choose the correct model in terms of the number of latent states $Q$, as the true model has the smallest cDIC values among $Q=2,3,4$ even when the MVAR order $m$ is misspecified.
On the other hand, the cDIC is not sensitive to varying the MVAR order $m$. 
In addition, the $95\%$ confidence intervals of the correlation between the posterior means of $\mathbf{M}_{rijk}(t)$ for the true model and misspecified models (with $Q=3$, $m=1$ and $3$) for all subjects, time segments, and latent states are $(0.993, 1.000)$ and $(0.999, 1.000)$, respectively. Both cDIC and correlation metrics suggest that the model's performance is robust and not significantly affected by the MVAR order $m$.

\section{Application to EMBARC}
\label{sec:real}

We applied the proposed model to the EMBARC study, a large multi-site randomized clinical trial \citep{trivedi2016establishing}. Participants in the studies are recruited at four study centers: the University of Texas Southwestern Medical Center (TX), Columbia University/Stony Brook (CU), Massachusetts General Hospital (MG), and the University of Michigan, University of Pittsburgh, and McLean Hospital (UM). The data acquisition involved four two-minute blocks, recorded in a specific sequence: eyes open, eyes closed, eyes closed, and eyes open.
Our data analysis included pre-treatment resting-state EEG recordings of subjects diagnosed with major depressive disorder (MDD) and a control group comprising healthy subjects from sites TX and CU (sites MG and UM were excluded due to inadequate participant numbers). As a result, the data is divided into four sub-groups. Site TX comprises 75 subjects in the MDD group and ten subjects in the Control group, while CU consists of 53 subjects in the MDD group and nine subjects in the Control group.

To account for site and equipment differences of the EEG recording, we used the EEG preprocessing procedure proposed in \cite{yang2023learning}, where a site-specific automatic EEG data preprocessing pipeline is employed to minimize the biases and achieve standardization of the EEG data. 
The detailed preprocessing procedure can be found in Web Appendix D.1.
After preprocessing, we included the 54 EEG electrodes common to both the TX and CU sites. 
We also down-sampled the EEG signals to a rate of $125$ Hz per second. 
For our analysis, we used the first 40 segments after removing the noisy segments with high-amplitude artifacts (see Web Appendix D.1 preprocessing step (7)) from the first 2-minute block characterized by the `eye-open' condition. 
For the MCMC computation, we used 10,000 iterations, with 3,000 iterations as burn-in and applied thinning by retaining every $10$-th sample. Recall that in Section \ref{sec:sensitivity}, we demonstrated that the model's performance is minimally impacted by the specification of the MVAR order $m$. Therefore, for the sake of simplicity, we set $m=1$.
To estimate the latent states $Q$, we fit the model using $Q$ from 2 to 6. The cDIC values decreased as $Q$ increased. However, the difference in cDIC between $Q=5$ and $Q=6$ was small. Considering  interpretability and computational efficiency, we selected $Q=5$.

\begin{figure}[h]
    \begin{subfigure}{0.495\textwidth}
        \centering
        \includegraphics[width=1.0\linewidth]{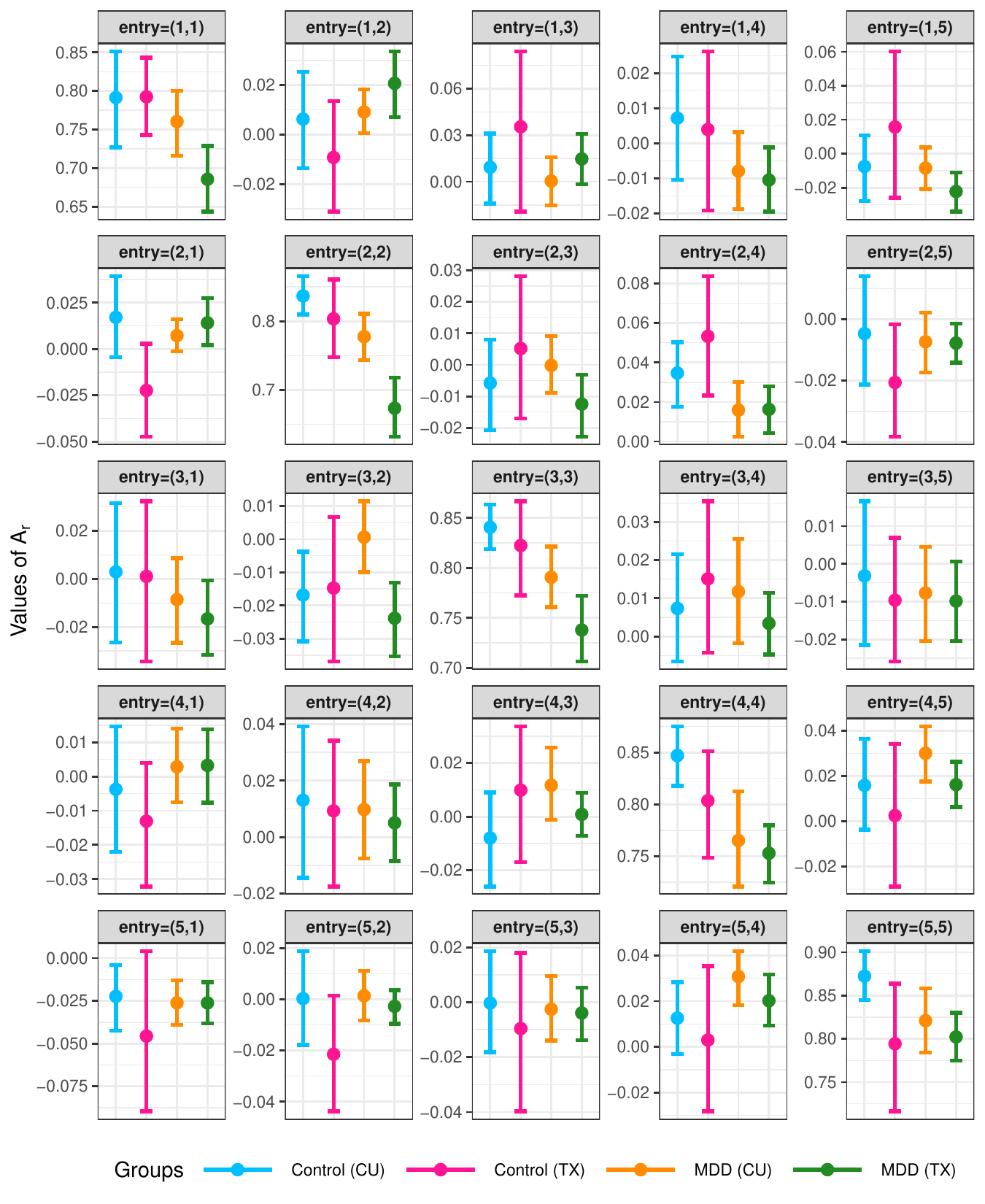}
    \end{subfigure}
    \begin{subfigure}{0.495\textwidth}
        \centering
        \includegraphics[width=1.0\linewidth]{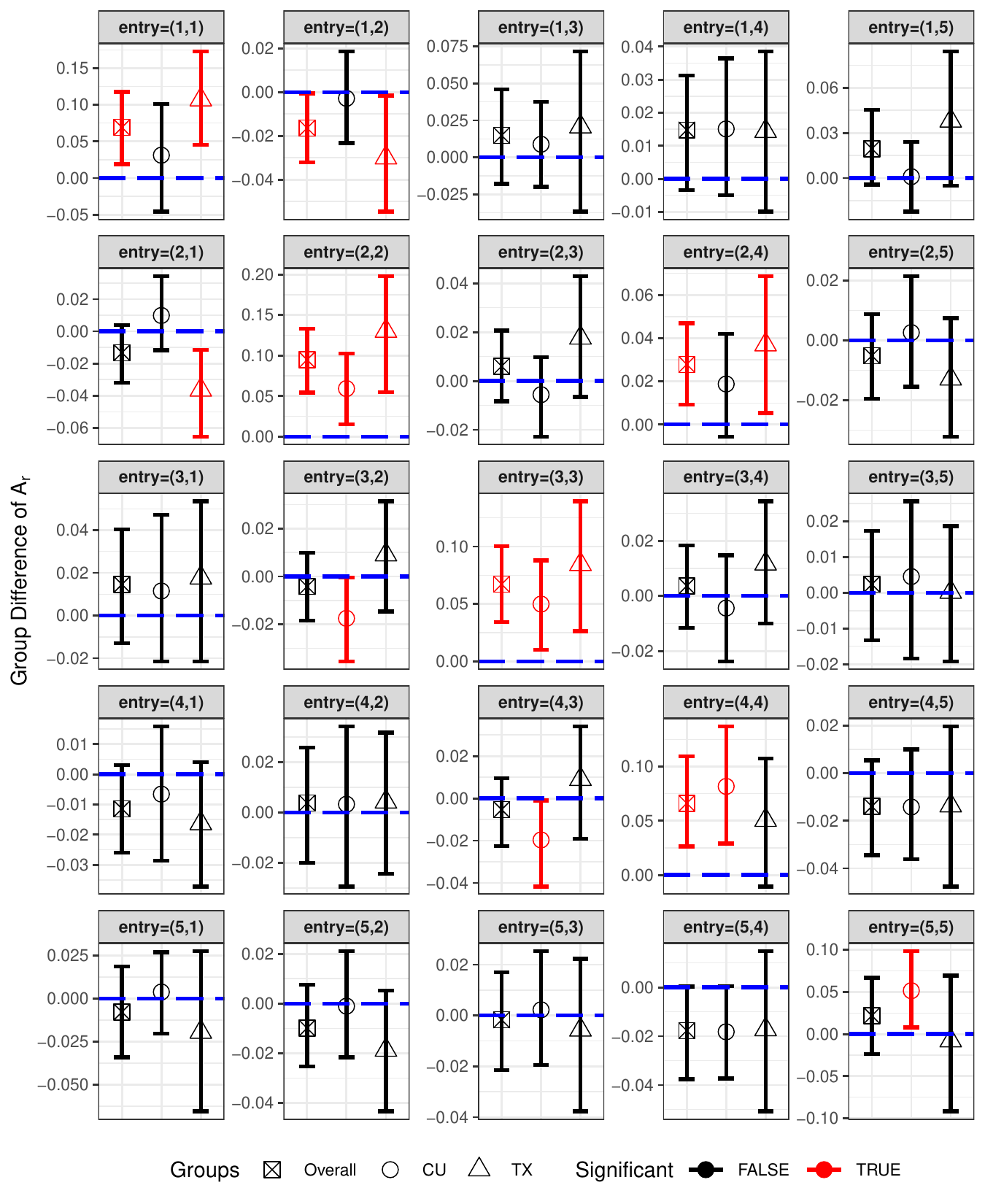}
    \end{subfigure}
    \caption{ Left Panel: the posterior means (dots) and $95\%$ credible intervals (bars) for each entry of the group-level temporal dynamical matrices $\mathbf{A}_{r}(q_1, q_2)$, ($r=1,2, 3, 4, \ q_1, q_2 \in \{1,2,3,4,5\}$) from the analysis EMBARC data; Right Panel: the posterior means and $95\%$ credible intervals of the group differences of the temporal dynamical matrices from the analysis of EMBARC data. The dashed blue lines indicate $y = 0$ and the red error bars are credible intervals that fail to cover $0$ (significant group differences).}
  	\label{img:real_A}
\end{figure}

The first goal is to compare the differences between MDD and controls regarding the temporal dynamical matrices $\mathbf{A}$ (which has dimension $5 \times 5$). The posterior mean (dots) and $95\%$ credible intervals (bars) of the group level temporal dynamical matrices $\mathbf{A}_{r}$, $r=1, 2, 3, 4$ are shown in the left panel of Figure \ref{img:real_A}.  
The $95\%$ credible intervals of $\mathbf{A}_{r}$ are constructed using the $2.5\%$ and $97.5\%$ quantiles of the MCMC samples.
We found that the control groups for both sites have larger diagonal values of $\mathbf{A}_{r}$ than the MDD groups. The off-diagonal components were mostly close to zero, implying little interaction between latent states. Furthermore, we constructed $95\%$ credible intervals on all elements of the group differences of the temporal dynamical matrices. Specifically, we calculated the differences $\mathbf{A}_{1} - \mathbf{A}_{3}$ to assess the contrast between the control and MDD groups at site CU, $\mathbf{A}_{2} - \mathbf{A}_{4}$ to assess the contrast between the control and MDD groups at site TX, and $(\mathbf{A}_{1} + \mathbf{A}_{2} - \mathbf{A}_{3} - \mathbf{A}_{4}) / 2$ to measure the overall difference. Results in the right panel of Figure \ref{img:real_A} 
show that the first four diagonal values of matrix $\mathbf{A}_{r}$ exhibit statistically significant overall differences between the control and MDD groups. This suggests that the autocorrelation of the latent states is stronger in the control groups compared to the MDD groups.
The MCMC diagnostics for $\mathbf{A}_r$ are presented in Web Figure S.9, demonstrating adequate mixing and convergence of the MCMC posteriors, thereby validating our inference results.

\begin{figure}[h]
    \begin{subfigure}{0.3\textwidth}
        \centering
        \includegraphics[width=.9\linewidth]{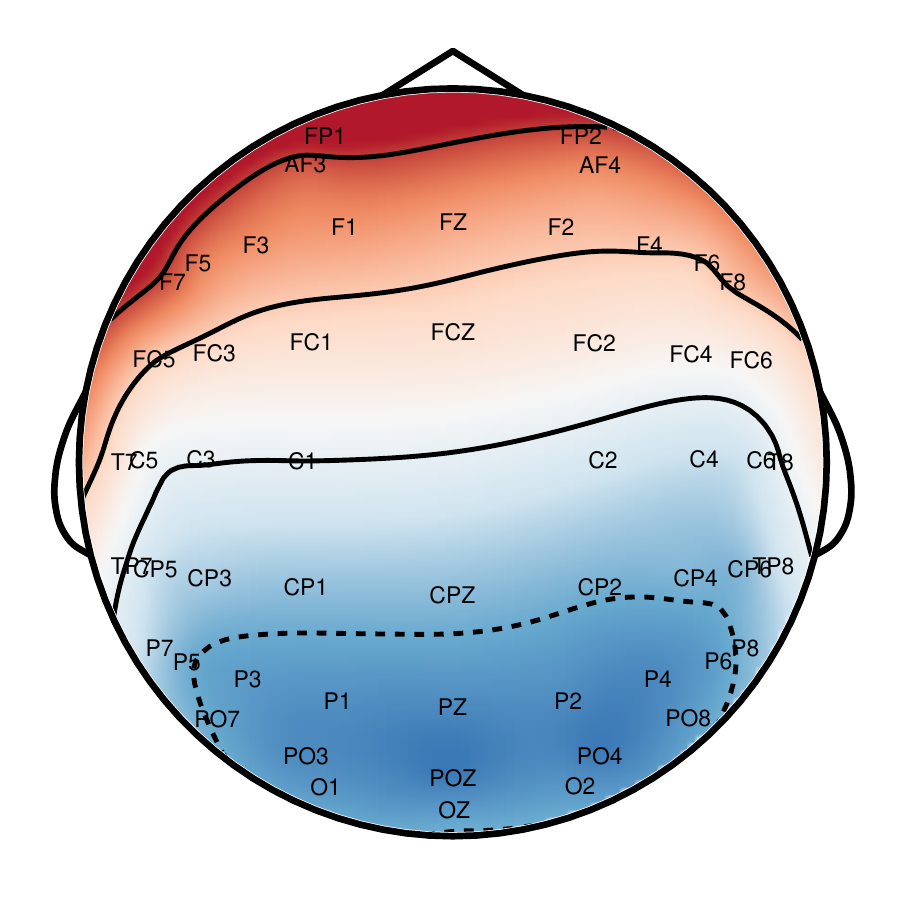}
    \end{subfigure}
    \begin{subfigure}{0.3\textwidth}
    \centering
    \includegraphics[width=.9\linewidth]{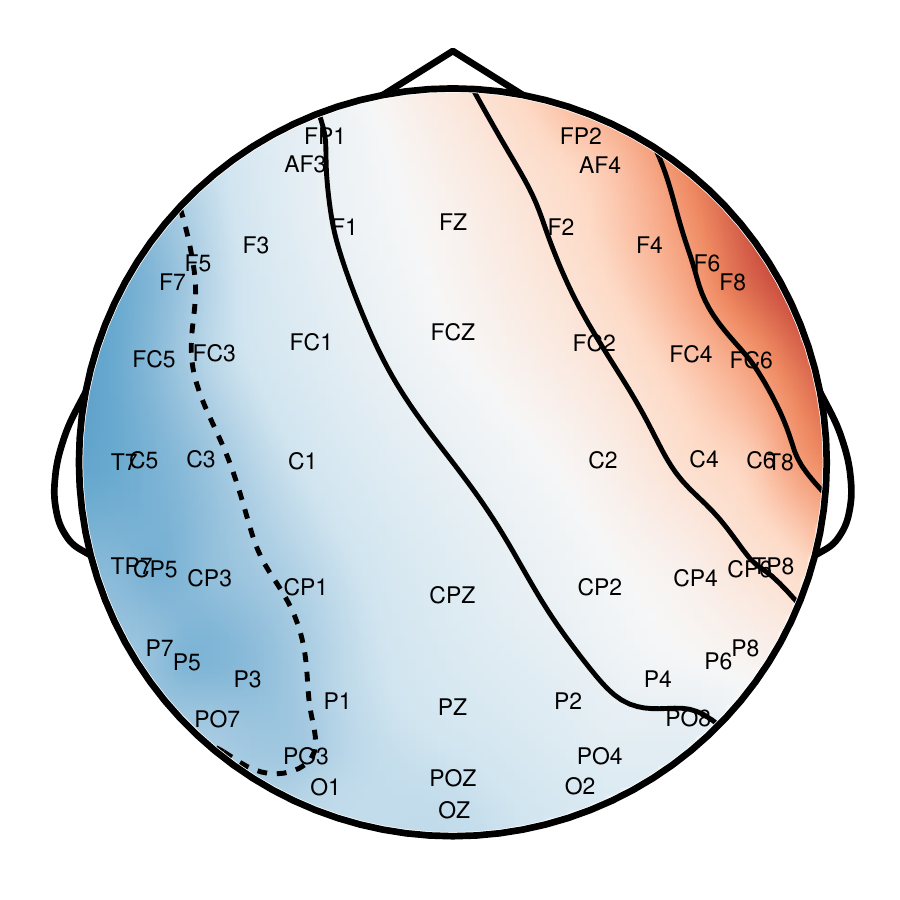}
  \end{subfigure}
    \begin{subfigure}{0.3\textwidth}
    \centering
    \includegraphics[width=.9\linewidth]{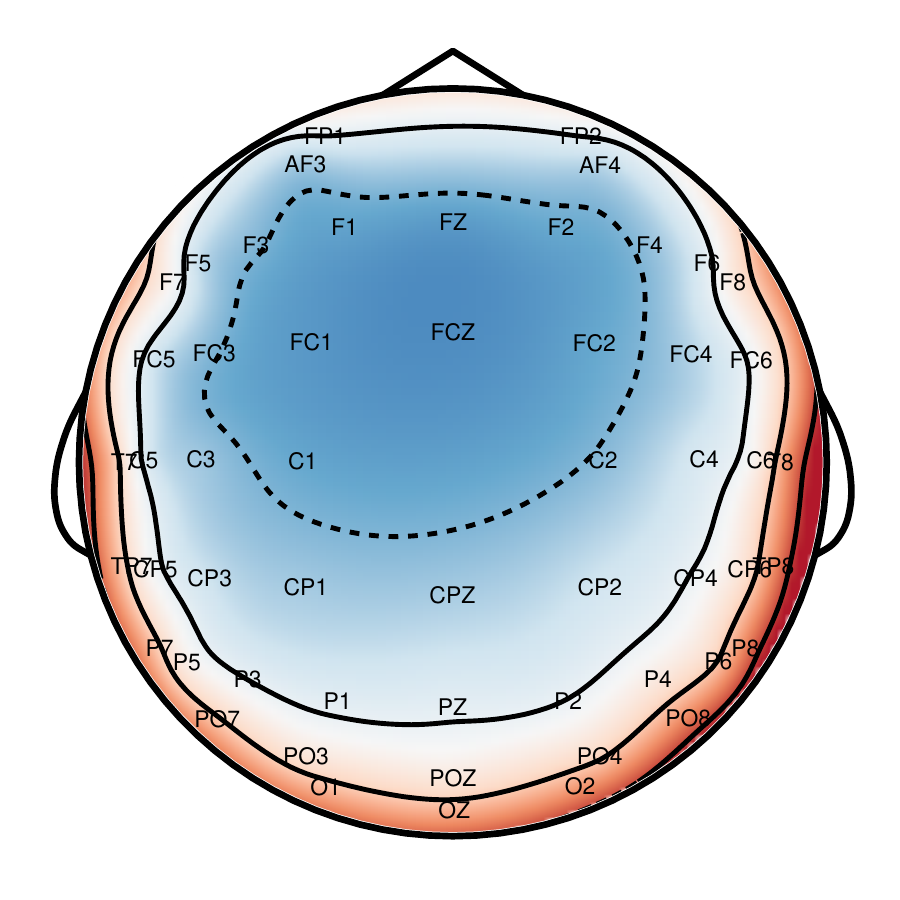}
  \end{subfigure}
      \begin{subfigure}{0.46\textwidth}
    \centering
    \includegraphics[width=.6\linewidth]{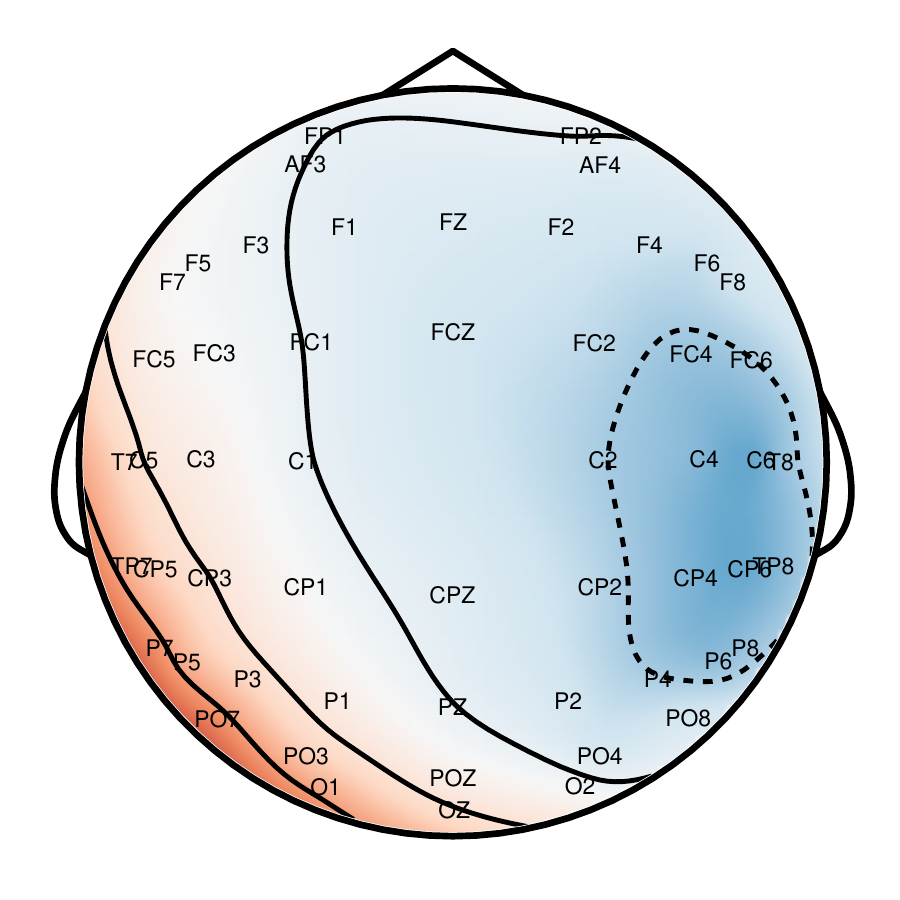}
  \end{subfigure}
      \begin{subfigure}{0.53\textwidth}
    \centering
    \includegraphics[width=.63\linewidth]{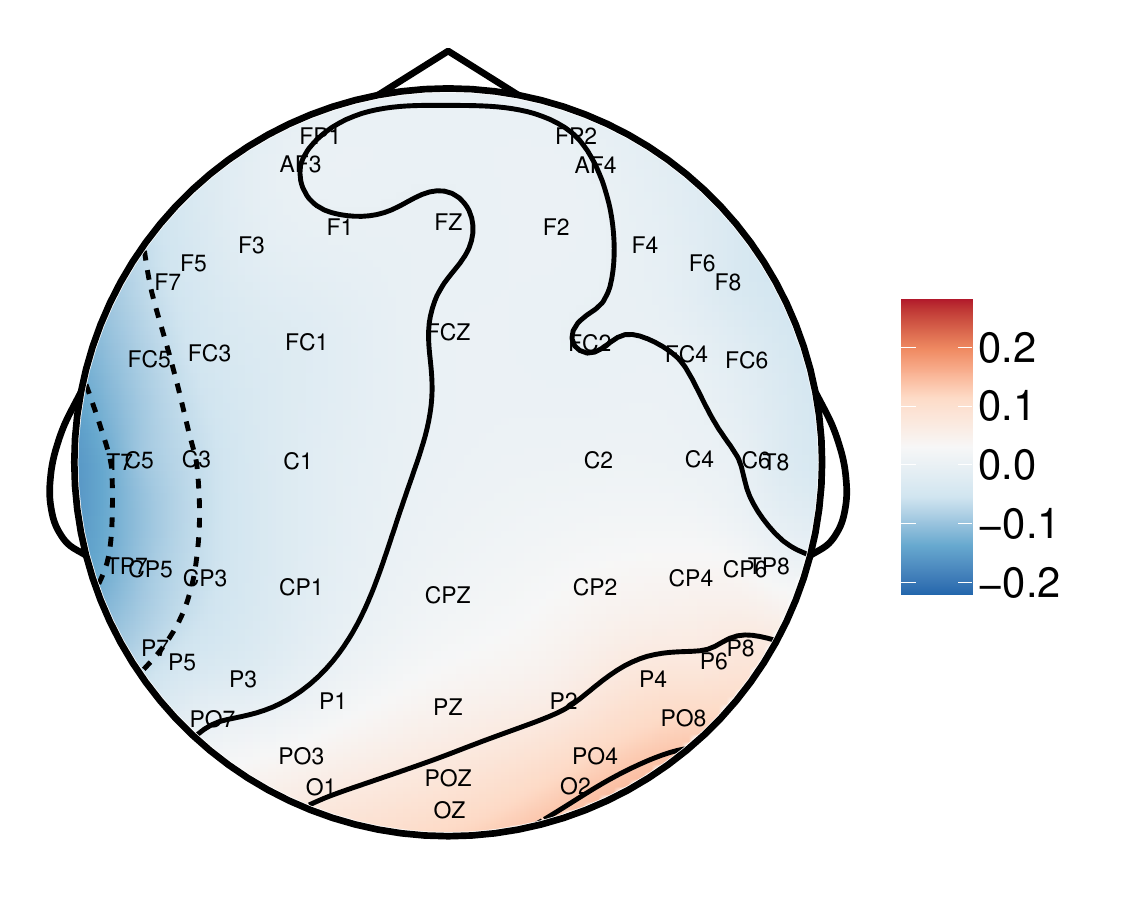}
  \end{subfigure}
  \vspace{-0.25em}
  	\caption{Topographies of the posterior means of the population-level spatial mapping matrix $\mathbf{\Theta}$ obtained from the analysis of EMBARC data. Each topography figure represents one $\mathbf{\Theta}$  column corresponding to a latent state.}
  	\label{img:topo_theta}
\end{figure}

Next, we compared the spatial mapping matrices $\mathbf{\Theta}$ (with dimension $54 \times 5$) between  MDD and controls. We computed $95\%$ credible intervals for all elements of the group differences in the mapping matrices. We used the Bonferroni correction on the $270$ credible intervals to account for multiple comparisons at $0.05$. Our analysis revealed no statistically significant differences in the spatial mapping matrices between  MDD and control at the individual site level (CU, TX) and with two sites combined. Figure \ref{img:topo_theta} shows the topographies of the posterior mean of the population level spatial mapping matrix $\mathbf{\Theta}$. Each topography graph represents a column of $\mathbf{\Theta}$, which corresponds to a latent state. The figure shows that each latent state corresponds to specific brain regions. For instance, the first latent state primarily corresponds to the pre-frontal and frontal lobes, the second latent state to the frontal lobes and central line of the right hemisphere of the brain, the third latent state to the temporal lobe, and the fourth and fifth latent states to the parietal and occipital lobes in the left and right hemispheres of the brain, respectively. This organized pattern of topographies demonstrates the interpretability of the latent states.

\begin{figure}[h]
    \begin{subfigure}{0.49\textwidth}
        \centering
        \includegraphics[width=\linewidth]{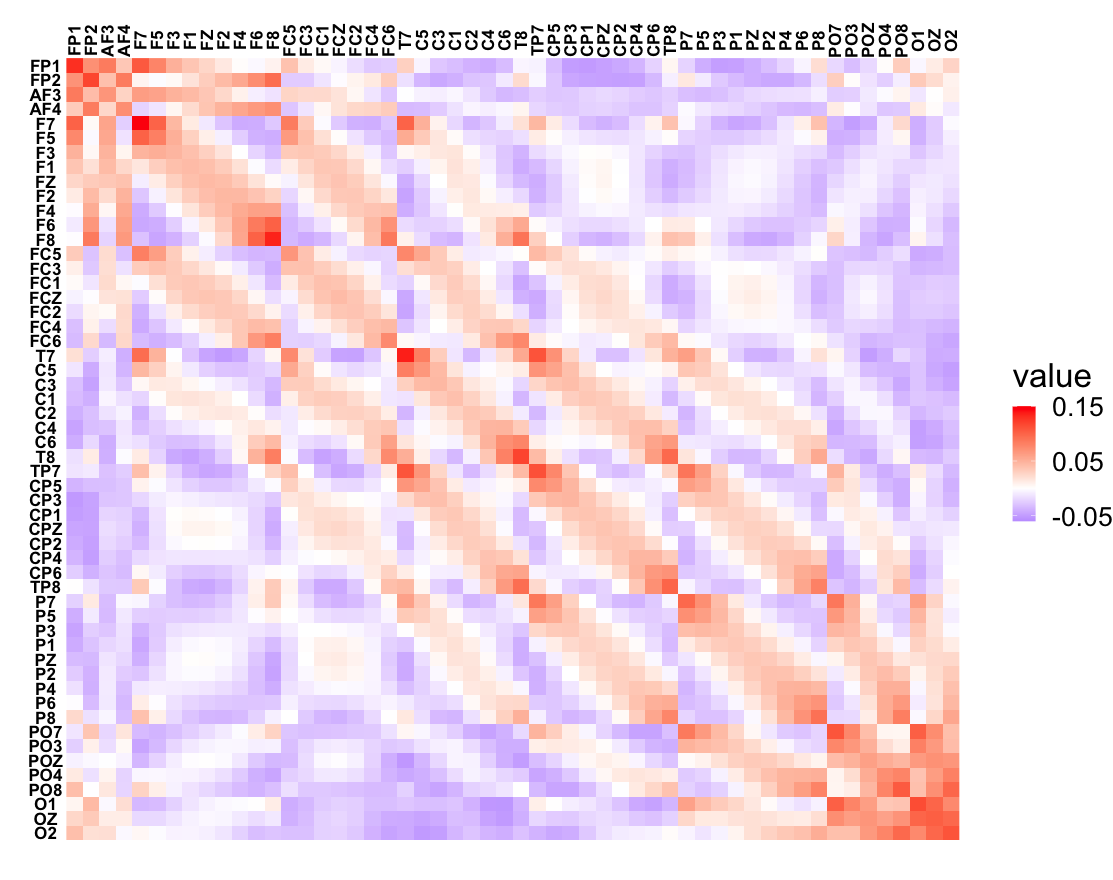}
        \caption{MDD}
    \end{subfigure}
    \begin{subfigure}{0.49\textwidth}
    \centering
    \includegraphics[width=1\linewidth]{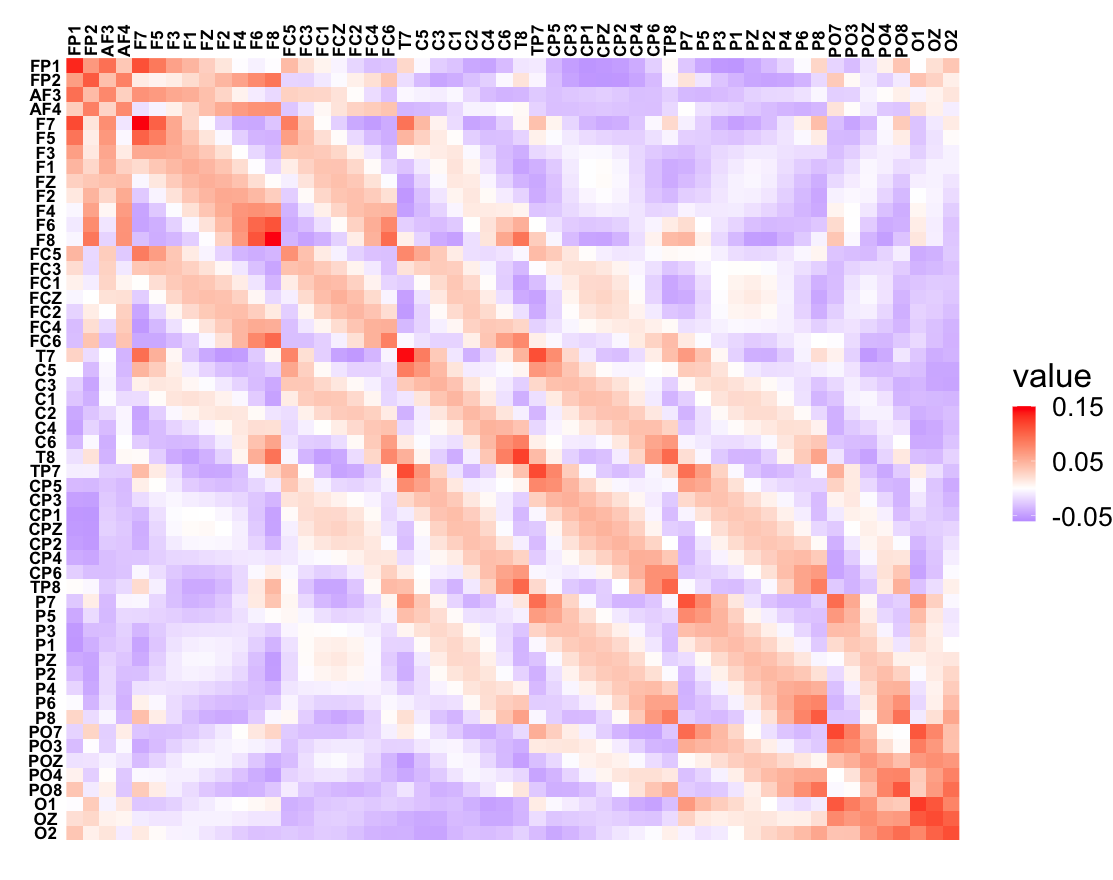}
    \caption{Control}
  \end{subfigure}
      \begin{subfigure}{1.0\textwidth}
    \centering
    \includegraphics[width=.49\linewidth]{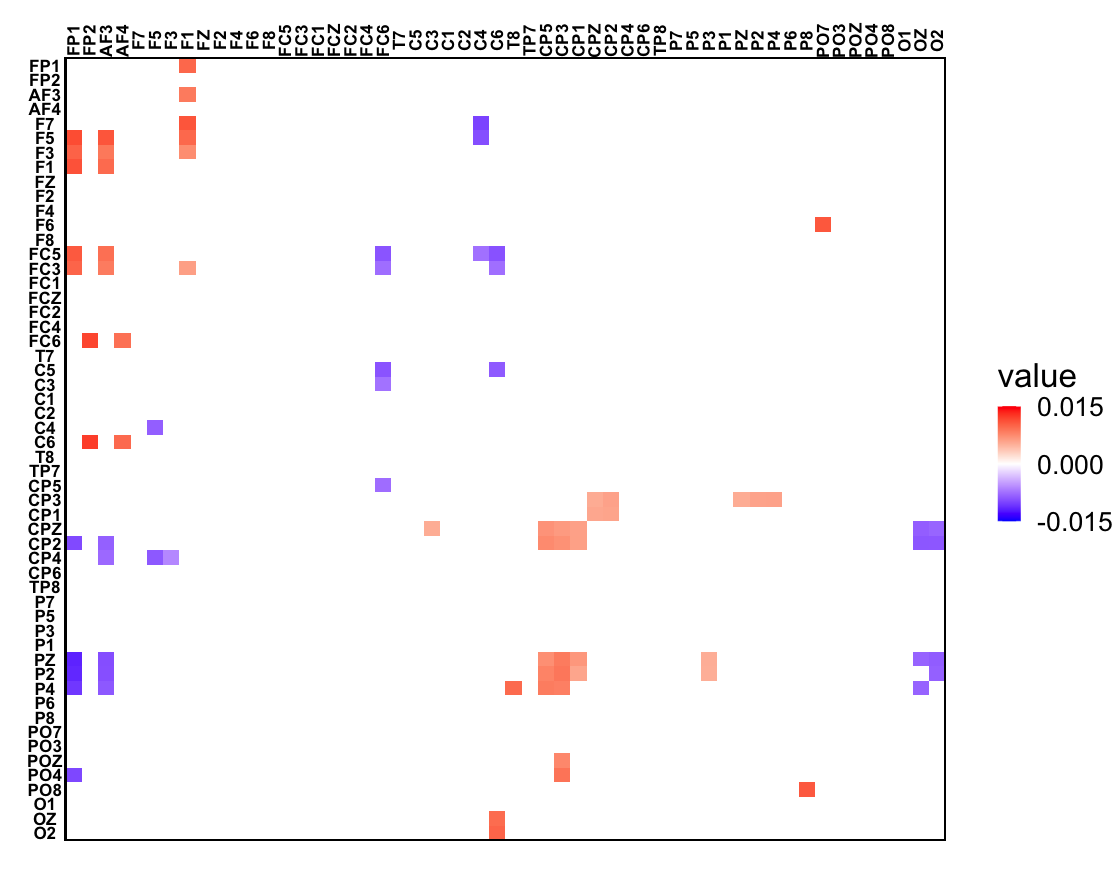}
    \caption{Control - MDD}
  \end{subfigure}
  	\caption{Panel (a) and (b): The posterior means of the directional connectivity matrices for the MDD and control groups averaged across two sites in the EMBARC study. 
   Panel (c): The posterior mean differences of the directional connectivities between MDD and control groups. Only statistically significant elements (i.e., $95\%$ credible intervals do not include zero) are shown.
   }
  	\label{img:DCM}
\end{figure}

Additionally, the posterior means of the group-level directional connectivity matrices (refer to a group-level version of \eqref{equ:directional_connectivity_matrix}) for the MDD and control groups (averaged across the two sites) and the difference of the directional connectivity matrices between the Control and MDD groups are shown in Figure \ref{img:DCM}.  
The 54 EEG nodes in the directional connectivity matrices are 
arranged in a sequential manner from the frontal to the posterior regions. It is interesting to note that the connectivity matrices exhibit a block-banded nature such that positive (excitatory) connectivities and negative (inhibitory) connectivities are organized into several bands. 
Positive connectivities tend to occur more frequently between EEG nodes that are closer in distance. On the other hand, negative connectivities are more commonly observed between EEG nodes located on opposite sides of the brain (left and right) or between two distinct brain regions, such as between the frontal/fronto-central and posterior/parieto-occipital regions. Furthermore, the presence of symmetry in the directional connectivity matrices indicate the connections are mostly bi-directional.
These general patterns of the connectivities are similar between MDD and controls.  However, we observed some significant differences in the connections between several brain regions. For instance, in the frontal region, there is a large positive difference (control greater than MDD) in the directed connection from F1 to F7; while the directed connection from FP1 to PZ shows a large negative difference.
These differential connections indicate distinct patterns of resting-state neuronal connections  in  MDD patients compared to the controls. Further discussion on directional connectivities can be found in Web Appendix D.2.

\begin{figure}[h]
    \begin{subfigure}{0.325\textwidth}
        \centering
        \includegraphics[width=.99\linewidth]{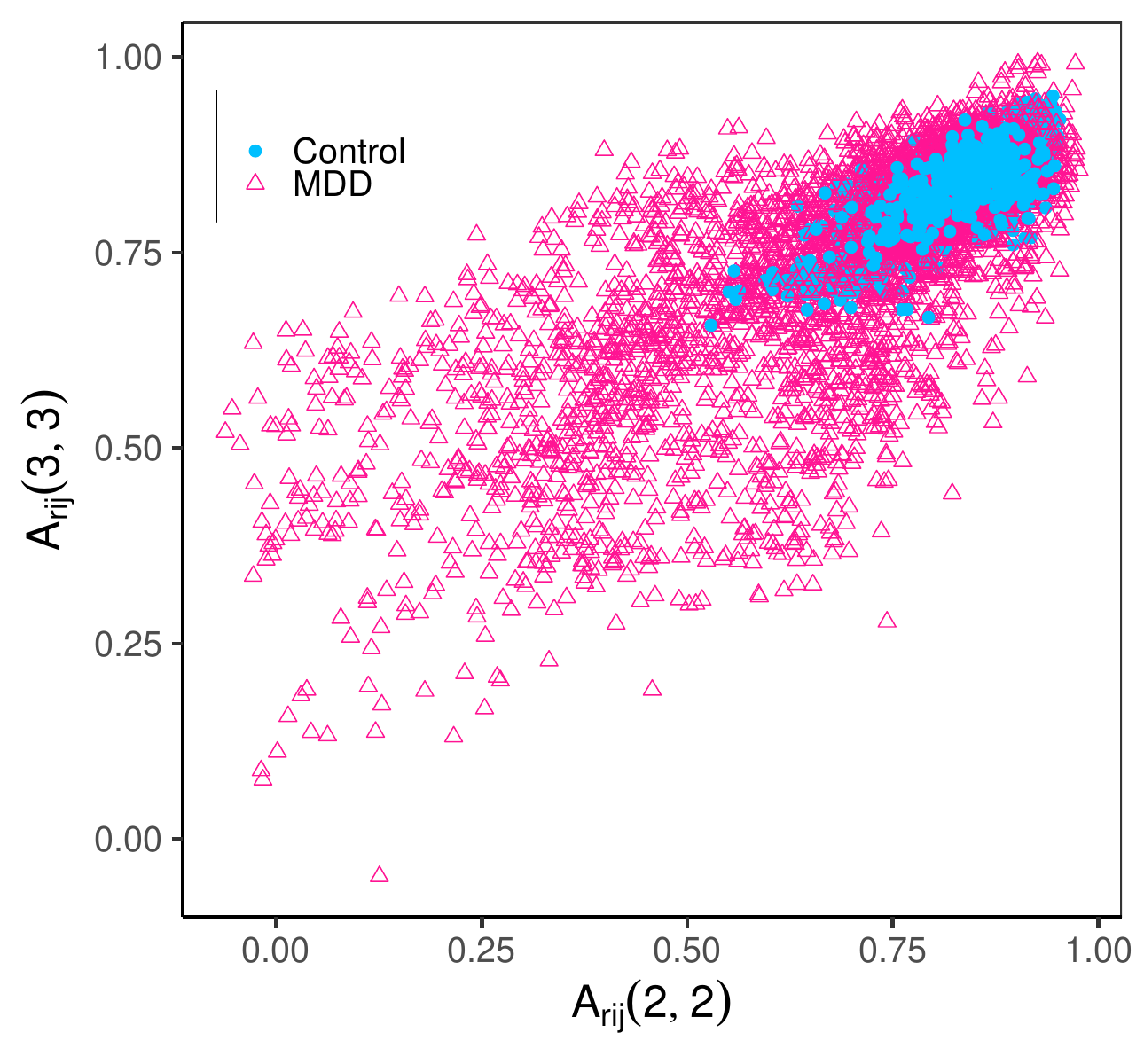}
        \caption{All participants}
    \end{subfigure}
    \begin{subfigure}{0.325\textwidth}
    \centering
    \includegraphics[width=.99\linewidth]{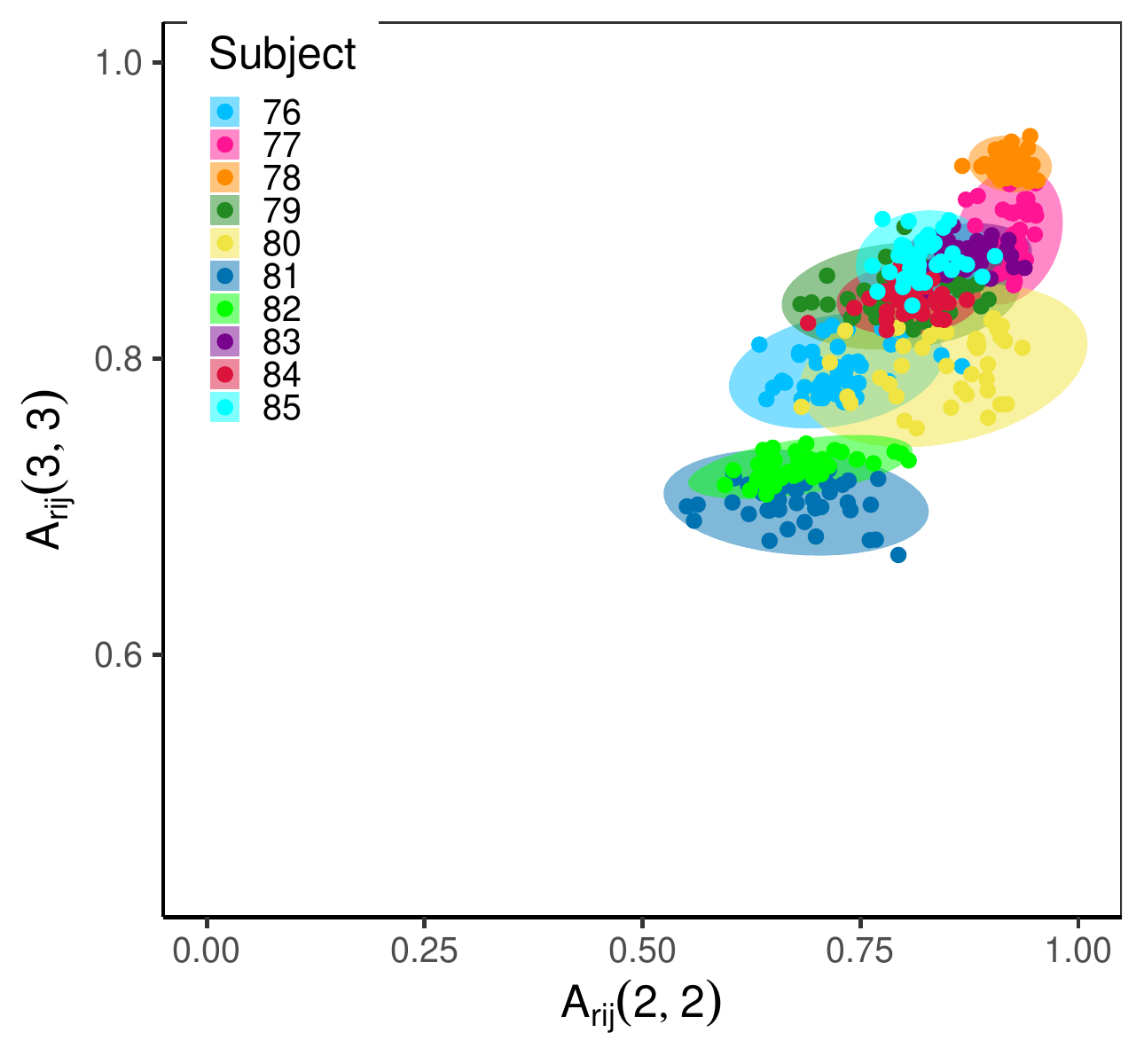}
    \caption{10 Control participants}
  \end{subfigure}
    \begin{subfigure}{0.325\textwidth}
    \centering
    \includegraphics[width=.99\linewidth]{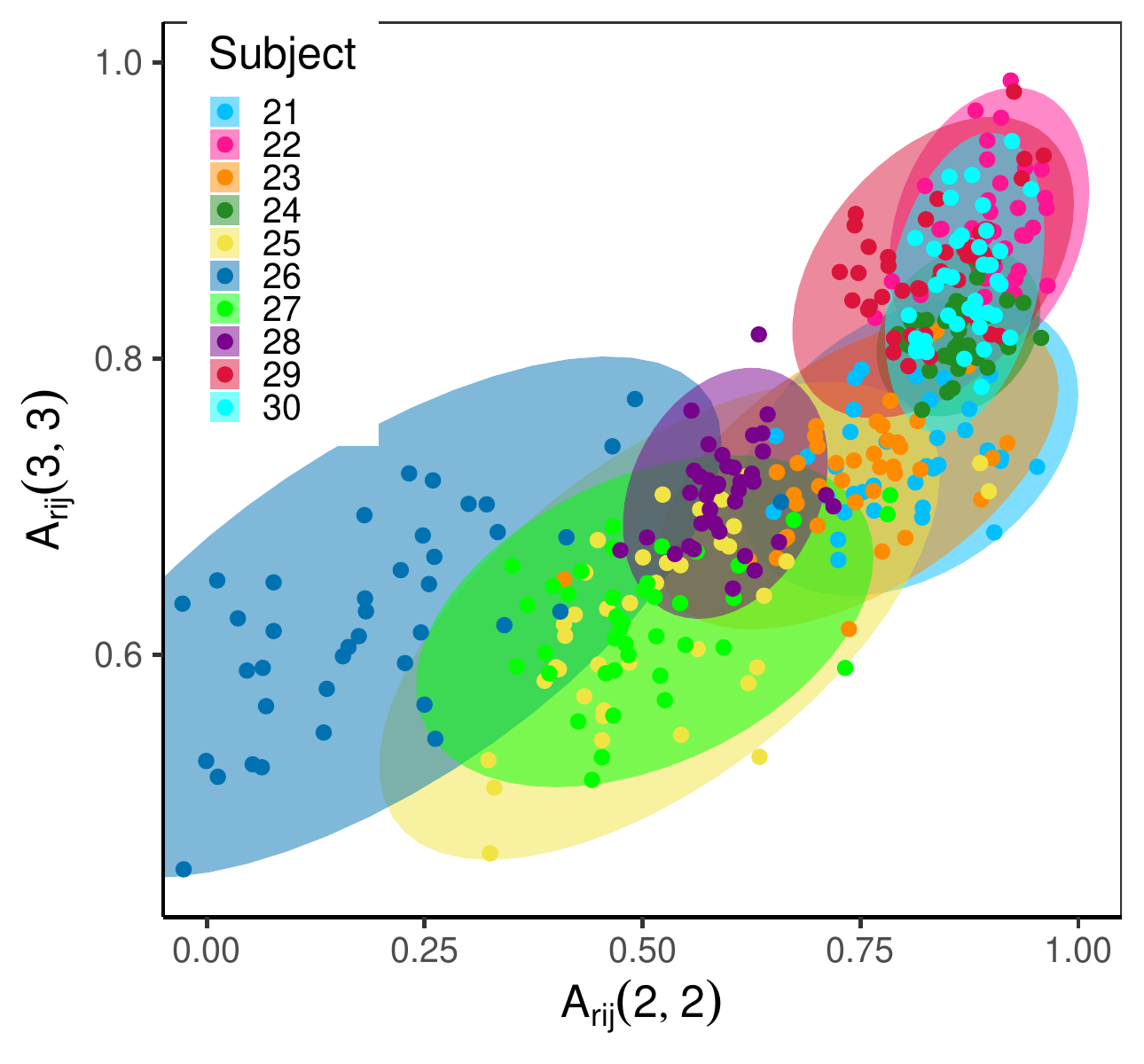}
    \caption{10 MDD participants}
  \end{subfigure}
  \vspace{-0.25em}
  	\caption{The scatterplots of the posterior means of segment-level $\mathbf{A}_{rij}(3,3)$ versus $\mathbf{A}_{rij}(2,2)$ obtained from the analysis of EMBARC data. Different colors represent different subjects in subfigure  (b) and (c).   }
  	\label{img:A_trial}
\end{figure}

We consider the subject- and segment-level RESSM-extracted EEG features within the four subgroups. In Figure \ref{img:A_trial}(a), the scatterplot illustrates the relationship between the posterior mean of segment-level $\mathbf{A}_{rij}(2,2)$ and $\mathbf{A}_{rij}(3,3)$ for all individuals in the MDD and control groups across the two sites. We note that the group means of $\mathbf{A}_{r}(2,2)$ and $\mathbf{A}_{r}(3,3)$ are higher in the control group compared to the MDD group, aligning with the findings shown in Figure \ref{img:real_A}. Moreover, the control group has a lower group variance in $\mathbf{A}_{r}(2,2)$ and $\mathbf{A}_{r}(3,3)$ compared to the MDD group. 
Figure \ref{img:A_trial}(b) and \ref{img:A_trial}(c) are the scatterplots of the posterior mean of segment-level $\mathbf{A}_{rij}(2,2)$ versus $\mathbf{A}_{rij}(3,3)$ for ten randomly chosen control participants and ten randomly chosen MDD participants. The $95\%$ confidence ellipses are fitted for each subject. These figures indicate that the segment-level temporal dynamical matrices, $\mathbf{A}_{rij}$, exhibit a smaller mean and a larger variance among segments for a subject in the MDD groups compared to a subject in the control group.
For subject-level spatial mapping matrices, we compute the norm of the $54$ rows for each $\mathbf{\Theta}_{ri}$. A comparison of the subject-level spatial mapping matrices between MDD and control groups is shown in Web Figure S.10. We observe that the MDD group exhibits higher group variance with more outliers compared to the control group. Moreover, we found significant heterogeneity at both the segment- and subject-level. Thus, methods not addressing these sources may result in an underestimation of uncertainty and incorrect inference.

After the baseline resting-state EEG measurements, participants diagnosed with MDD were randomly assigned to either sertraline, an SSRI antidepressant, or a placebo in a randomized trial. We explored the utilities of the subject-level EEG parameters in predicting heterogeneous treatment effects (HTE) and constructing optimal treatment decision rules. We used the subject-level mean and standard deviation of the 5 diagonal entries of $\mathbf{A}_{ri}$, norm of the 54 row vectors of spatial maps $\mathbf{\Theta}_{ri}$ corresponding to the 54 channels, as well as the clinical and demographic variables from EMBARC to estimate the conditional average treatment effect (CATE). We used multiple imputations to address missing values in the clinical and demographic variables. We randomly divided the dataset into 80\% training and 20\%  testing. Treatment outcomes are the change of the Hamilton Rating Scale for Depression (HAMD17) scores \citep{hamilton1960rating}, remission status (last observed HAMD17 score is less than or equal to 7), and response at exit \citep[i.e., HAMD17 score reduction of 50\% or greater;][]{trivedi2016establishing}. 

To evaluate the potential of RESSM features as biomarkers for predicting treatment response, we conducted several analyses. First,  we trained a causal forest  \citep{wager2018estimation} to capture potential nonlinear interactions among the covariates, estimate the conditional average treatment effect (CATE) of sertraline, and determine the variable importance. Additionally, we used the policy learning algorithm  \citep{athey2021policy} to construct the optimal treatment rule that maximizes the expected outcome or utility. Second, to evaluate the predictive value of the subject-level EEG features for the CATE using causal forest, we compared the root mean squared error (RMSE) of the transformed outcome (see definition in Web Appendix D.2) via ten repeats of 5-fold cross-validation. We compared three scenarios: (1) using 6 clinical and demographic variables only (age, sex, education years, baseline HAMD, chronicity, duration of illness), (2)  clinical + 64 features extracted by RESSM,  and (3)  clinical + 216 measures of relative alpha/theta EEG band powers based on previous literature \citep[see descriptions in][]{yang2023learning}.  Third, we fit a ridge regression and averaged ten repeats of 5-fold cross-validation to compare the R-squared of predicting external treatment outcomes under the same three sets of feature variables as the second analysis. 

As illustrated in Web Figure S.13, subject-level RESSM features, such as the mean of temporal dynamics (i.e., $\mathbf{A}_{ri}(1,1)$ and so on), along with the norm of channels including F7 and FP2, are identified as among the most important features from the causal forest. This result shows the important roles of these EEG features in moderating the treatment effect and predicting individual treatment responses. As indicated by the resulting decision trees (Web Figure S.14), for response at study exit, the optimal treatment rule depends on the temporal dynamic parameters. As shown in Table \ref{t:comparison_table}, compared to using clinical variables or EEG band powers, RESSM-extracted temporal and spatial features achieve much improved performance with fewer features, yielding an increased AUC, R-squared and a decreased RMSE. 
The performance gap between the features extracted using RESSM and those from \cite{yang2023learning} could be partially attributed to the omission of considering cross-spectra dependency between channels in the latter method.




\begin{table}[h]
\small
    \vspace*{-6pt}
    \caption{Comparing utilities of RESSM-extracted EEG features on external treatment response outcomes in EMBARC study analysis.}
    \centering
    \def\~{\hphantom{0}}
    \begin{minipage}{155mm}
    \begin{tabular*}{\columnwidth}{@{}l@{\extracolsep{\fill}}l@{\extracolsep{\fill}}c@{\extracolsep{\fill}}c@{\extracolsep{\fill}}c@{\extracolsep{\fill}}@{}}
    \toprule
    \multicolumn{2}{c}{} & \multicolumn{3}{c}{Outcome Measures}  \\[0.2em]
    \cline{3-5} 
    \\[-1em]
    &  & response at exit &  remission status & change of HAMD score \\
    \\
    \multirow{2}{*}{RMSE$^1$} & clinical + RESSM$^2$& $1.317 \ (0.0446)$ &$1.261 \ (0.0463)$ &  $21.114 \ (0.609)$ \\
    \\[-1em]
    & clinical + band power$^3$ & $1.322 \ (0.0335)$ & $1.263 \ (0.0366)$ &  $21.359 \ (0.588)$  \\
    & clinical variables only & $1.326 \ (0.0353)$ & $1.266 \ (0.0364)$ & $21.387 \ (0.579)$
    \\
    \midrule
     & & AUC & AUC & R-squared\\
    \multirow{3}{*}{{AUC or R-squared$^4$}} & clinical + RESSM  &$0. 677 \ (0.036)$ & $0.654 \ (0.039)  $ & $17.9\% \ (4.6\%)$ \\
    \\[-1em]
    & clinical + band power & $0.621 \ (0.022)$ & $0.584 \ (0.041)$ & $12.0\% \ (3.7\%)$  \\    
    & clinical variables only & $0.575 \ (0.045)$ & $0.517 \ (0.029) $&  $8.6\% \ (2.3\%)$ 
    \\
    \bottomrule\\\\
    \end{tabular*}
    \label{t:comparison_table}
    \vskip .05in
    \footnotesize{$^1$: mean and standard deviation for the RMSE of predicting the CATE}. The mean and standard deviation of RMSE are obtained by taking the mean and standard deviation of the mean RMSE of each of the ten repeated samples.\\
    \footnotesize{$^2$: Model includes 6 baseline demographic and clinical variables and 64 subject-specific EEG temporal dynamics and spatial maps extracted from our RESSM.} \\
    \footnotesize{$^3$:  Model includes 6 baseline demographic and clinical variables and 216 EEG theta and alpha band powers obtained from spectral analysis.} \\
    \footnotesize{$^4$: R-squared is obtained by fitting ridge regression for the change of HAMD score after treatment on different sets of feature variables; AUC is obtained by fitting ridge regression for the response at exit and remission status after treatment on different sets of feature variables. The mean and standard deviation of AUC and R-squared are obtained by taking the mean and standard deviation of the mean AUC and R-squared of each of the ten repeated samples.}
    \end{minipage}
\end{table}



\section{Discussion}
\label{sec:discussion}

This paper introduces a new approach to analyzing intensive multi-channel resting-state EEG signals, taking into account the heterogeneity in brain connectivity patterns across different clinical groups, subjects, and time segments by introducing random effects to the temporal dynamical and spatial mapping matrices. We overcome the limitations of previous mixed-effects state-space models by relaxing assumptions on model structure and identifiability. The model is fitted using a Bayesian hierarchical framework with a Gibbs sampler. 
Our analysis reveals a significant difference in the temporal dynamics of resting-state brain activity between individuals with MDD and healthy groups in the EMBARC study. Furthermore, our analysis uncovers important utilities of the extracted subject-level EEG parameters in predicting HTE and constructing treatment decision rules for prescribing antidepressants, suggesting that EEG may serve as a valuable  biomarker for MDD. 
An extension of our work is to integrate outcome data and the HTE model with latent state-space models. The goal is to jointly learn a lower-dimensional space that can capture EEG signals while contributing to the HTE on the outcome.



Furthermore, \cite{michel2018eeg} suggested that a small set of characteristic microstates of resting-state EEG signals can be consistently identified across participants. These microstates occur in a repetitive sequence over time. Research suggests that these microstates may be fundamental components of the sequential spontaneous conscious mental processes. Their presence and temporal dynamics play a crucial role in determining the quality of cognitive experiences.
Several studies have explored state-switching dynamical systems for single-subjects \citep{ombao2018statistical, quinn2018task}. It is of interest to extend the concept of state switching to multi-group, multi-subject state-space models. This extension would allow for testing differences in microstate switching patterns among different groups or individual subjects.
It is also of interest to extend our model to partition the high-dimensional EEG/fMRI data defined over the complex brain networks into a finite set of regional clusters using hierarchical latent factor models \citep{ting2018multi}, and to model the multi-level power spectral density (PSD) of EEG signals with functional data analysis \citep{campos2022multilevel}.



\section*{Acknowledgements}

This research is supported by U.S. NIH grants  MH123487, NS073671 and GM124104.

\vspace{-1.5em}
\section*{Supporting Information}

Web Appendices and Figures referenced in Sections 2-5 are available with this paper at the Biometrics website on Wiley Online Library.
The R code for the MCMC algorithm and data analysis is available on GitHub repository: \url{https://github.com/xingcheg/RESSM-EEG}.


\vspace{-1em}
\section*{Data Availability Statement}

The EMBARC data used in this paper can be obtained through an application to NIH data archive at: \url{https://nda.nih.gov/edit_collection.html?id=2199}

\vspace{-1em}



\bibliographystyle{apalike}
\bibliography{refs}


\clearpage\pagebreak\newpage
\setcounter{section}{0}
\renewcommand{\thesection}{S.\arabic{section}} 
\renewcommand{\thesubsection}{S.\arabic{section}.\arabic{subsection}}
\setcounter{page}{1}
\renewcommand{\theequation}{S.\arabic{equation}}
\renewcommand{\thefigure}{S.\arabic{figure}}
\renewcommand{\thetable}{S.\arabic{table}}
\setcounter{table}{0}
\setcounter{equation}{0} \setcounter{figure}{0} 
		
\renewcommand{\theenumi}{\arabic{enumi}}

\begin{center}
{\Large{\bf Supplementary Materials for 
  ``A Hierarchical Random Effects State-space Model for Modeling Brain Activities from Electroencephalogram Data''}}\\ 
			\vskip1cm
   Xingche Guo, Bin Yang, Ji Meng Loh, Qinxia Wang, and Yuanjia Wang
\end{center}

\vskip1cm

\section*{Web Appendix A}

This section provides additional details on the RESSM and discusses its identifiability.

\subsection*{A.1. Reformulation of MVAR(m) to MVAR(1)}

Define $\widetilde{\mathbf{M}}_{rij}(t_k) = \left(\mathbf{M}_{rij}^{\top}\left(t_{k}\right), \mathbf{M}_{rij}^{\top}\left(t_{k-1}\right), \dots, \mathbf{M}^{\top}_{rij}\left(t_{k-m+1}\right) \right)^{\top}$, Equation (2) in the paper can be re-written in an $mQ$-dimensional MVAR(1) form
\begin{align*}
    \widetilde{\mathbf{M}}_{rij}(t_k) = \widetilde{\mathbf{A}}_{rij} \widetilde{\mathbf{M}}_{rij}(t_{k-1}) + \widetilde{\mathbf{W}}_{rij}(t_{k}),
\end{align*}
where 
\begin{align*}
    \widetilde{\mathbf{A}}_{rij} & =
    \left[\begin{array}{ccccc}
    \mathbf{A}_{rij1} & \mathbf{A}_{rij2} & \ldots & \mathbf{A}_{rijm-1} & \mathbf{A}_{rijm} \\ 
    \mathbf{I}_{Q} & \mathbf{0} & \ldots & \mathbf{0} & \mathbf{0} \\ 
    \mathbf{0} & \mathbf{I}_{Q} & & \mathbf{0} & \mathbf{0} \\ 
    \vdots & & \ddots & \vdots & \vdots \\ 
    \mathbf{0} & \mathbf{0} & \ldots & \mathbf{I}_{Q} & \mathbf{0} \end{array}\right], \quad
    \widetilde{\mathbf{W}}_{rij}(t_{k}) =\left[\begin{array}{c}
    \mathbf{W}_{rij}\left(t_{k}\right) \\ 
    \mathbf{0} \\ 
    \vdots \\ 
    \mathbf{0}
    \end{array}\right]
\end{align*}

\subsection*{A.2. Details on model identifiability}

\subsubsection*{Choice of $\boldsymbol{\Sigma}_{w}$}
According to \cite{murphy2012machine}, $\boldsymbol{\Sigma}_{w}$ can be set to an identity matrix through appropriate reparameterization. To illustrate this, let $\boldsymbol{\Sigma}_{w}$ be any positive definite matrix with an eigenvalue decomposition $\mathbf{P}_{w} \boldsymbol{\Lambda}_{w} \mathbf{P}_{w}^{\top}$, where $\boldsymbol{\Lambda}_{w}$ is the diagonal matrix with each diagonal element representing an eigenvalue, and $\mathbf{P}_{w}$ is the orthogonal eigenvector matrix.
Define
\begin{align*}
\widetilde{\mathbf{M}}_{rij}\left(t_{k}\right)= \boldsymbol{\Lambda}_{w}^{-1/2} \mathbf{P}_{w}^{\top}  \mathbf{M}_{rij}\left(t_{k}\right), \quad \widetilde{\mathbf{\Theta}}_{rij} = \mathbf{\Theta}_{rij}  \mathbf{P}_{w} \boldsymbol{\Lambda}_{w}^{1/2}, \quad \widetilde{\mathbf{A}}_{rijh} = \boldsymbol{\Lambda}_{w}^{-1/2} \mathbf{P}_{w}^{\top} \mathbf{A}_{rijh} \mathbf{P}_{w} \boldsymbol{\Lambda}_{w}^{1/2}. 
\end{align*}
The proposed RESSM becomes
\begin{align*}
\mathbf{Y}_{rij}\left(t_{k}\right) &= \widetilde{\mathbf{\Theta}}_{rij} \widetilde{\mathbf{M}}_{rij}\left(t_{k}\right)+\boldsymbol{\epsilon}_{rij}\left(t_{k}\right), \quad \boldsymbol{\epsilon}_{rij}\left(t_{k}\right) \sim \mathcal{N}\left(\mathbf{0}, \boldsymbol{\Sigma}_{rij}\right) \\
\widetilde{\mathbf{M}}_{rij}\left(t_{k}\right) &= \sum_{h=1}^m \widetilde{\mathbf{A}}_{rijh}   \widetilde{\mathbf{M}}_{rij}\left(t_{k-h}\right) + \widetilde{\mathbf{W}}_{rij}\left(t_{k}\right), \quad 
\widetilde{\mathbf{W}}_{rij}\left(t_{k}\right) \sim \mathcal{N}\left(\mathbf{0}, \mathbf{I} \right).
\end{align*}
Hence, we fix $\boldsymbol{\Sigma}_{w} = \mathbf{I}$ to ensure identifiability. 
In Section 2.2 of the paper, we constrain the spatial mapping matrices to be lower-triangular. This can be accomplished through QR decomposition, wherein there exists an orthogonal matrix $\widecheck{\mathbf{R}}$ such that $\widecheck{\mathbf{\Theta}}_{rij} = \widetilde{\mathbf{\Theta}}_{rij} \widecheck{\mathbf{R}}^{\top}$ forms a lower-triangular matrix. Correspondingly,
\begin{align*}
\widecheck{\mathbf{M}}_{rij}\left(t_{k}\right)=\widecheck{\mathbf{R}} \widetilde{\mathbf{M}}_{rij}\left(t_{k}\right), \quad \widecheck{\mathbf{A}}_{rijh} = \widecheck{\mathbf{R}} \widetilde{\mathbf{ A}}_{rijh} \widecheck{\mathbf{R}}^{\top}, \quad \widecheck{\mathbf{W}}_{rij}\left(t_{k}\right)=\widecheck{\mathbf{R}} \widetilde{\mathbf{W}}_{rij}\left(t_{k}\right).
\end{align*}
The proposed RESSM becomes
\begin{align*}
\mathbf{Y}_{rij}\left(t_{k}\right) &= \widecheck{\mathbf{\Theta}}_{rij} \widecheck{\mathbf{M}}_{rij}\left(t_{k}\right)+\boldsymbol{\epsilon}_{rij}\left(t_{k}\right), \quad \boldsymbol{\epsilon}_{rij}\left(t_{k}\right) \sim \mathcal{N}\left(\mathbf{0}, \boldsymbol{\Sigma}_{rij}\right) \\
\widecheck{\mathbf{M}}\left(t_{k}\right) &= \sum_{h=1}^m \widecheck{\mathbf{A}}_{rijh}   \widecheck{\mathbf{M}}_{rij}\left(t_{k-h}\right) + \widecheck{\mathbf{W}}_{rij}\left(t_{k}\right), \quad 
\widecheck{\mathbf{W}}_{rij}\left(t_{k}\right) \sim \mathcal{N}\left(\mathbf{0}, \mathbf{I} \right).
\end{align*}
Note that the covariance matrix of $\widecheck{\mathbf{W}}_{rij}\left(t_{k}\right)$ remains an identity matrix, which is a desirable property.

\subsubsection*{Choice of $\boldsymbol{\Sigma}_{rij}$}

When $(Q < P)$, \cite{lutkepohl2005new} refers to the sensor model of RESSM as Factor Analytic Models, where each component of $\boldsymbol{\epsilon}_{rij}\left(t_{k}\right)$ is assumed to be uncorrelated, that is, $\boldsymbol{\Sigma}_{rij}$ is a diagonal matrix. \cite{murphy2012machine} also suggests using a diagonal covariance matrix $\boldsymbol{\Sigma}_{rij}$ to reduce the number of free parameters and enhance numerical stability. 

Furthermore, the diagonal constraint ensures identifiability. To see this,
assume that $\boldsymbol{\Sigma}_{rij} = \mathbf{R}_{rij} + \mathbf{\Lambda}_{rij}$, where $\mathbf{R}_{rij}$ is a non-negative definite matrix containing supplementary spatial correlation information to $\mathbf{\Theta}_{rij}$, and $\mathbf{\Lambda}_{rij}$ is a diagonal matrix with all diagonal values greater than $0$, representing the nugget effect covariance matrix. Then
\begin{align*}
    \boldsymbol{\epsilon}_{rij}\left(t_{k}\right) \overset{d}{=} \mathbf{P}_{rij} \mathbf{D}_{rij}^{1/2} \mathbf{U}_{rij}(t_k) + \boldsymbol{\epsilon}_{rij}^{*}\left(t_{k}\right),
\end{align*}
where $\mathbf{U}_{rij}(t_k) \overset{i.i.d.}{\sim} \mathcal{N}(\mathbf{0}, \mathbf{I}_{Q'})$, $Q' \le P$ is the rank of $\mathbf{R}_{rij}$; $\mathbf{P}_{rij}$ and $\mathbf{D}_{rij}$ are the $P \times Q'$ eigenvector and $Q' \times Q'$ eigenvalue matrices of $\mathbf{R}_{rij}$, respectively; $\boldsymbol{\epsilon}_{rij}^{*}\left(t_{k}\right) \overset{i.i.d.}{\sim} \mathcal{N}(\mathbf{0}, \mathbf{\Lambda}_{rij})$.
The proposed RESSM can be reformulated into an equivalent RESSM featuring more latent states: 
\begin{align*}
\mathbf{Y}_{rij}\left(t_{k}\right) &= \mathbf{\Theta}_{rij}^{*} \mathbf{M}_{rij}^{*}\left(t_{k}\right)+\boldsymbol{\epsilon}_{rij}^{*}\left(t_{k}\right), \quad \boldsymbol{\epsilon}_{rij}^{*}\left(t_{k}\right) \sim \mathcal{N}\left(\mathbf{0}, \boldsymbol{\Lambda}_{rij}\right) \\
\mathbf{M}_{rij}^{*}\left(t_{k}\right) &= \sum_{h=1}^m \mathbf{A}_{rijh}^{*}   \mathbf{M}_{rij}^{*}\left(t_{k-h}\right) + \mathbf{W}_{rij}^{*}\left(t_{k}\right), \quad 
\mathbf{W}_{rij}^{*}\left(t_{k}\right) \sim \mathcal{N}\left(\mathbf{0}, \mathbf{I}_{Q+Q'} \right).
\end{align*}
where 
\begin{align*}
    &\mathbf{\Theta}_{rij}^*  =
    \left[\begin{array}{cc}
    \mathbf{\Theta}_{rij} & \mathbf{P}_{rij} \mathbf{D}_{rij}^{1/2}
    \end{array}\right],
    \quad
    \mathbf{M}_{rij}^*(t_k)  =
    \left[\begin{array}{c}
    \mathbf{M}_{rij}(t_k) \\
    \mathbf{U}_{rij}(t_k) 
    \end{array}\right], \\
    &\mathbf{A}_{rijh}^*  =
    \left[\begin{array}{cc}
    \mathbf{A}_{rijh} & \mathbf{0} \\
    \mathbf{0} & \mathbf{0}
    \end{array}\right], \quad
    \mathbf{W}_{rij}^*(t_k)  =
    \left[\begin{array}{c}
    \mathbf{W}_{rij}(t_k) \\
    \mathbf{U}_{rij}(t_k) 
    \end{array}\right], \\
\end{align*}
Therefore, we fix the covariance matrix of $\boldsymbol{\epsilon}_{rij}\left(t_{k}\right)$ as a diagonal matrix to ensure identifiability.

\subsubsection*{Sign identifiability}

We implement a two-stage procedure to achieve sign identifiability. 

\noindent \textbf{First stage (initialization stage)}: During the initialization stage, we assume that subjects in each time segment across all subgroups share the same spatial mapping matrix, denoted as $\mathbf{\Theta}_{0} = \mathbf{\Theta}_{rij} = \mathbf{\Theta}_{ri} = \mathbf{\Theta}_{r}$. An MCMC is employed for this simplified RESSM, and the computation of the full conditional distribution of $\mathbf{\Theta}_{0}$ mirrors that of $\mathbf{\Theta}_{rij}$ in Section 3 of the paper. By setting an improper prior to $\mathbf{\Theta}_{0}$, we have
\begin{align*}
\left[ \mbox{low} \left( \mathbf{\Theta}_{0} \right) \Big| \cdots \right] &\propto
\prod_{r=1}^R \prod_{i=1}^{N_r} \prod_{j=1}^{J_{ri}} \prod_{k=1}^K p\Big( \mathbf{Y}_{rij}\left(t_{k}\right) \Big| \mathbf{M}_{rij}\left(t_{k}\right), \mathbf{\Theta}_{0} \Big)   
\sim \mathcal{N}_C \left( \boldsymbol{b}_{rij}^{(\Theta_0)}, \boldsymbol{Q}_{rij}^{(\Theta_0)} \right),
\end{align*}
where
\begin{gather*}
\boldsymbol{Q}_{rij}^{(\Theta_0)} =  \sigma_{rij}^{-2} \left[ \left\{ \sum_{r=1}^R \sum_{i=1}^{N_r} \sum_{j=1}^{J_{ri}} \sum_{k=1}^K  \mathbf{M}_{rij}\left(t_{k}\right) \mathbf{M}_{rij}^{\top}\left(t_{k}\right) \right\} \otimes \mathbf{I}_P \right]_{\mathcal{F}, \mathcal{F}}, \\
\boldsymbol{b}_{rij}^{(\Theta_0)} = \sigma_{rij}^{-2} \left[ \sum_{r=1}^R \sum_{i=1}^{N_r} \sum_{j=1}^{J_{ri}} \sum_{k=1}^K   \mathbf{M}_{rij}\left(t_{k}\right) \otimes  \mathbf{Y}_{rij}\left(t_{k}\right) \right]_{\mathcal{F}}.
\end{gather*}
Assuming a shared $\mathbf{\Theta}_0$ across all subjects in every time segment reduces dimensionality, facilitating faster convergence of the MCMC algorithm. 
The posterior mean of $\mathbf{\Theta}_0$ obtained from the first step is utilized as the initial values for the spatial mapping matrices at each level in the proposed method.
The initialization stage ensures an identical initial value for spatial mapping matrices at all levels that is close to the truth, thereby mitigating the concentration of random effects around two means with opposite signs to a significant extent. However, there might still be a small portion that has incorrectly identified signs, especially when the sign-to-noise ratio is low for some latent states.

\noindent \textbf{Second stage (sign-tracking stage)}: 
During the sign-tracking stage, we monitor the signs for each column of spatial mapping matrices at each iteration of the MCMC, and adjust the signs if potential errors are identified.
We first check whether the sign for each column of $\mathbf{\Theta}_{rij}$ aligns with the sign for the corresponding column of $\mathbf{\Theta}_{ri}$. Define $\boldsymbol{\theta}_{rij(q)}^{(l)}$ and $\boldsymbol{\theta}_{ri(q)}^{(l)}$ as the $q$-th column for $\mathbf{\Theta}_{rij}$ and $\mathbf{\Theta}_{ri}$ at the $l$-th MCMC iteration,
we monitor the sign by computing the cosine correlation:
\begin{align*}
    \rho_{rij(q)}^{(l)} := \cos \left( \boldsymbol{\theta}_{rij(q)}^{(l)}, \boldsymbol{\theta}_{ri(q)}^{(l)} \right) = \frac{ \left\langle \boldsymbol{\theta}_{rij(q)}^{(l)}, \boldsymbol{\theta}_{ri(q)}^{(l)} \right\rangle_2}{\| \boldsymbol{\theta}_{rij(q)}^{(l)} \|_2 \| \boldsymbol{\theta}_{ri(q)}^{(l)} \|_2 }
\end{align*}
Note that the cosine correlation always falls within the range of $-1$ and $1$. A substantially negative value indicates an incorrect sign identification. If $\rho_{rij(q)}^{(l)} < \rho_0$, where $\rho_0 \le 0$ is a user-defined threshold, then we alter the sign of $\boldsymbol{\theta}_{rij(q)}^{(l)}$ and the corresponding latent signal $M_{rijq}(t)$.
A similar sign-tracking procedure is subsequently applied to the subject- and group-level spatial mapping matrices by examining each $\cos \left( \boldsymbol{\theta}_{ri(q)}^{(l)}, \boldsymbol{\theta}_{r(q)}^{(l)} \right)$ and $\cos \left( \boldsymbol{\theta}_{r(q)}^{(l)}, \boldsymbol{\theta}_{(q)}^{(l)} \right)$. 

It is worth noting that \citet{conti2014bayesian} employ a similar sign-tracking procedure for factor models. They use one element in each column of the spatial mapping matrices as a benchmark and monitor the sign. In comparison, our method is more stable as the computation of cosine correlation is robust and considers all the elements. A simulation study that shows the usefulness for both stages of the proposed two-stage MCMC algorithm are provided in Web Appendix C.3.

\section*{Web Appendix B}

This section shows more details of priors and full conditional distributions.
For posterior computation of the temporal dynamical matrices and spatial mapping matrices, we use the canonical form of the Gaussian distribution
$\mathcal{N}_C(\boldsymbol{b}, \boldsymbol{Q})$ to represent a Gaussian distribution in the form $\mathcal{N}(\boldsymbol{Q}^{-1} \boldsymbol{b}, \boldsymbol{Q}^{-1})$. 
It is easy to see that if $\boldsymbol{x} \sim \mathcal{N}_C(\boldsymbol{b}, \boldsymbol{Q})$, then the density function $\log p(\boldsymbol{x}) \propto -\frac{1}{2} \boldsymbol{x}^{\top} \boldsymbol{Q} \boldsymbol{x} + \boldsymbol{b}^{\top} \boldsymbol{x}$.
The detailed computation of the full conditional distributions and the computation complexity of the MCMC algorithm are provided below.

\subsection*{B.1. Detailed computation of full conditional posteriors}

\subsubsection*{Sampling $\mathbf{M}_{rij}\left(t_{k}\right)$}
Let $\propto_M$ denote the proportionality function for $\mathbf{M}_{rij}\left(t_{k}\right)$, and $\|\cdot \|$ represent the Euclidean norm. 
Denote the $Q \times Q$ matrix $\boldsymbol{A}_{rij0}$ as $- \mathbf{I}$ to simplify the expressions. 
For the $r$-th subgroup, $i$-th subject in the $j$-th segment, at each time $t_k$, we sample the latent EEG signal $\mathbf{M}_{rij}\left(t_{k}\right)$ from its full conditional posterior:
\begin{align*}
& \log p \left( \mathbf{M}_{rij}\left(t_{k}\right) \mid \cdots \right) \ \propto_M \ 
\log \left\{ p\Big( \mathbf{Y}_{rij}\left(t_{k}\right) \Big| \mathbf{M}_{rij}\left(t_{k}\right) \Big)  \prod_{s=k}^{k+m} p\Big( \mathbf{M}_{rij}\left(t_{s}\right) \Big| \left\{ \mathbf{M}_{rij}\left(t_{s-h}\right) \right\}_{h=1}^m \Big) \right\} \\
& \propto_{M} -\frac{1}{2 \sigma_{rij}^2} \left\| \mathbf{Y}_{rij}(t_k) - \mathbf{\Theta}_{rij} \mathbf{M}_{rij}(t_k) \right\|^2 
- \frac{1}{2} \sum_{s=k}^{k+m} \left\| \mathbf{M}_{rij}(t_s) - \sum_{h=1}^m \mathbf{A}_{rijh} \mathbf{M}_{rij}(t_{s-h}) \right\|^2 \\
& \propto_{M} -\frac{1}{2 \sigma_{rij}^2} \left\| \mathbf{Y}_{rij}(t_k) - \mathbf{\Theta}_{rij} \mathbf{M}_{rij}(t_k) \right\|^2 
- \frac{1}{2} \sum_{s=k}^{k+m} \left\| \sum_{h=0}^m \mathbf{A}_{rijh} \mathbf{M}_{rij}(t_{s-h}) \right\|^2 \\
& \propto_{M} -\frac{1}{2 \sigma_{rij}^2} \left\| \mathbf{Y}_{rij}(t_k) - \mathbf{\Theta}_{rij} \mathbf{M}_{rij}(t_k) \right\|^2 
- \frac{1}{2} \sum_{h_1=0}^{m} \left\| \sum_{h_2=0}^m \mathbf{A}_{rijh_2} \mathbf{M}_{rij}(t_{k+h_{1}-h_{2}}) \right\|^2 \\
& \propto_{M} -\frac{1}{2 \sigma_{rij}^2} \left\| \mathbf{Y}_{rij}(t_k) - \mathbf{\Theta}_{rij} \mathbf{M}_{rij}(t_k) \right\|^2 
- \frac{1}{2} \sum_{h_1=0}^{m} \left\| \mathbf{A}_{rijh_{1}} \mathbf{M}_{rij}(t_{k}) + \sum_{h_2=0, h_2 \neq h_1}^m \mathbf{A}_{rijh_2} \mathbf{M}_{rij}(t_{k+h_{1}-h_{2}}) \right\|^2 \\
&\propto_{M} -\frac{1}{2} \mathbf{M}_{rij}\left(t_{k}\right)^{\top} \boldsymbol{Q}_{rijk}^{(M)} \ \mathbf{M}_{rij}\left(t_{k}\right) + \boldsymbol{b}_{rijk}^{(M) \top} \mathbf{M}_{rij}\left(t_{k}\right) \\
&\sim \mathcal{N}_C \left( \boldsymbol{b}_{rijk}^{(M)}, \boldsymbol{Q}_{rijk}^{(M)}  \right),
\end{align*}
where
\begin{gather*}
\boldsymbol{Q}_{rijk}^{(M)} = \sigma_{rij}^{-2} \left(\mathbf{\Theta}_{rij} \right)^{\top}  \mathbf{\Theta}_{rij} + \sum_{h=0}^{m} \mathbf{A}_{rijh}^{\top} \mathbf{A}_{rijh}, \\
\boldsymbol{b}_{rijk}^{(M)} = \sigma_{rij}^{-2} \left(\mathbf{\Theta}_{rij} \right)^{\top} \mathbf{Y}_{rij}\left(t_{k}\right) -  \sum_{h_1=0}^m \mathbf{A}_{rijh_1}^{\top} \sum_{h_2=0, h_2 \neq h_1}^m \mathbf{A}_{rijh_2}
\mathbf{M}_{rij}\left(t_{k+h_1-h_2}\right),
\end{gather*}

\subsubsection*{Sampling $\mathbf{A}_{rij}$}
Recall
$\mathbf{A}_{rij} = \left[ \mathbf{A}_{rij 1}, \dots, \mathbf{A}_{rij m}  \right]$.
Define 
$\mathbf{M}_{rijk}^{*} = \left( \mathbf{M}_{rij}^{\top}\left(t_{k-1}\right), \dots,  \mathbf{M}_{rij}^{\top}\left(t_{k-m}\right) \right)^{\top}$. Let $\propto_A$ denote the proportionality function for $\mathbf{A}_{rij}$.
First note that
\begin{align*}
    &\log p\Big( \mathbf{M}_{rij}\left(t_{k}\right) \Big| \mathbf{M}_{rijk}^{*}, \mathbf{A}_{rij} \Big) \propto_A - \frac{1}{2} \left\| \mathbf{M}_{rij}(t_k) - \sum_{h=1}^m \mathbf{A}_{rijh} \mathbf{M}_{rij}(t_{k-h}) \right\|^2 \\
    &\propto_A - \frac{1}{2} \left\| \mathbf{M}_{rij}(t_k) - \mathbf{A}_{rij} \mathbf{M}_{rijk}^{*} \right\|^2 \\
    &\propto_A - \frac{1}{2} \left\| \mathbf{M}_{rij}(t_k) - \left( \left(\mathbf{M}_{rijk}^*\right)^{\top} \otimes \mathbf{I}_Q \right) \mbox{vec} \left( \mathbf{A}_{rij} \right) \right\|^2 \\
    &\propto_A - \frac{1}{2} \mbox{vec} \left( \mathbf{A}_{rij} \right)^{\top}
    \Big( \mathbf{M}_{rijk}^* \otimes \mathbf{I}_Q \Big)
    \Big( \left(\mathbf{M}_{rijk}^*\right)^{\top} \otimes \mathbf{I}_Q \Big)
    \mbox{vec} \left( \mathbf{A}_{rij} \right) \\
    & \quad\quad +  \mbox{vec} \left( \mathbf{A}_{rij} \right)^{\top}
    \left( \mathbf{M}_{rijk}^* \otimes \mathbf{I}_Q \right) 
    \mathbf{M}_{rij}(t_k) \\
    &\propto_A - \frac{1}{2} \mbox{vec} \left( \mathbf{A}_{rij} \right)^{\top}
    \left[\left\{ \mathbf{M}_{rijk}^* \left(\mathbf{M}_{rijk}^*\right)^{\top} \right\}
     \otimes \mathbf{I}_Q  \right]
    \mbox{vec} \left( \mathbf{A}_{rij} \right) \\
    & \quad\quad +  \mbox{vec} \left( \mathbf{A}_{rij} \right)^{\top}
    \left( \mathbf{M}_{rijk}^* \otimes \mathbf{M}_{rij}(t_k) \right) 
\end{align*}
We sample the temporal dynamical matrices $\mathbf{A}_{rij}$ from its full conditional posterior:
  \begin{align*}
\left[ \mbox{vec} \left( \mathbf{A}_{rij} \right) \Big| \cdots \right] & \ \propto_A \
p\left( \mbox{vec} \left( \mathbf{A}_{rij} \right)  \Big| \mbox{vec} \left( \mathbf{A}_{ri} \right) \right)
\prod_{k=m+1}^K p\Big( \mathbf{M}_{rij}\left(t_{k}\right) \Big| \mathbf{M}_{rijk}^{*}, \mathbf{A}_{rij} \Big)  
\sim \mathcal{N}_C \left( \boldsymbol{b}_{rij}^{(A)}, \boldsymbol{Q}_{rij}^{(A)} \right),
\end{align*}
where
\begin{gather*}
\boldsymbol{Q}_{rij}^{(A)} = \boldsymbol{\Sigma}_{v,r}^{-1} + 
\left\{\sum_{k=m+1}^K \mathbf{M}_{rijk}^* \left(\mathbf{M}_{rijk}^*\right)^{\top} \right\} \otimes \mathbf{I}_Q, \\
\boldsymbol{b}_{rij}^{(A)} = \boldsymbol{\Sigma}_{v,r}^{-1} \ \mbox{vec}\left( \mathbf{A}_{ri} \right) + \sum_{k=m+1}^K  \mathbf{M}_{rijk}^* \otimes  \mathbf{M}_{rij}\left(t_{k}\right).
\end{gather*}

\subsubsection*{Sampling $\mathbf{\Theta}_{rij}$}
Let $\propto_{\Theta}$ denote the proportionality function for $\mathbf{\Theta}_{rij}$.
Define $\mathcal{F}$ to be the index mapping from the elements of $\mbox{vec} \left( \mathbf{\Theta}_{rij} \right)$ to the elements of $\mbox{low} \left( \mathbf{\Theta}_{rij} \right)$ in $\mbox{vec} \left( \mathbf{\Theta}_{rij} \right)$, i.e. $\left[ \mbox{vec} \left( \mathbf{\Theta}_{rij} \right) \right]_{\mathcal{F}} = \mbox{low} \left( \mathbf{\Theta}_{rij} \right)$. 
First note that
\vspace{-0.8em}
\begin{align*}
    &\log p\Big( \mathbf{Y}_{rij}\left(t_{k}\right) \Big| \mathbf{M}_{rij}\left(t_{k}\right), \mathbf{\Theta}_{rij} \Big) \propto_{\Theta}   -\frac{1}{2 \sigma_{rij}^2} \left\| \mathbf{Y}_{rij}(t_k) - \mathbf{\Theta}_{rij} \mathbf{M}_{rij}(t_k) \right\|^2  \\
    & \propto_{\Theta}   -\frac{1}{2 \sigma_{rij}^2} \left\| \mathbf{Y}_{rij}(t_k) - \left( \mathbf{M}_{rij}^{\top}\left(t_{k}\right) \otimes \mathbf{I}_P \right) \mbox{vec} \left( \mathbf{\Theta}_{rij} \right) \right\|^2  \\
    &\propto_{\Theta} - \frac{1}{2 \sigma_{rij}^2} \mbox{vec} \left( \mathbf{\Theta}_{rij} \right)^{\top}
    \Big( \mathbf{M}_{rij}(t_k) \otimes \mathbf{I}_P \Big)
    \Big( \mathbf{M}_{rij}^{\top}(t_k) \otimes \mathbf{I}_P \Big)
    \mbox{vec} \left( \mathbf{\Theta}_{rij} \right) \\
    & \quad\quad +  \frac{1}{\sigma_{rij}^2 }\mbox{vec} \left( \mathbf{\Theta}_{rij} \right)^{\top}
    \Big( \mathbf{M}_{rij}(t_k) \otimes \mathbf{I}_P \Big)
    \mathbf{Y}_{rij}(t_k) \\
    &\propto_{\Theta} - \frac{1}{2 \sigma_{rij}^2} \mbox{vec} \left( \mathbf{\Theta}_{rij} \right)^{\top}
    \left[ \left\{  \mathbf{M}_{rij}\left(t_{k}\right) \mathbf{M}_{rij}^{\top}\left(t_{k}\right) \right\} \otimes \mathbf{I}_P \right]
    \mbox{vec} \left( \mathbf{\Theta}_{rij} \right) \\
    & \quad\quad +  \frac{1}{\sigma_{rij}^2 }\mbox{vec} \left( \mathbf{\Theta}_{rij} \right)^{\top}
    \Big( \mathbf{M}_{rij}(t_k) \otimes \mathbf{Y}_{rij}(t_k) \Big) \\
    &\propto_{\Theta} - \frac{1}{2 \sigma_{rij}^2} \mbox{low} \left( \mathbf{\Theta}_{rij} \right)^{\top}
    \left[ \left\{  \mathbf{M}_{rij}\left(t_{k}\right) \mathbf{M}_{rij}^{\top}\left(t_{k}\right) \right\} \otimes \mathbf{I}_P \right]_{\mathcal{F}, \mathcal{F}}
    \mbox{low} \left( \mathbf{\Theta}_{rij} \right) \\
    & \quad\quad +  \frac{1}{\sigma_{rij}^2 }\mbox{low} \left( \mathbf{\Theta}_{rij} \right)^{\top}
    \Big[ \mathbf{M}_{rij}(t_k) \otimes \mathbf{Y}_{rij}(t_k) \Big]_{\mathcal{F}}. 
\end{align*}
For the $r$-th subgroup, $i$-th subject in the $j$-th segment, we sample the spatial mapping matrices $\mathbf{\Theta}_{rij}$ from its full conditional posterior:
\vspace{-0.8em}
  \begin{align*}
\left[ \mbox{low} \left( \mathbf{\Theta}_{rij} \right) \Big| \cdots \right] &\propto
p\left( \mbox{low} \left( \mathbf{\Theta}_{rij} \right)  \Big| \mbox{low} \left( \mathbf{\Theta}_{ri} \right) \right) 
\prod_{k=1}^K p\Big( \mathbf{Y}_{rij}\left(t_{k}\right) \Big| \mathbf{M}_{rij}\left(t_{k}\right), \mathbf{\Theta}_{rij} \Big)   
\sim \mathcal{N}_C \left( \boldsymbol{b}_{rij}^{(\Theta)}, \boldsymbol{Q}_{rij}^{(\Theta)} \right),
\end{align*}

\vspace{-1.2em}
\begin{gather*}
\boldsymbol{Q}_{rij}^{(\Theta)} = \left( \boldsymbol{\Sigma}_{u,r} \right)^{-1} + \sigma_{rij}^{-2} \left[ \left\{ \sum_{k=1}^K  \mathbf{M}_{rij}\left(t_{k}\right) \mathbf{M}_{rij}^{\top}\left(t_{k}\right) \right\} \otimes \mathbf{I}_P \right]_{\mathcal{F}, \mathcal{F}}, \\
\boldsymbol{b}_{rij}^{(\Theta)} = \left( \boldsymbol{\Sigma}_{u,r} \right)^{-1} \mbox{low}\left( \mathbf{\Theta}_{ri} \right) + \sigma_{rij}^{-2} \left[ \sum_{k=1}^K  \mathbf{M}_{rij}\left(t_{k}\right) \otimes  \mathbf{Y}_{rij}\left(t_{k}\right) \right]_{\mathcal{F}}.
\end{gather*}

\subsubsection*{Sampling parent parameters of $\mathbf{A}_{rij}$ and $\mathbf{\Theta}_{rij}$}

Given the hierarchical structure and prior distributions in Section 2.3,
We give the full conditional distributions of $\mathbf{A}_{ri}$, $\mathbf{A}_{r}$, and $\mathbf{A}$:
\begin{align*}
&\left[ \mbox{vec} \left( \mathbf{A}_{ri} \right) \Big| \cdots \right] 
\propto 
p\left( \mbox{vec} \left( \mathbf{A}_{ri} \right)  \Big| \mbox{vec} \left( \mathbf{A}_r \right) \right)
\prod_{j=1}^{J_{ri}} p\left( \mbox{vec} \left( \mathbf{A}_{rij} \right)  \Big| \mbox{vec} \left( \mathbf{A}_{ri} \right) \right)
\sim \mathcal{N}_C \left( \boldsymbol{b}_{ri}^{(A)}, \boldsymbol{Q}_{ri}^{(A)} \right), \\
&\boldsymbol{Q}_{ri}^{(A)} =  \boldsymbol{\Sigma}_{\gamma,r}^{-1} + J_{ri}  \boldsymbol{\Sigma}_{v,r}^{-1} \quad,\quad
\boldsymbol{b}_{ri}^{(A)} = \boldsymbol{\Sigma}_{\gamma,r}^{-1} \mbox{vec}\left( \mathbf{A}_r \right) + \boldsymbol{\Sigma}_{v,r}^{-1} \sum_{j=1}^{J_{ri}}  \mbox{vec}\left( \mathbf{A}_{rij} \right); \\
& \left[ \mbox{vec} \left( \mathbf{A}_{r} \right) \Big| \cdots \right] 
\propto 
p\left( \mbox{vec} \left( \mathbf{A}_{r} \right)  \Big| \mbox{vec} \left( \mathbf{A} \right) \right)
\prod_{i=1}^{N_{r}} p\left( \mbox{vec} \left( \mathbf{A}_{ri} \right)  \Big| \mbox{vec} \left( \mathbf{A}_{r} \right) \right)
\sim \mathcal{N}_C \left( \boldsymbol{b}_{r}^{(A)}, \boldsymbol{Q}_{r}^{(A)} \right), \\
&\boldsymbol{Q}_{r}^{(A)} = \boldsymbol{\Sigma}_{a}^{-1} + N_{r} \boldsymbol{\Sigma}_{\gamma,r}^{-1} \quad,\quad
\boldsymbol{b}_{r}^{(A)} = \boldsymbol{\Sigma}_{a}^{-1} \mbox{vec}\left( \mathbf{A} \right) +  \boldsymbol{\Sigma}_{\gamma,r}^{-1} \sum_{i=1}^{N_{r}}  \mbox{vec}\left( \mathbf{A}_{ri} \right); \\
&\left[ \mbox{vec} \left( \mathbf{A} \right) \Big| \cdots \right] 
\propto 
p\left( \mbox{vec} \left( \mathbf{A} \right) \right)
\prod_{r=1}^{R} p\left( \mbox{vec} \left( \mathbf{A}_{r} \right)  \Big| \mbox{vec} \left( \mathbf{A} \right) \right)
\sim \mathcal{N} \left( R^{-1} \sum_{r=1}^{R}  \mbox{vec}\left( \mathbf{A}_{r} \right), R^{-1} \boldsymbol{\Sigma}_{a} \right); 
\end{align*}
The full conditional distributions of $\mathbf{\Theta}_{ri}$, $\mathbf{\Theta}_{r}$, and $\mathbf{\Theta}$ are given below:
\begin{align*}
&\left[ \mbox{low} \left( \mathbf{\Theta}_{ri} \right) \Big| \cdots \right] 
\propto 
p\left( \mbox{low} \left( \mathbf{\Theta}_{ri} \right)  \Big| \mbox{low} \left( \mathbf{\Theta}_{r} \right) \right)
\prod_{j=1}^{J_{ri}} p\left( \mbox{low} \left( \mathbf{\Theta}_{rij} \right)  \Big| \mbox{low} \left( \mathbf{\Theta}_{ri} \right) \right)
\sim \mathcal{N}_C \left( \boldsymbol{b}_{ri}^{(\Theta)}, \boldsymbol{Q}_{ri}^{(\Theta)} \right), \\
&\boldsymbol{Q}_{ri}^{(\Theta)} = \boldsymbol{\Sigma}_{\psi,r} ^{-1} + J_{ri}  \boldsymbol{\Sigma}_{u,r}^{-1} \quad,\quad
\boldsymbol{b}_{ri}^{(\Theta)} = \boldsymbol{\Sigma}_{\psi,r}^{-1} \mbox{low}\left( \mathbf{\Theta}_r \right) + \boldsymbol{\Sigma}_{u,r}^{-1} \sum_{j=1}^{J_{ri}}  \mbox{low}\left( \mathbf{\Theta}_{rij} \right); \\
&\left[ \mbox{low} \left( \mathbf{\Theta}_{r} \right) \Big| \cdots \right] 
\propto 
p\left( \mbox{low} \left( \mathbf{\Theta}_{r} \right)  \Big| \mbox{low} \left( \mathbf{\Theta} \right) \right)
\prod_{i=1}^{N_{r}} p\left( \mbox{low} \left( \mathbf{\Theta}_{ri} \right)  \Big| \mbox{low} \left( \mathbf{\Theta}_{r} \right) \right)
\sim \mathcal{N}_C \left( \boldsymbol{b}_{r}^{(\Theta)}, \boldsymbol{Q}_{r}^{(\Theta)} \right), \\
&\boldsymbol{Q}_{r}^{(\Theta)} = \boldsymbol{\Sigma}_{\theta}^{-1} + N_{r}  \boldsymbol{\Sigma}_{\psi,r}^{-1}\quad,\quad
\boldsymbol{b}_{r}^{(\Theta)} = \boldsymbol{\Sigma}_{\theta}^{-1} \mbox{low}\left( \mathbf{\Theta} \right) + \boldsymbol{\Sigma}_{\psi,r}^{-1} \sum_{i=1}^{N_{r}}  \mbox{low}\left( \mathbf{\Theta}_{ri} \right); \\
&\left[ \mbox{low} \left( \mathbf{\Theta} \right) \Big| \cdots \right] 
\propto 
p\left( \mbox{low} \left( \mathbf{\Theta} \right) \right)
\prod_{r=1}^{R} p\left( \mbox{low} \left( \mathbf{\Theta}_{r} \right)  \Big| \mbox{low} \left( \mathbf{\Theta} \right) \right)
\sim \mathcal{N} \left( R^{-1} \sum_{r=1}^{R}  \mbox{low}\left( \mathbf{\Theta}_{r} \right), R^{-1} \boldsymbol{\Sigma}_{\theta} \right).
\end{align*}

\subsection*{B.2. Sampling variance components}

\subsubsection*{Priors for the variance components}
We choose conjugate inverse Wishart priors $\mathcal{IW}(\nu, \mathbf{H})$ for the variance components with $\nu$ degrees of freedom and scale matrix $\mathbf{H}$. 
From Section 2.3 of the paper, we know that $\boldsymbol{\Sigma}_{v,r}$, $\boldsymbol{\Sigma}_{\gamma,r}$, and $\boldsymbol{\Sigma}_{a}$ are the covariance matrices of the (vectorized) temporal dynamical matrices for the segment-, subject-, and group-levels;
$\boldsymbol{\Sigma}_{u,r}$, $\boldsymbol{\Sigma}_{\psi,r}$, and $\boldsymbol{\Sigma}_{\theta}$ are the covariance matrices of the (vectorized lower-trangular) spatial mapping matrices for the segment-, subject-, and group-levels. We assign inverse Wishart priors:
\begin{align*}
& \boldsymbol{\Sigma}_{u,r} \sim \mathcal{IW}\Big(\nu_{u}, \mathbf{H}_u  \Big),
\quad
\boldsymbol{\Sigma}_{v, r} \sim \mathcal{IW}\Big(\nu_{v}, \mathbf{H}_v \Big), \\
& \boldsymbol{\Sigma}_{\psi,r} \sim \mathcal{IW}\Big( \nu_{\psi}, \mathbf{H}_{\psi} \Big),
\quad
\boldsymbol{\Sigma}_{\gamma, r} \sim \mathcal{IW}\Big( \nu_{\gamma}, \mathbf{H}_{\gamma} \Big),   \\
& \boldsymbol{\Sigma}_{\theta} \sim \mathcal{IW}\Big(\nu_{\theta}, \mathbf{H}_{\theta} \Big), \quad
\boldsymbol{\Sigma}_{a} \sim \mathcal{IW}\Big(\nu_{a}, \mathbf{H}_{a} \Big).
\end{align*}
A potential concern arises as demonstrated by \cite{alvarez2014bayesian}, who illustrated a significant prior dependency between correlation and variances when simulating covariance matrices using the inverse Wishart prior. To mitigate this limitation, one may consider exploring more flexible priors, such as the scaled inverse Wishart prior \citep{gelman2006data}, the hierarchical Half-t prior \citep{huang2013simple}, or the separation strategy \citep{barnard2000modeling}. These alternative priors offer the advantage of relaxing constraints and providing greater flexibility in modeling covariance matrices. However, it is essential to note that the discussion of these alternative priors is beyond the scope of the current study.
On the other hand, if we have prior knowledge of spatial information, such as an appropriate head model describing the geometrical characteristics of the head and the positions of the EEG electrodes \citep{baillet2001electromagnetic}, we can establish an informative prior on the variance components of spatial mapping matrices by incorporating the spatial information through $\mathbf{H}$. However, we should allow for enough flexibility in estimating the variance components using the data. This is crucial because two adjacent electrodes of EEG might represent two entirely different brain networks and may not be strongly correlated.

\subsubsection*{Full conditional posteriors for the variance components}

For an i.i.d.\ data sample $\{ \mathbf{x}_i \}_{i=1}^n$ with a multivariate normal distribution $\mathbf{x}_i \overset{i.i.d.}{\sim} \mathcal{N}(\boldsymbol{\mu}, \boldsymbol{\Sigma})$, if we assign an inverse Wishart prior $\mathcal{IW}(\nu, \mathbf{H})$ to the covariance matrix $\boldsymbol{\Sigma}$, the full conditional posterior of $\boldsymbol{\Sigma}$ is a conjugate inverse Wishart distribution $\mathcal{IW}(\nu + n, \mathbf{H} + \mathbf{S})$, where $\mathbf{S} = \sum_{i=1}^n (\mathbf{x}_i - \boldsymbol{\mu}) (\mathbf{x}_i - \boldsymbol{\mu})^{\top}$.
Equivalently, we can sample from the precision matrix $\mathbf{P}$ (the inverse of $\boldsymbol{\Sigma}$), which has a prior following a Wishart distribution $\mathcal{W}(\nu, \mathbf{H}^{-1})$. The full conditional posterior of $\mathbf{P}$ is a conjugate Wishart distribution $\mathcal{W}(\nu + n, (\mathbf{H} + \mathbf{S})^{-1})$ \citep[see e.g.][]{zhang2021note}.
Utilizing the precision matrix has the computational advantage of avoiding matrix inversions in density computations. The full conditional distributions for the inverses of $\boldsymbol{\Sigma}_{u,r}$,  $\boldsymbol{\Sigma}_{v,r}$, $\boldsymbol{\Sigma}_{\psi,r}$, $\boldsymbol{\Sigma}_{\gamma,r}$, $\boldsymbol{\Sigma}_{a}$, and $\boldsymbol{\Sigma}_{\theta}$ are given below

\begin{align*}
&\left[ \boldsymbol{\Sigma}_{u,r}^{-1} \Big| \cdots \right] 
\sim \mathcal{W}\Big( \nu_{u} + \sum_{i=1}^{N_r} J_{ri}, \  \left(\mathbf{H}_u + \mathbf{S}_{u,r} \right)^{-1} \Big), \\
&\mathbf{S}_{u,r} = \sum_{i=1}^{N_r} \sum_{j=1}^{J_{ri}} 
\bigg( \mbox{low} \left( \mathbf{\Theta}_{rij} \right) - \mbox{low} \left( \mathbf{\Theta}_{ri} \right) \bigg)  
\bigg( \mbox{low} \left( \mathbf{\Theta}_{rij} \right) - \mbox{low} \left( \mathbf{\Theta}_{ri} \right) \bigg)^{\top}; \\
&\left[ \boldsymbol{\Sigma}_{v,r}^{-1} \Big| \cdots \right] 
\sim \mathcal{W}\Big( \nu_{v} + \sum_{i=1}^{N_r} J_{ri}, \  \left(\mathbf{H}_v + \mathbf{S}_{v,r} \right)^{-1} \Big), \\
&\mathbf{S}_{v,r} = \sum_{i=1}^{N_r} \sum_{j=1}^{J_{ri}} 
\bigg( \mbox{vec} \left( \mathbf{A}_{rij} \right) - \mbox{vec} \left( \mathbf{A}_{ri} \right) \bigg)  
\bigg( \mbox{vec} \left( \mathbf{A}_{rij} \right) - \mbox{vec} \left( \mathbf{A}_{ri} \right) \bigg)^{\top}. 
\end{align*}
\begin{align*}
&\left[ \boldsymbol{\Sigma}_{\psi,r}^{-1} \Big| \cdots \right] 
\sim \mathcal{W}\Big( \nu_{\psi} + N_{r}, \  \left(\mathbf{H}_{\psi} +  \mathbf{S}_{\psi,r}\right)^{-1} \Big), \\
&\mathbf{S}_{\psi,r} = \sum_{i=1}^{N_{r}} 
\bigg( \mbox{low} \left( \mathbf{\Theta}_{ri} \right) - \mbox{low} \left( \mathbf{\Theta}_{r} \right) \bigg)  
\bigg( \mbox{low} \left( \mathbf{\Theta}_{ri} \right) - \mbox{low} \left( \mathbf{\Theta}_{r} \right) \bigg)^{\top}; \\
&\left[ \boldsymbol{\Sigma}_{\gamma,r}^{-1} \Big| \cdots \right] 
\sim \mathcal{W}\Big( \nu_{\gamma} + N_{r}, \left(\mathbf{H}_{\gamma} +  \mathbf{S}_{\gamma,r} \right)^{-1} \Big), \\
&\mathbf{S}_{\gamma,r} = \sum_{i=1}^{N_{r}} 
\bigg( \mbox{vec} \left( \mathbf{A}_{ri} \right) - \mbox{vec} \left( \mathbf{A}_{r} \right) \bigg)  
\bigg( \mbox{vec} \left( \mathbf{A}_{ri} \right) - \mbox{vec} \left( \mathbf{A}_{r} \right) \bigg)^{\top}; 
\end{align*}
\begin{align*}
&\left[ \boldsymbol{\Sigma}_{\theta}^{-1} \Big| \cdots \right] 
\sim \mathcal{W}\Big( \nu_{\theta} + R, \ \left(\mathbf{H}_{\theta} +  \mathbf{S}_{\theta} \right)^{-1} \Big), \\
&\mathbf{S}_{\theta} = \sum_{r=1}^{R} 
\bigg( \mbox{low} \left( \mathbf{\Theta}_{r} \right) - \mbox{low} \left( \mathbf{\Theta} \right) \bigg)  
\bigg( \mbox{low} \left( \mathbf{\Theta}_{r} \right) - \mbox{low} \left( \mathbf{\Theta} \right) \bigg)^{\top};\\
&\left[ \boldsymbol{\Sigma}_{a}^{-1} \Big| \cdots \right] 
\sim \mathcal{W}\Big( \nu_{a} + R, \ \left(\mathbf{H}_{a} +  \mathbf{S}_{a} \right)^{-1}  \Big), \\
&\mathbf{S}_{a} =  \sum_{r=1}^{R} 
\bigg( \mbox{vec} \left( \mathbf{A}_{r} \right) - \mbox{vec} \left( \mathbf{A} \right) \bigg)  
\bigg( \mbox{vec} \left( \mathbf{A}_{r} \right) - \mbox{vec} \left( \mathbf{A} \right) \bigg)^{\top}.
\end{align*}

\subsubsection*{Prior hyperparameter selection}

For $[\boldsymbol{\Sigma} \mid \mathbf{D} ]$ following a conjugate inverse Wishart distribution $\mathcal{IW}(\nu + n, \mathbf{H} + \mathbf{S})$, where we define $\mathbf{D} = \{\mathbf{x}_{i}\}_{i=1}^n$, the posterior mean of $\boldsymbol{\Sigma}$ (with dimension $p \times p$) has the form: $\mathrm{E}\left( \boldsymbol{\Sigma} \mid \mathbf{D} \right) = (\nu + n - p - 1)^{-1}(\mathbf{H} + \mathbf{S})$, where $\mathbf{S} = \sum_{i=1}^n (\mathbf{x}_i - \boldsymbol{\mu}) (\mathbf{x}_i - \boldsymbol{\mu})^{\top}$.  
Let $\nu = p$, $\mathbf{H} = \kappa \mathbf{I}$, 
the posterior mean of $\boldsymbol{\Sigma}$ becomes
\begin{align}
    \mathrm{E}\left( \boldsymbol{\Sigma} \mid \mathbf{D} \right) = \frac{1}{n-1} \sum_{i=1}^n (\mathbf{x}_i - \boldsymbol{\mu} ) (\mathbf{x}_i - \boldsymbol{\mu} )^{\top} +  \frac{\kappa}{n-1} \mathbf{I}. \label{equ:post_mean_sigma}
\end{align}
When substituting $\boldsymbol{\mu}$ with $\bar{\mathbf{x}}_n$ (i.e., the sample average) in \eqref{equ:post_mean_sigma}, the first term corresponds to the restricted maximum likelihood (REML) estimator of $\boldsymbol{\Sigma}$, while the identity matrix in the second term ensures that the posterior mean of $\boldsymbol{\Sigma}$ is positive definite, especially when $p > n-1$.
On the other hand, we require $\nu + n - p > 3$ to ensure the existence of the posterior covariance of $\boldsymbol{\Sigma}$,

In this paper, we denote the dimension of $\mbox{vec} \left( \mathbf{A}_{rij} \right)$ as $L_a = mQ^2$ and the dimension of $\mbox{low} \left( \mathbf{\Theta}_{rij} \right)$ as $L_{\theta} = (2P-Q+1) Q/2$. Assuming $N_r > 3$, we set $\nu_{v} = \nu_{\gamma} = L_a$ and $\nu_{u} = \nu_{\psi} = L_{\theta}$ as the degrees of freedom for the inverse Wishart prior at the segment- and subject-level. We further assign $\nu_a = L_a + 3$ and $\nu_{\theta} = L_{\theta} + 3$ to ensure the existence of the posterior covariance for $\boldsymbol{\Sigma}_a$ and $\boldsymbol{\Sigma}_{\theta}$, even when $R=2$ (noting that for $R=1$, the group-level hierarchy is unnecessary).

We set the scale matrices to all have the form $\kappa \mathbf{I}$. It is important to highlight that $\mathbf{S}_{u,r}$ and $\mathbf{S}_{v,r}$ are computed using all segment-level random effects for all subjects within a group, enabling them to estimate the true covariance matrix with relatively low variability. To ensure small estimation bias, it is crucial for $\kappa_u$ and $\kappa_v$ to be set to small values.
On the contrary, $\mathbf{S}_{\theta}$ and $\mathbf{S}_{a}$ are computed using only $R$ group-level random effects, necessitating $\kappa_{\theta}$ and $\kappa_{a}$ to be relatively large for computational robustness. Meanwhile, small values of $\kappa_{\theta}$ and $\kappa_{a}$ will cause group-level effects to be closer to each other, making it more challenging to detect group differences.
Alternatively, if we have prior information indicating that different subgroups share the same group-level effects (for instance, identical group-level spatial mapping matrices) and our interest lies in detecting differences only for the group-level temporal dynamical matrices, we can set $\kappa_a$ to be large and $\kappa_{\theta}$ to be small, allowing different groups to borrow spatial mapping information from each other.
In the paper, we opted for large values of $\kappa_{a}$ and $\kappa_{\theta}$ (using $\kappa_a = \kappa_{\theta} = 100$ in simulations and real data analysis) because (i) $R$ is small, and (ii) we aim to detect group differences for both spatial mapping and temporal dynamical matrices.
In Web Appendix C.4, we provide a detailed sensitivity analysis for selecting the values of $\kappa_u$, $\kappa_v$, $\kappa_{\psi}$, and $\kappa_{\gamma}$.


\subsection*{B.3. Computation time complexity}

For simplicity, in this section, we assume $N_r \equiv N$ and $J_{ri} \equiv J$. It is important to note that the major computation time costs arise from simulating $\left\{ \mathbf{M}_{rij}\left(t_{k}\right) \right\}_{k=1}^K$, $\mathbf{A}_{rij}$, and $\mathbf{\Theta}_{rij}$ for each group, subject, and time-segment.

For subgroup $r$, subject $i$, segment $j$,  since $\boldsymbol{Q}_{rijk}^{(M)}$ is identical at all time points $t_k \ (k=1,\dots, K)$, we only need to compute it once. Hence the computation time complexity for simulating $\left\{ \mathbf{M}_{rij}\left(t_{k}\right) \right\}_{k=1}^K$ is $\mathcal{O}\left( P^2 Q + mQ^3 + KPQ + K m^2 Q^2 + K Q^3  \right)$.
The computation time complexity for simulating $\mathbf{A}_{rij}$ is $\mathcal{O}\left( K m^2 Q^2 + m^3 Q^6  \right)$.
The computation time complexity for simulating $\mathbf{\Theta}_{rij}$ is $\mathcal{O}\left( K P Q + P^3 Q^3  \right)$.
Overall, the computational complexity per iteration of the proposed MCMC algorithm is essentially 
\begin{align*}
    \mathcal{O}\left( RNJ \left\{ m^3 Q^6 + P^3 Q^3 + KPQ + K Q^3 + K m^2 Q^2 \right\} \right).
\end{align*}
It is worth noting that we can compute each subject or segment in parallel. If the computation node has $C$ cores, the computation time will be reduced by a factor of $C$.
Specifically, for an MCMC run with ($R=1$, $N=75$, $J=30$, $P=54$, $Q=2$, $m=2$), each MCMC iteration takes approximately $4.8$ seconds when executed on a SLURM server with one computation node equipped with 25 CPU cores.

\section*{Web Appendix C}

This section shows additional simulation results for Section 4 in the paper, and additional simulation studies on MCMC diagnostics, sign identifiability, prior sensitivity, model selection, and model comparison with \cite{wang2022latent}.

\subsection*{C.1. Additional simulation results}

First of all, we provide detailed information on the assignments of variance components used in Section 4.1 of the paper.
For the variance components of the spatial mapping matrices for both scenarios, we set $\boldsymbol{\Sigma}_{u,1} = \boldsymbol{\Sigma}_{u,2} = \sigma_u^2 \mathbf{I}_{L_{\theta}}$, $\boldsymbol{\Sigma}_{\psi,1} = \boldsymbol{\Sigma}_{\psi,2} = \sigma_{\psi}^2 \mathbf{I}_{L_{\theta}}$, $\sigma_u = 0.04$, $\sigma_{\psi} = 0.075$, where $L_{\theta} = (2P-Q+1) Q/2$. 
For the variance components of the temporal dynamical matrices for both scenarios,
we assume $\boldsymbol{\Sigma}_{v,1} = \boldsymbol{\Sigma}_{v,2}$ to be diagnoal matrices. For the diagonal values, define $A_{rij1}(q_1, q_2)$ as the $(q_1, q_2)$ element of matrix $\mathbf{A}_{rij1}$,
we set $\mathrm{Var}\left\{A_{rij1}(q_1, q_1) \right\} = 0.08^2$, $\mathrm{Var}\left\{A_{rij2}(q_1, q_1) \right\} = 0.04^2$, $\mathrm{Var}\left\{A_{rijh}(q_1, q_2) \right\} = 0.03^2$ $(q_1 \neq q_2, h=1,2)$. 
Similarly, we assume $\boldsymbol{\Sigma}_{\gamma,1} = \boldsymbol{\Sigma}_{\gamma,2}$ to be diagonal matrices. The diagonal elements are set to be $\mathrm{Var}\left\{A_{ri1}(q_1, q_1) \right\} = 0.10^2$, $\mathrm{Var}\left\{A_{ri2}(q_1, q_1) \right\} = 0.05^2$, $\mathrm{Var}\left\{A_{rih}(q_1, q_2) \right\} = 0.03^2$ $(q_1 \neq q_2, h=1,2)$.

\begin{figure}[h]
 \centerline{\includegraphics[width=1.0\linewidth]{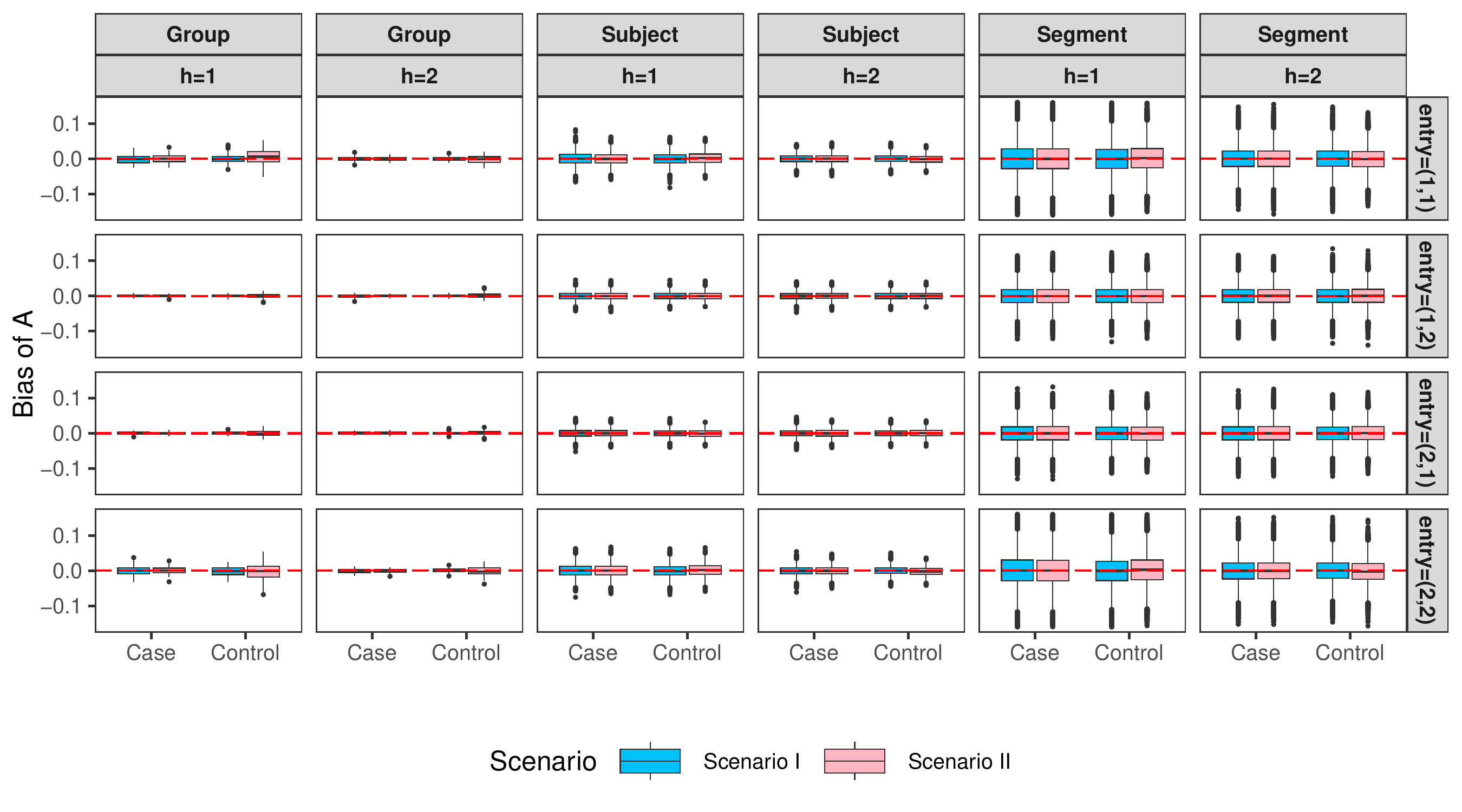}}
\caption{The boxplots of the biases of the group-, subject-, and segment-level temporal dynamical matrices $\mathbf{A}$ for each element $(q_1, q_2) \in \{1,2\}\times\{1,2\}$ at each time lag ($h=1,2$) for both the case and control groups ($r=1,2$) under the two sample size scenarios.}
\label{img:simu_error_A}
\end{figure}
\begin{figure}[h]
 \centerline{\includegraphics[width=0.8\linewidth]{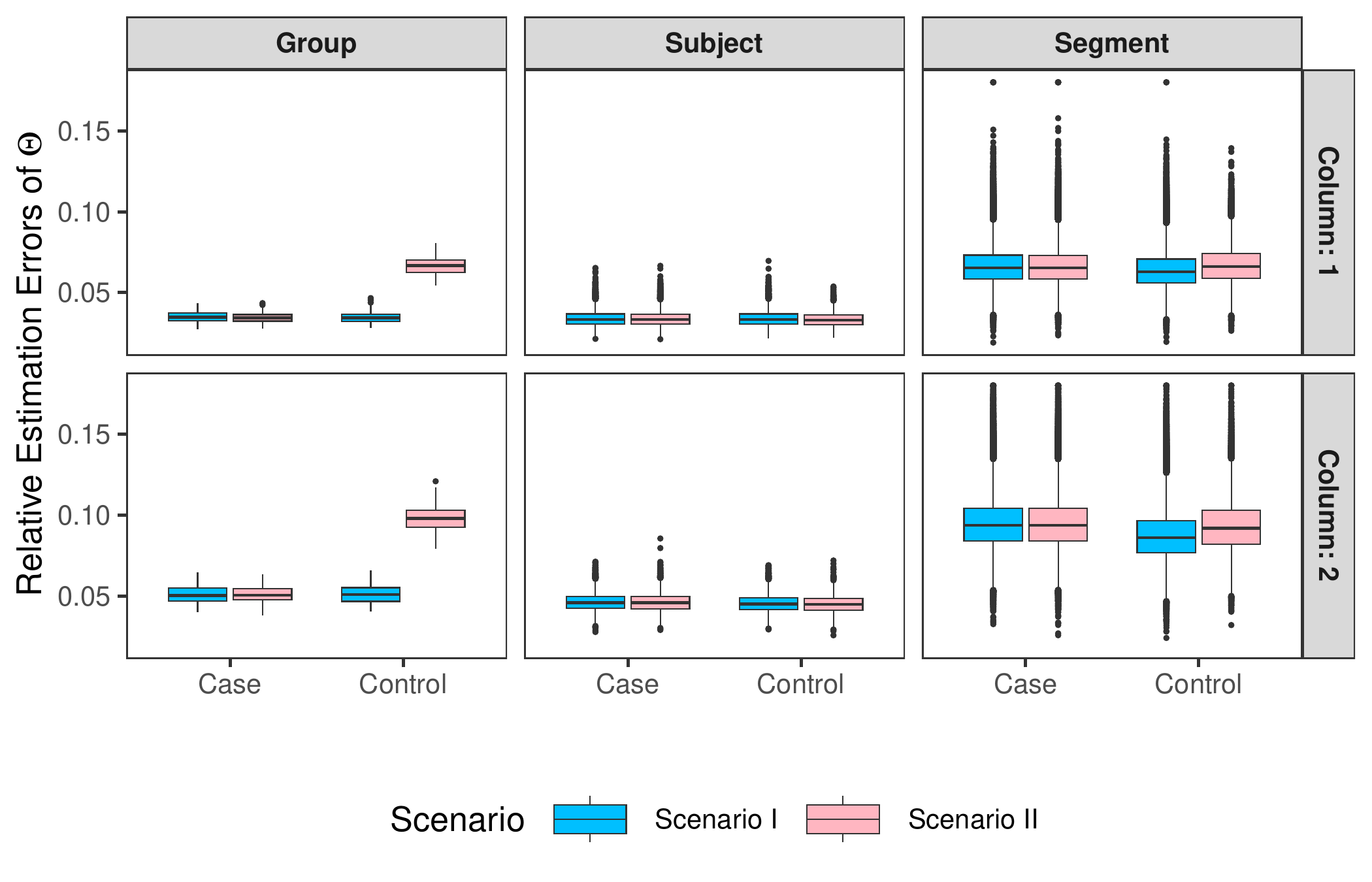}}
\caption{The boxplots of the relative estimation error of the group-, subject-, and segment-level of (each column of) the spatial mapping matrices $\mathbf{\Theta}$.}
\label{img:simu_error_Theta}
\end{figure}
\begin{figure}[h]
    \begin{subfigure}{0.57\textwidth}
        \centering
        \includegraphics[width=0.93\linewidth]{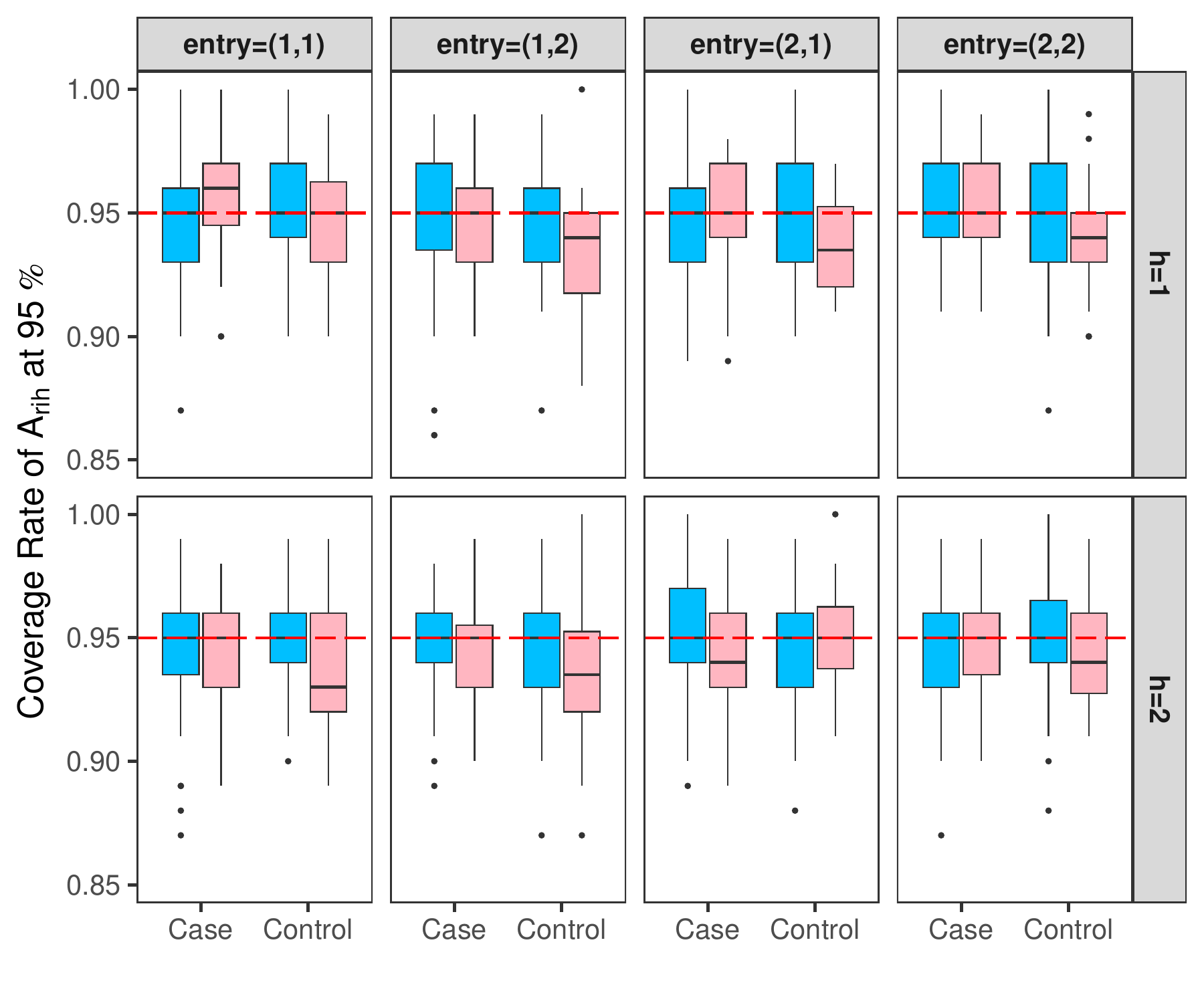}
        \caption{Coverage rate for $\mathbf{A}_{rih}$}
    \end{subfigure}
    \begin{subfigure}{0.42\textwidth}
        \centering
        \includegraphics[width=0.93\linewidth]{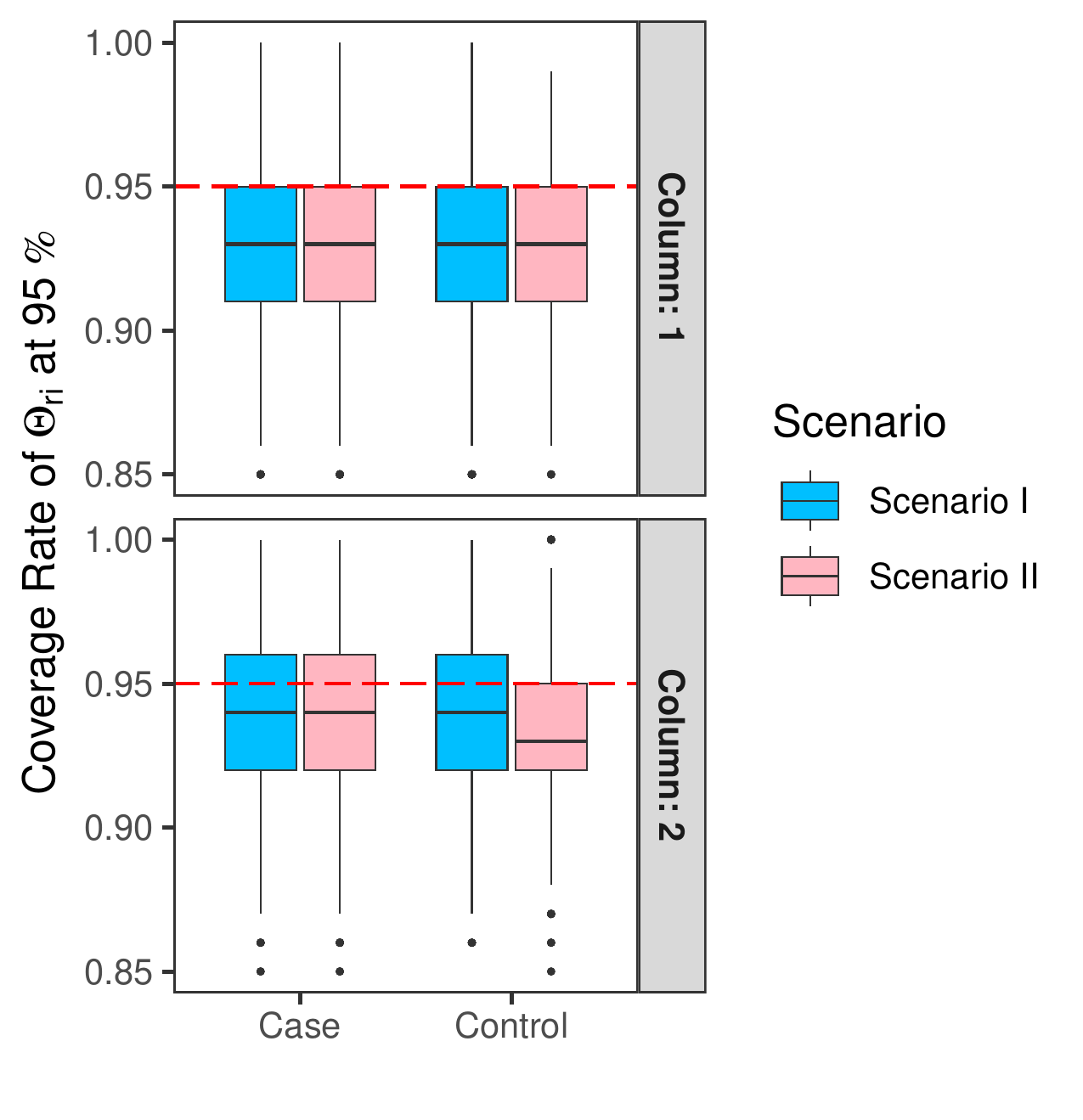}
        \caption{Coverage rate for $\mathbf{\Theta}_{ri}$}
    \end{subfigure}
    \caption{(a): the boxplots of the coverage rates of (each element of) the subject-level temporal dynamical matrices $\mathbf{A}_{rih}$. (b): the boxplots of the coverage rates of (each element of) the subject-level of spatial mapping matrices $\mathbf{\Theta}_{ri}$. Entries in each column are summarized by a box.   }
  	\label{img:simu_coverage}
\end{figure}

\begin{figure}[h]
 \centerline{\includegraphics[width=0.35\linewidth]{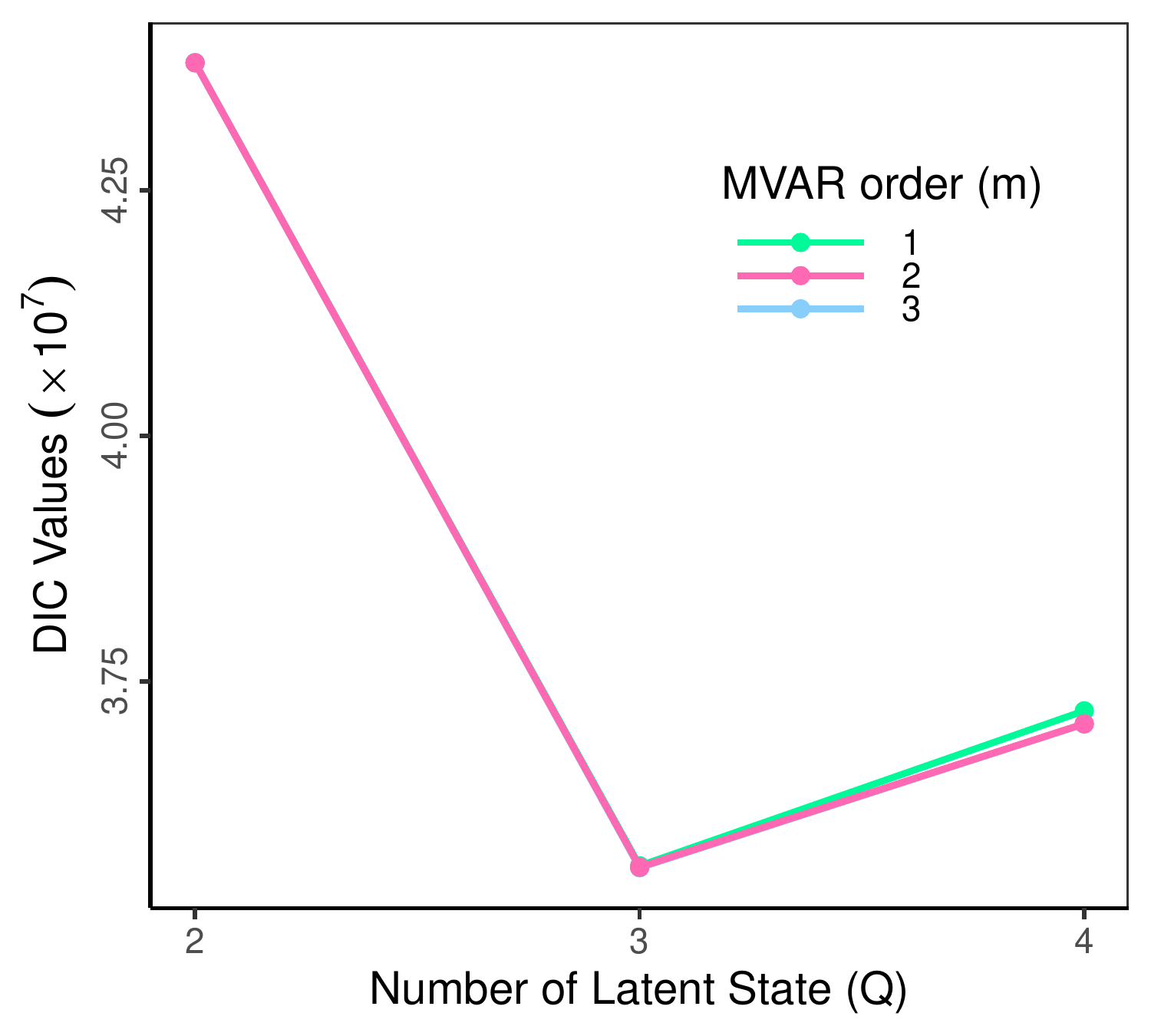}}
\caption{The cDIC values for the 9 models, where $Q$ varies in $\{2,3,4\}$ and $m$ varies in $\{1,2,3\}$, with the true model being $(Q=3, m=2)$.} 
\label{img:sensitivity}
\end{figure}

Next, we show additional figures for simulations in Section 4 in the paper.
Web Figure \ref{img:simu_error_A} shows the biases of the group-, subject-, and segment-level temporal dynamical matrices $\mathbf{A}$ for each element at each MVAR time lag (i.e., $h=1,2$) for both the case and control groups ($r=1,2$) under the two sample size scenarios. 
Web Figure \ref{img:simu_error_Theta} shows the relative estimation errors (REEs) of the group-, subject- and segment-level of (each column of) the spatial mapping matrices $\mathbf{\Theta}$. 
In the $d$-th replication of the simulated dataset ($d=1,\dots,100$), the REEs for the $q$th column of spatial mapping matrices at group-, subject-, and segment-levels (with true values as $\boldsymbol{\theta}_{r(q)}^{(d)}$, $\boldsymbol{\theta}_{ri(q)}^{(d)}$, $\boldsymbol{\theta}_{rij(q)}^{(d)}$) are defined as 
\begin{align*}
    \frac{\left\| \boldsymbol{\theta}_{r(q)}^{(d)} - \widehat{\boldsymbol{\theta}}_{r(q)}^{(d)} \right\|_2}{\left\| \boldsymbol{\theta}_{r(q)}^{(d)} \right\|_2},  \quad \quad
    \frac{\left\| \boldsymbol{\theta}_{ri(q)}^{(d)} - \widehat{\boldsymbol{\theta}}_{ri(q)}^{(d)} \right\|_2}{\left\| \boldsymbol{\theta}_{ri(q)}^{(d)} \right\|_2}, \quad \mbox{and} \quad
    \frac{\left\| \boldsymbol{\theta}_{rij(q)}^{(d)} - \widehat{\boldsymbol{\theta}}_{rij(q)}^{(d)} \right\|_2}{\left\| \boldsymbol{\theta}_{rij(q)}^{(d)} \right\|_2}, \quad \mbox{respectively},
\end{align*}
where $\widehat{\boldsymbol{\theta}}_{r(q)}^{(d)}$, $\widehat{\boldsymbol{\theta}}_{ri(q)}^{(d)}$, and $\widehat{\boldsymbol{\theta}}_{rij(q)}^{(d)}$ are the corresponding posterior mean, $\|\cdot\|_2$ is the Euclidean norm.
Web Figure \ref{img:simu_coverage} shows the coverage rates (CR) for the subject-level temporal dynamical matrices $\mathbf{A}_{rih}$ and spatial mapping matrices $\mathbf{\Theta}_{ri}$ at $95\%$. 
Specifically, the CR for $\mathbf{A}_{rih}$ and $\mathbf{\Theta}_{ri}$ are computed by 
\begin{align*}
    \mbox{CR}(A_{rih}(q_1, q_2)) &= \frac{1}{100} \sum_{d=1}^{100} I\Big\{ A_{rih}^{(d)}(q_1, q_2) \in \mathcal{CI}^{(d)} \Big\} \\
    \mbox{CR}(\Theta_{ri}(p, q)) &= \frac{1}{100} \sum_{d=1}^{100}   I\Big\{ \Theta_{ri}^{(d)}(p, q) \in \mathcal{CI}^{(d)} \Big\}.
\end{align*}
Here, $\mathbf{A}_{rih}^{(d)}$ and $\boldsymbol{\Theta}_{ri}^{(d)}$ represent the simulated true parameters in the $d$-th replication of the simulated dataset, and $\mathcal{CI}^{(d)}$ denotes the corresponding collection of $95\%$ credible intervals for all parameters. 
Note that the coverage rate (CR) for the subject-level temporal dynamical matrices are concentrated around $95\%$, while the CR for the subject-level spatial mapping matrices are concentrated around $93\%-94\%$. 
To bring the CR closer to the nominal level for subject-level spatial mapping matrices, as evident from \eqref{equ:post_mean_sigma} in Web Appendix B.2, an increase in $\kappa_{\psi}$ leads to an increase in the posterior mean of $\mathbf{\Sigma}_{\psi,r}$ (i.e., the variance components of $\mathbf{\Theta}_{ri}$). Consequently, this results in wider credible intervals and larger coverage rates.
Web Figure \ref{img:sensitivity} shows the cDIC values for the 9 models (1 true, 8 misspecified) discussed in Section 4.3 in the paper, where $Q$ varies in $\{2,3,4\}$ and $m$ varies in $\{1,2,3\}$, with the true model being $(Q=3, m=2)$.
We observed that cDIC can choose the correct model in terms of the number of latent states $Q$, as the true model has the smallest cDIC values among $Q=2,3,4$ even when the MVAR order $m$ is misspecified. The figure also suggests that the model's performance is not significantly affected by the MVAR order $m$.

In the remainder of Web Appendix C, we use the first 20 EEG channels ($P=20$) to shorten computation time. The number of latent states $Q$ may vary from Web Appendix C.2 to Web Appendix C.6 as needed.  
We set all the parameters using the same manner as in Section 4 of the paper.

\subsection*{C.2. MCMC diagnostics}


This simulation assesses the MCMC chains mixing and convergence for the spatial mapping matrices and temporal dynamical matrices at each level. We use a single group (the control group, i.e. $r=2$) with ($n=50$, $J=30$, $P=20$, $Q=5$, $m=2$).  

\begin{figure}[h]
    \begin{subfigure}{0.496\textwidth}
        \centering
        \includegraphics[width=1.01\linewidth]{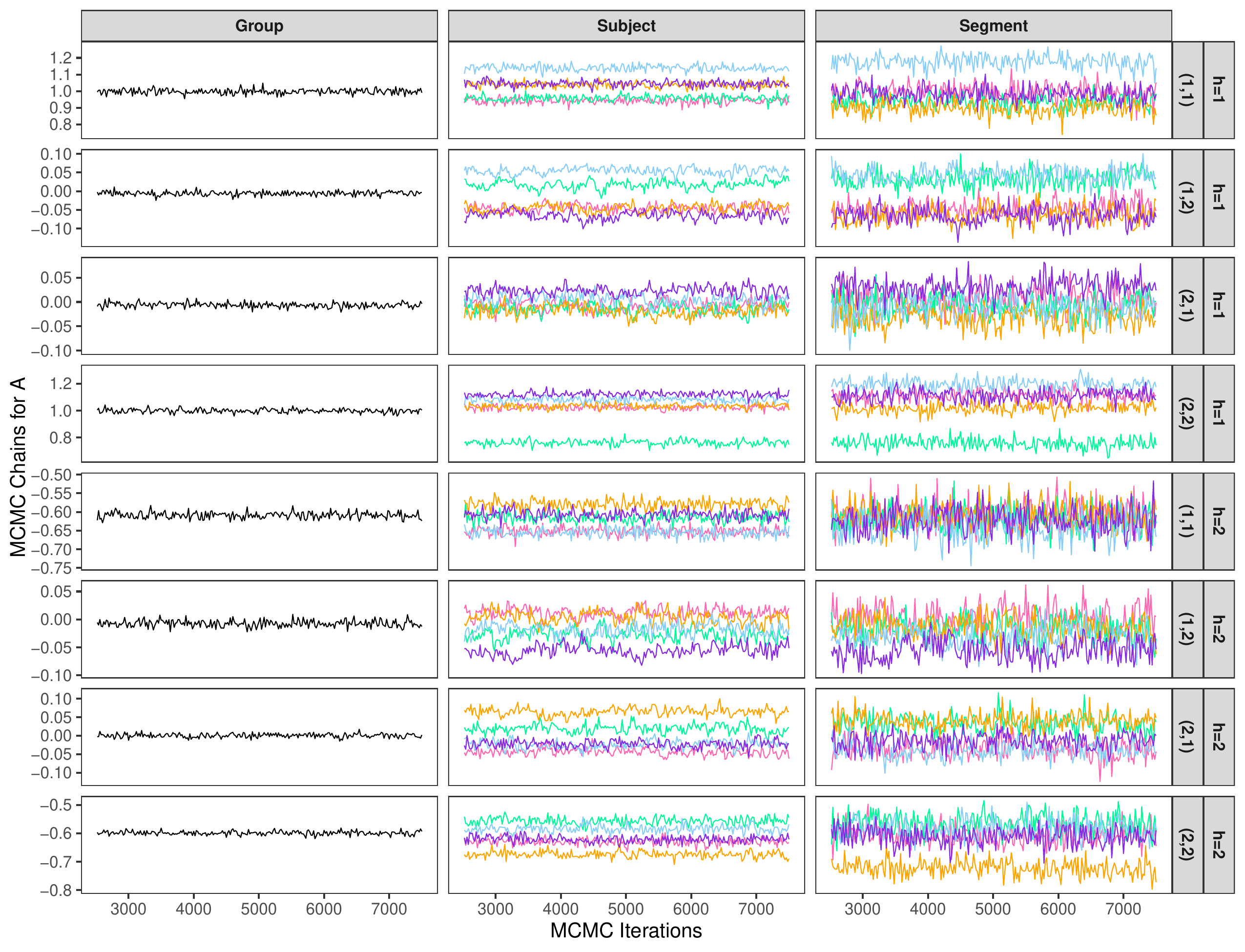}
        \caption{Trace plots for $\mathbf{A}$ matrices}
    \end{subfigure}
    \begin{subfigure}{0.496\textwidth}
        \centering
        \includegraphics[width=1.01\linewidth]{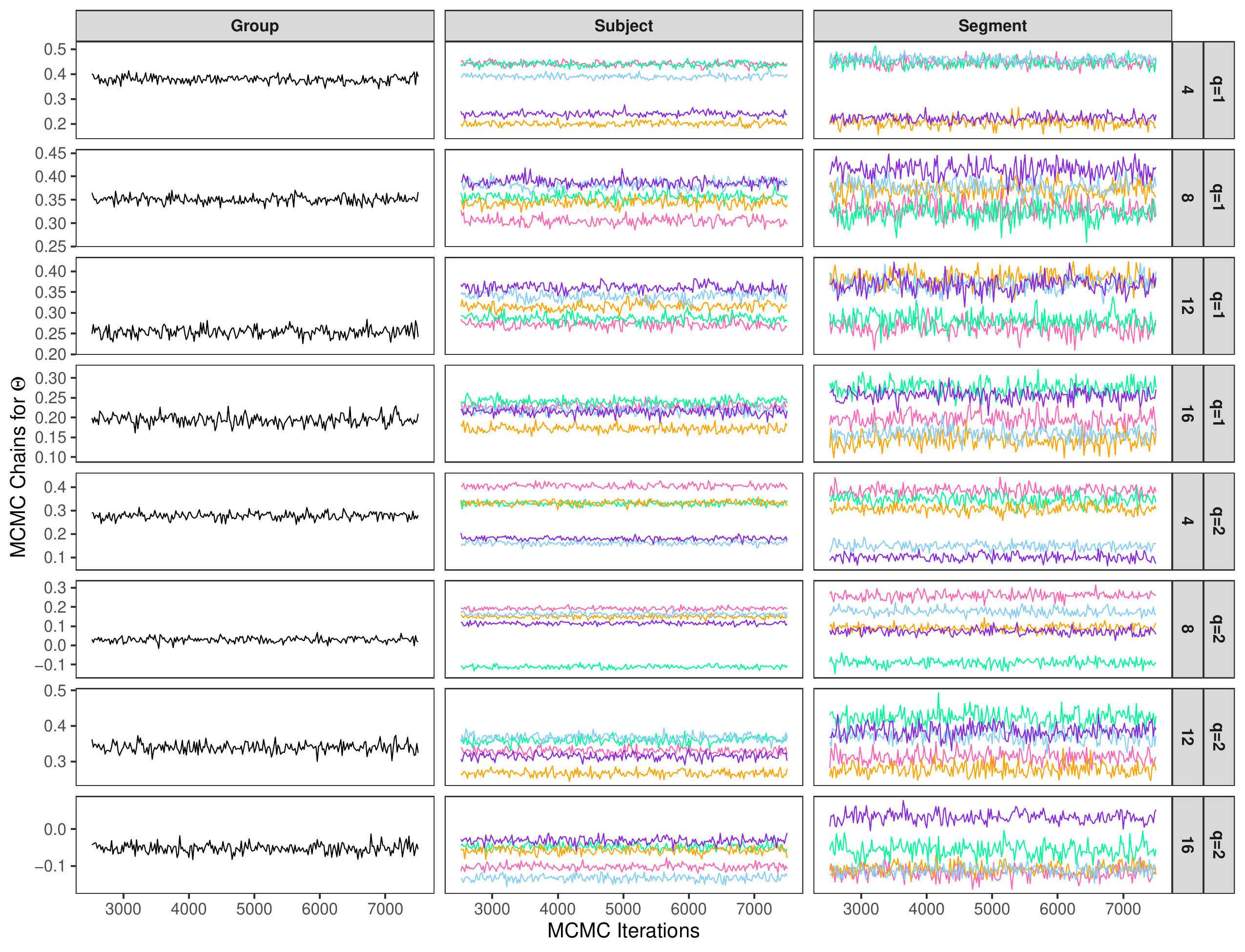}
        \caption{Trace plots for $\mathbf{\Theta}$ matrices}
    \end{subfigure}
    \begin{subfigure}{0.496\textwidth}
        \centering
        \includegraphics[width=1.01\linewidth]{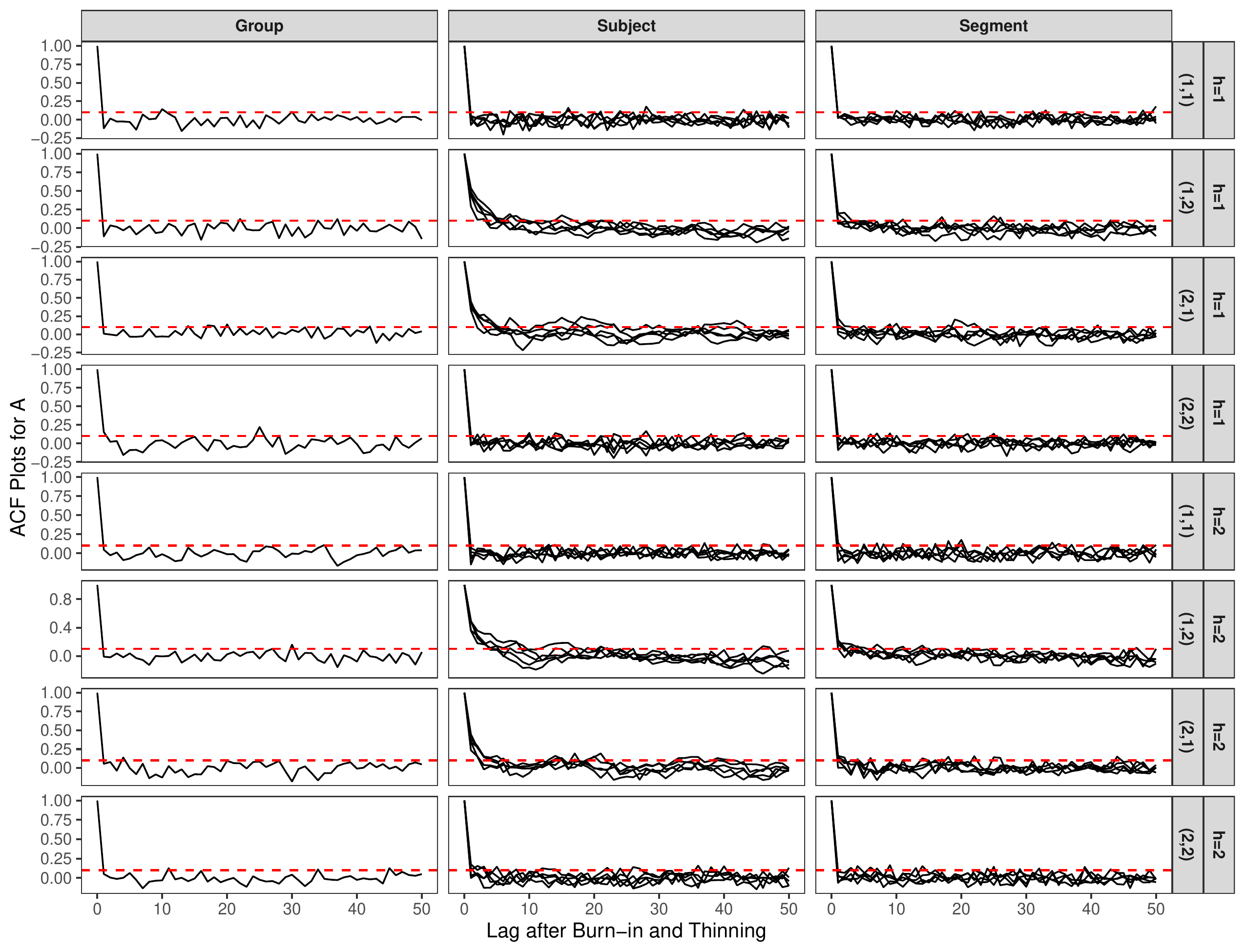}
        \caption{ACF plots for $\mathbf{A}$ matrices}
    \end{subfigure}
    \begin{subfigure}{0.496\textwidth}
        \centering
        \includegraphics[width=1.01\linewidth]{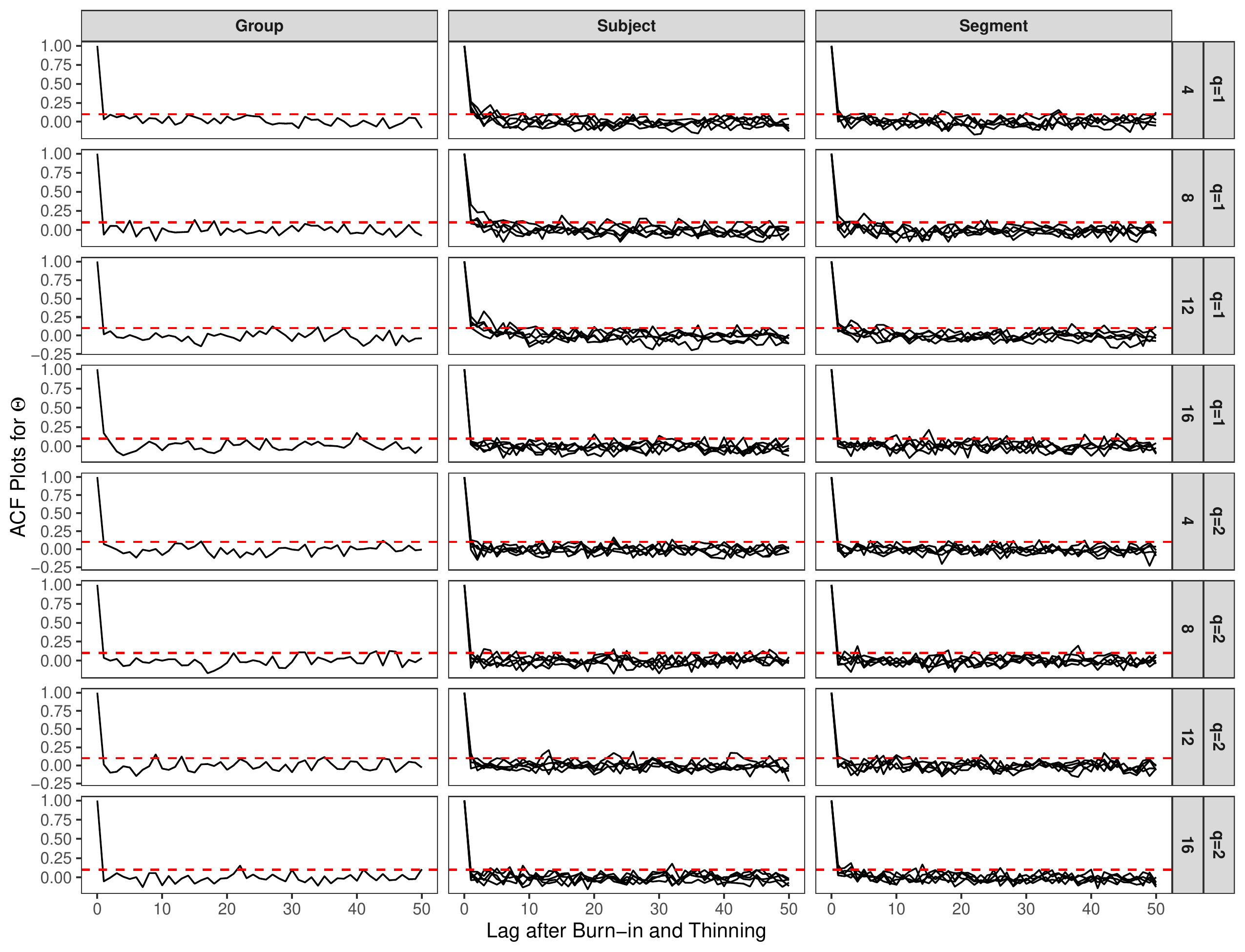}
        \caption{ACF plots for $\mathbf{\Theta}$ matrices}
    \end{subfigure}
    \caption{Trace plots and autocorrelation function (ACF) plots for group-, subject-, and segment-level $\mathbf{A}$ and $\mathbf{\Theta}$ matrices to assess MCMC mixing and convergence. The red dashed lines in the ACF plots are $y=0.1$.}
  	\label{img:trace_acf}
\end{figure}

Web Figure \ref{img:trace_acf} displays the MCMC trace plots and autocorrelation function (ACF) plots for group-, subject-, and segment-level $\mathbf{A}$ and $\mathbf{\Theta}$ matrices. 
For $\mathbf{A}$ matrices, each element in the two MVAR lags are presented; for $\mathbf{\Theta}$ matrices, the $4$th, $8$th, $12$th, anf $16$th elements in the two columns are presented.
For subject-level $\mathbf{A}_{rih}$ and $\mathbf{\Theta}_{ri}$, the MCMC chains for the five subjects ($i=1,\dots,5$) are presented. For segment-level $\mathbf{A}_{rijh}$ and $\mathbf{\Theta}_{rij}$, the MCMC chains for the five subjects ($i=1,\dots,5$) at the 28s-30s segment ($j=15$) are presented. The red dashed lines in the ACF plots indicate $y=0.1$. From the trace plots, we observe that the MCMC chains for all parameters stabilize after 2500 burn-ins. From the ACF plots, we note that most parameters have ACF values smaller than $0.1$ after $10$ lags. These figures demonstrate good mixing and convergence of the MCMC sampler using the proposed algorithm.

In the context of any hierarchical models, research on the convergence of the Gibbs sampler has explored factors such as the tail behavior of error distributions \citep{papaspiliopoulos2008stability, auger2016state} and the impact of choosing between centered and non-centered parametrizations \citep{papaspiliopoulos2007general}. An extension of this work may involve relaxing the Gaussian constraint and incorporating non-Gaussian distributions for errors in both sensor and MVAR models. Consideration of a mixture of centered and non-centered parameterizations could be explored to ensure faster convergence.

\subsection*{C.3. Sign identifiability}

This simulation study examines the effectiveness of our proposed two-stage MCMC algorithm in addressing the sign identifiability issue. We compared our algorithm (Two-stage), which includes both initialization and sign-tracking stages, with three alternative MCMC algorithms. The first alternative involves using only the initialization stage (Initialize-only), the second alternative involves using only the sign-tracking stage (Tracking-only)
while the third alternative employs MCMC without initialization or sign-tracking (No-treatment).
We use a single group (the control group, i.e. $r=2$) with ($n=50$, $J=30$, $P=20$, $Q=5$, $m=2$).
The MCMC computation consisted of 10,000 iterations, with the first 5,000 iterations treated as burn-in and a thinning rate of 10. For both our proposed Two-stage algorithm and the Initialize-only algorithm, we added an extra 5,000 iterations for initialization before the main MCMC step.
For both our proposed Two-stage algorithm and the Tracking-only algorithm, we change the signs of the latent signals if the cosine correlations (as defined in Web Appendix A) are smaller than $0$ (i.e., $\rho_0 = 0$). The sign-tracking steps are exclusively implemented during the 2,500 to 5,000 iterations (check once per 10 iterations) within the burn-in step of the MCMC process. Finally, we do not assume any prior knowledge of each level of the spatial mapping matrices and assign initial values to be all zero matrices for each of them.

For the segment- and trial-level spatial mapping matrices, we evaluate the correct sign identification for each column by computing the cosine correlation between the MCMC samples and the underlying truth. A cosine correlation greater than $0$ is considered correct identification. The Correct Sign Identification Rate (CSIR) over MCMC iterations and segments/subjects, as well as the Mean Estimation Error (MEE) between the posterior mean and the underlying truth over segments/subjects for the four methods are provided in Web Table \ref{table:sign}. 
To be more specific, consider the $q$-th column of the segment-level spatial mapping matrices. Let $\boldsymbol{\theta}_{rij(q)}$ represents the truth value for the $q$-th column of $\mathbf{\Theta}_{rij}$, $\widehat{\boldsymbol{\theta}}_{rij(q)}$ denotes the corresponding posterior mean, and $\boldsymbol{\theta}_{rij(q)}^{*(l)}$ denotes the corresponding MCMC sample at the $l$-th iteration. Subsequently, the CSIR and the MEE are defined as follows:
\begin{align*}
    \mbox{CSIR}(q, \mbox{segment}) &= \frac{1}{LNJ} \sum_{l=1}^L  \sum_{i=1}^N \sum_{j=1}^J I\left( \mbox{cos}\left(\boldsymbol{\theta}_{rij(q)}, \boldsymbol{\theta}_{rij(q)}^{*(l)} \right) \ge 0 \right) \\
    \mbox{MEE}(q, \mbox{segment}) &= \frac{1}{NJ}  \sum_{i=1}^N \sum_{j=1}^J  \left\| \boldsymbol{\theta}_{rij(q)} - \widehat{\boldsymbol{\theta}}_{rij(q)} \right\|_2, 
\end{align*}
where $\| \cdot \|_2$ denotes the Euclidean norm.

\begin{table}[t]
\caption{Summary of the sign identification performance among the four methods.}
\begin{center}\vskip -.15in
\scriptsize
\begin{tabular}{cccccccccccccc}
\toprule
 & &&  \multicolumn{2}{c}{No-treatment} && \multicolumn{2}{c}{Initialize-only} && \multicolumn{2}{c}{Tracking-only} && \multicolumn{2}{c}{Two-stage} \\
 \midrule
Level   & q  &&      MEE  & CSIR ($\%$)       && MEE  & CSIR ($\%$)    && MEE  & CSIR ($\%$) && MEE  & CSIR ($\%$) \\
                     \cline{4-5}          \cline{7-8}              \cline{10-11}    \cline{13-14}\\
Subject &  1 &&    1.398 & 53.1    &&  0.050 & 100.0  &&  0.050 & 100.0    && 0.050  & 100.0\\
        &  2 &&    0.897 & 81.1    &&  0.055 & 100.0  &&  0.051 & 100.0    && 0.050  & 100.0\\
        &  3 &&    0.790 & 80.6    &&  0.063 & 100.0  &&  0.055 & 100.0    && 0.054  & 100.0\\
        &  4 &&    0.453 & 64.2    &&  0.228 &  85.4  &&  0.120 & 100.0    && 0.070  & 100.0\\
        &  5 &&    0.422 & 53.7    &&  0.110 & 100.0  &&  0.129 &  97.2    && 0.070  & 100.0\\
Segment &  1 &&    1.487 & 50.1    &&  0.103 & 100.0  &&  0.103 & 100.0    && 0.103  & 100.0\\
        &  2 &&    0.987 & 52.3    &&  0.104 & 100.0  &&  0.101 & 100.0    && 0.100  & 100.0\\
        &  3 &&    0.876 & 53.3    &&  0.107 & 100.0  &&  0.101 & 100.0    && 0.100  & 100.0\\
        &  4 &&    0.543 & 56.6    &&  0.292 &  80.6  &&  0.161 &  99.9    && 0.113  & 100.0\\
        &  5 &&    0.488 & 55.6    &&  0.154 &  97.7  &&  0.170 &  97.2    && 0.115  &  99.1\\
\bottomrule
\end{tabular}
\end{center}
\label{table:sign}
\footnotesize \textcolor{black}{(MEE): mean estimation error; (CSIR): correct sign identification rate; (q): one of the five latent states; (No-treatment): the MCMC algorithm without initialization and sign-tracking.  (Initialize-only): the MCMC algorithm only having the initialization stage; (Tracking-only): the MCMC algorithm only having the sing-tracking stage; (Two-stage): our proposed method;}
\end{table}

From Web Table \ref{table:sign}, we observe that both the initialization stage and sign-tracking stage significantly mitigate the sign identification issue, especially for the first three latent states where the sign-to-noise ratio is relatively high. In these cases, the sign identification issue is effectively resolved. However, for the $4$th and $5$th latent states, which have a relatively low sign-to-noise ratio, there may still be a small portion with incorrectly identified signs, resulting in a larger estimation error.
The implementation of the both stages further addresses the sign identification issue, resulting in correct identification of the signs for all subject-level latent signals and almost all trial-level signals. This highlights the effectiveness of our proposed two-stage method.

\subsection*{C.4. Prior sensitivity and hyperparameter selection}

\begin{figure}[h]
 \centerline{\includegraphics[width=0.97\linewidth]{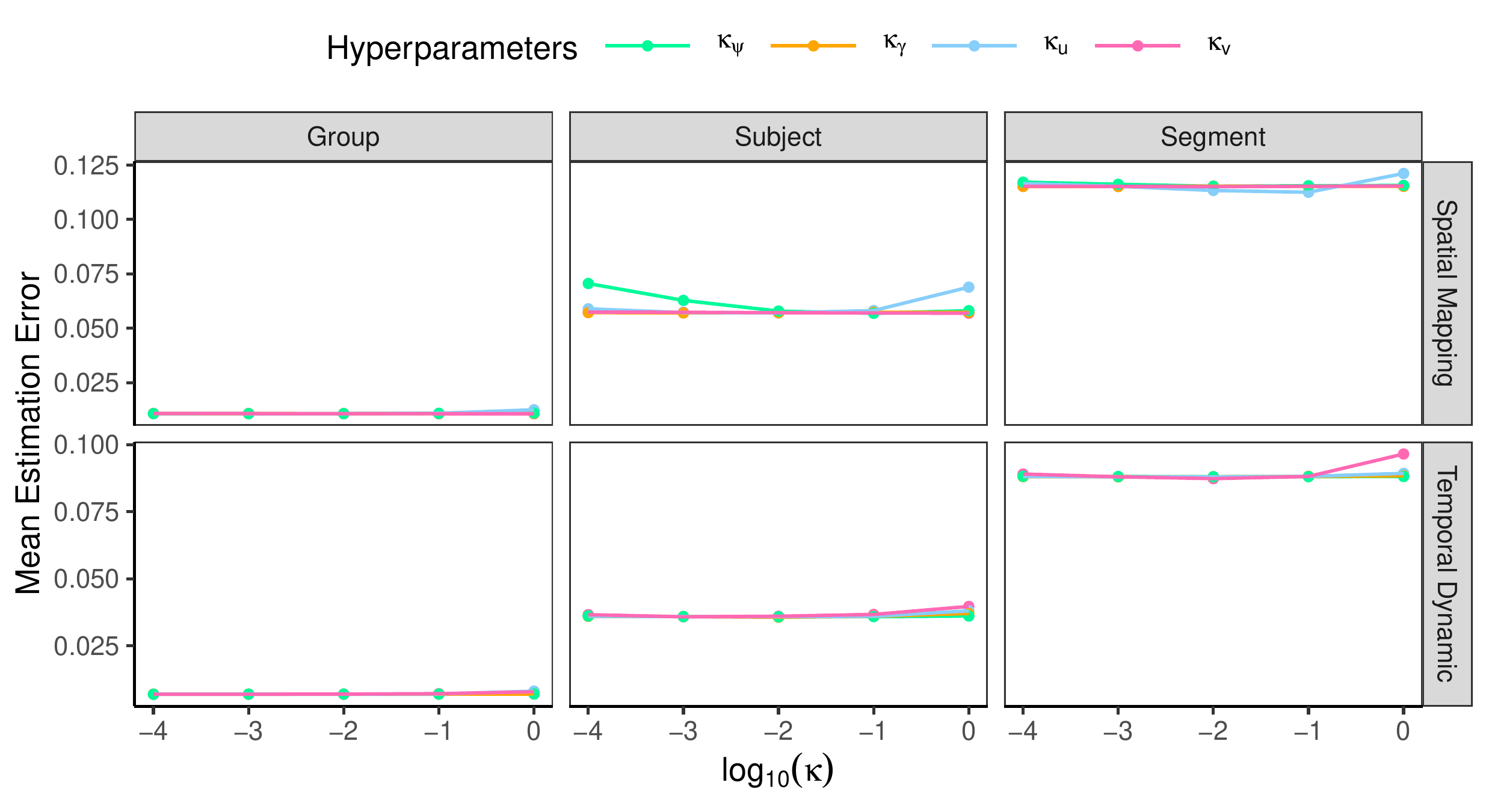}}
\caption{Sensitivity analysis for the selection of prior hyperparameters. Model performance is assessed using the MEE for group-, subject-, and segment-level spatial mapping matrices, as well as temporal dynamic matrices.}
\label{img_hyper_selec}
\end{figure}

This simulation study provides a detailed sensitivity analysis for selecting the values of $\kappa_u$, $\kappa_v$, $\kappa_{\psi}$, and $\kappa_{\gamma}$ in the priors of variance components.
We use a single group (the control group, i.e. $r=2$) with ($n=50$, $J=30$, $P=20$, $Q=2$, $m=2$). The EEG data is simulated for 10 replications. 
Separate MCMC runs are conducted, each with one of the values for $\kappa$ selected from the set $\{10^{-4}, 10^{-3}, 10^{-2}, 10^{-1}, 1\}$, while keeping the other three $\kappa$ values fixed at $10^{-3}$.
We compute the posterior mean of the parameters of interest from the MCMC samples. Model performance is then assessed using the Mean Estimation Error (MEE) for group-, subject-, and segment-level spatial mapping matrices, as well as temporal dynamic matrices.
For example, to evaluate the model's performance in modeling segment-level spatial mapping matrices, we compute the following MEE:
\begin{align*}
    \frac{1}{10 N J} \sum_{d=1}^{10} \sum_{i=1}^N \sum_{j=1}^J \left\| 
    \mbox{low}\left( \boldsymbol{\Theta}_{rij}^{(d)} \right) 
    - \mbox{low}\left( \widehat{\boldsymbol{\Theta}}_{rij}^{(d)} \right) \right\|_2,
\end{align*}
where $d = 1, \dots, 10$ denotes the simulation replicates, $\widehat{\boldsymbol{\Theta}}_{rij}^{(d)}$ denotes the posterior mean for the true value $\boldsymbol{\Theta}_{rij}^{(d)}$, $\| \cdot \|_2$ denotes the Euclidean norm.
The simulation results are shown in Web Figure \ref{img_hyper_selec}.

Analysis of Web Figure \ref{img_hyper_selec} reveals the overall robustness of the proposed model to the choices of $\kappa_{u}$, $\kappa_{v}$, $\kappa_{\psi}$, and $\kappa_{\gamma}$. However, a large value of $\kappa_u$ (e.g., $\kappa_u = 1$) can lead to higher MEE for the segment- and subject-level of spatial mapping matrices, while a large value of $\kappa_v$ (e.g., $\kappa_v = 1$) results in increased MEE specifically for segment-level temporal dynamic matrices. Conversely, a very small value of $\kappa_{\psi}$ (e.g., $\kappa_{\psi} = 10^{-4}$) can cause higher MEE for subject-level spatial mapping matrices, while the model's performance remains stable with variations in $\kappa_{\gamma}$.

\subsection*{C.5. Alternative model selection criteria versus cDIC}

\subsubsection*{Alternative model selection criteria}
This section introduces three alternative model selection methods based on DIC, along with a simulation comparing cDIC with the three methods.
First note that, in addition to the complete DIC, \cite{celeux2006deviance} proposed alternative observed DICs by using the observed likelihood $p(\mathbf{Y} \mid \mathbf{\Lambda}) =  \int p(\mathbf{Y}, \mathbf{M} \mid \mathbf{\Lambda}) d \mathbf{M}$, and conditional DICs by using the conditional likelihood $p(\mathbf{Y} \mid \mathbf{M}, \mathbf{\Lambda})$ in the DIC computation.
The observed likelihood for a state-space model is computed following \cite{de1988likelihood}, and obtaining it directly from MCMC samples is not feasible. In contrast, the conditional log-likelihood is easy to compute. 
The conditional DIC can be understood by treating the latent variable 
$\mathbf{M}$ as an additional parameter.

Denote $\widetilde{\mathbf{Y}}_{rij}\left(t_{k}\right) = \mathbf{\Theta}_{rij} \mathbf{M}_{rij}\left(t_{k}\right)$ as the EEG signal without random measurement error. The conditional likelihood can then be reparameterized as  $p(\mathbf{Y} \mid \mathbf{\Lambda}_1)$, where $\mathbf{\Lambda}_1 = \{\mathbf{\Lambda}, \widetilde{\mathbf{Y}}(\mathbf{M}, \mathbf{\Lambda}) \}$.
Define
\begin{align*}
    \mbox{D}(\widebar{\mathbf{\Lambda}}_1) = -2 \log p( \mathbf{Y} \mid \widehat{\mathbf{\Lambda}}_1 ) \quad \mbox{and} \quad
    \widebar{\mbox{D}(\mathbf{\Lambda}_1)} = -2 \mathbb{E}_{\mathbf{\Lambda}_1 \mid \mathbf{Y}} \left[ \log p( \mathbf{Y} \mid \mathbf{\Lambda}_1 )  \right],
\end{align*}
where $\widehat{\mathbf{\Lambda}}_1$ is the posterior mean of $\mathbf{\Lambda}_1$.
The conditional DIC is represented by
\begin{align}
    \mathrm{DIC}_1 = \mbox{D}(\widebar{\mathbf{\Lambda}}_1) + 2 p_D, \quad p_D = \widebar{\mbox{D}(\mathbf{\Lambda}_1)} - \mbox{D}(\widebar{\mathbf{\Lambda}}_1),
    \label{equ:DIC1}
\end{align}
where $p_D$ is the effective number of parameters. Equation \eqref{equ:DIC1} is esstentially the classical DIC proposed by \cite{spiegelhalter2002bayesian} with parameter $\mathbf{\Lambda}_1$. \cite{spiegelhalter2014deviance} concluded that $\mathrm{DIC}_1$ overfits because it employs a plug-in estimate in the predictive target rather than the proper predictive distribution. A better method would be to increase the penalty in the following: 
\begin{align*}
    \mathrm{DIC}_2 = \mbox{D}(\widebar{\mathbf{\Lambda}}_1) + 3 p_D.
\end{align*}
\cite{gelman2003bayesian} provide an alternative method for computing the effective number of parameters:
\begin{align}
    \mathrm{DIC}_3 = \mbox{D}(\widebar{\mathbf{\Lambda}}_1) + 2 p_V, \quad p_V = 2 \mathbb{V}_{\mathbf{\Lambda}_1 \mid \mathbf{Y}} \left[ \log p( \mathbf{Y} \mid \mathbf{\Lambda}_1 ) \right],
\end{align}
where $\mathbb{V}_{\mathbf{\Lambda}_1 \mid \mathbf{Y}}$ is the conditional variance. 
In addition to the three DIC-based methods, \cite{watanabe2010asymptotic} introduced a fully Bayesian method called WAIC, which has been demonstrated to be asymptotically equal to using Bayesian leave-one-out cross-validation. However, as our RESSM involves multi-level models, potentially leading to different definitions of WAIC, we do not delve into WAIC in the paper.

\subsubsection*{Comparative analysis of model selection criteria}

\begin{figure}[h]
 \centerline{\includegraphics[width=0.97\linewidth]{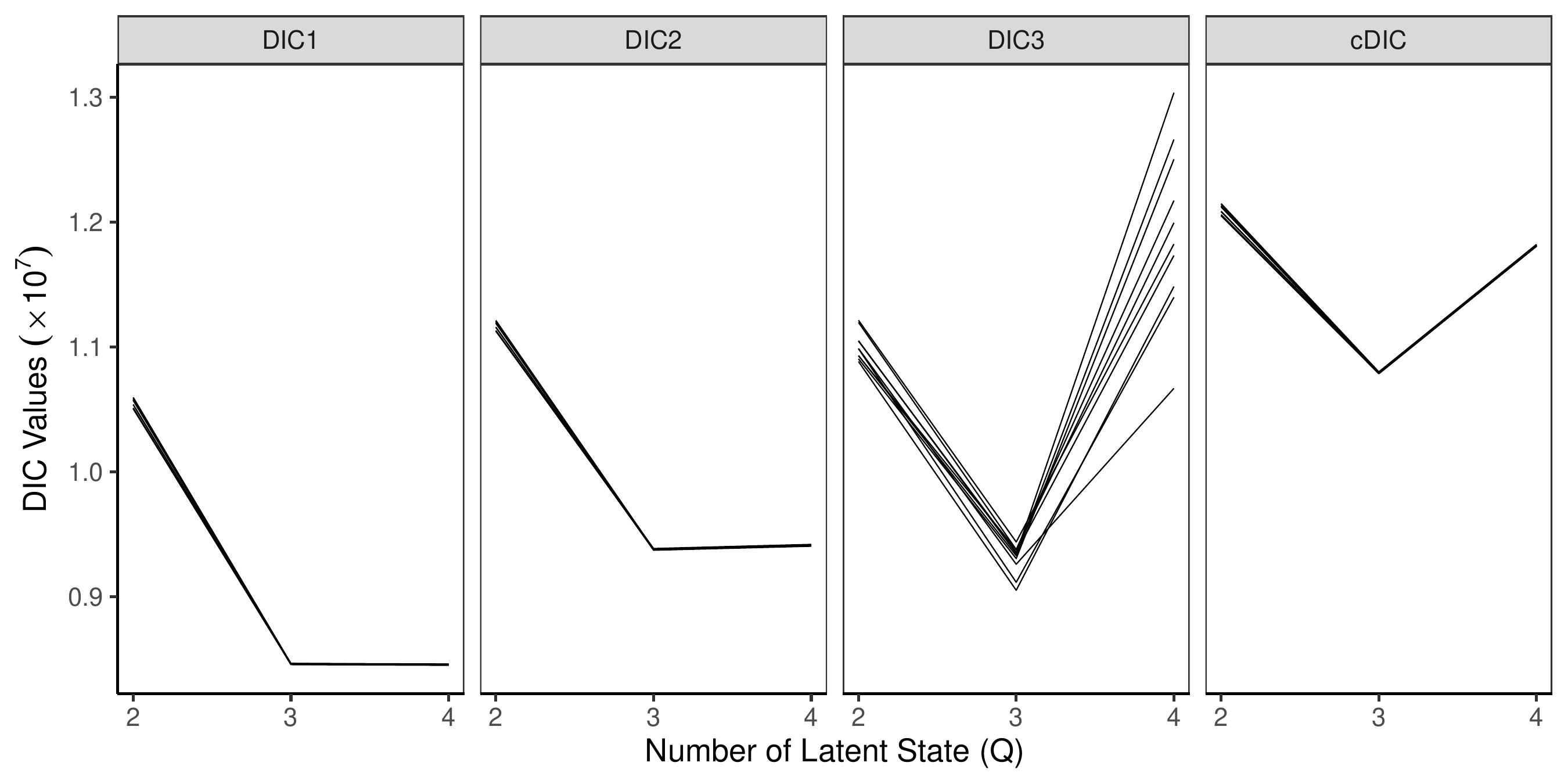}}
\caption{Comparative analysis of cDIC against three alternative model selection methods for identifying the true number of latent states ($Q=3$) among the possibilities $Q=2, 3, 4$. Each line represents one replication. }
\label{img:DIC}
\end{figure}

In Section 4.3 of this paper, we demonstrate that the proposed RESSM exhibits robustness in the face of misspecification of the MVAR order $m$. Consequently, this section only assesses the performance of cDIC and three alternative DIC methods in selecting the number of latent states $Q$.
We use a single group (the control group) with ($n=50$, $J=30$, $P=20$, $Q=3$, $m=2$). The EEG data is simulated for 10 replications. 
For the MCMC analysis, we held the MVAR order $m$ fixed at $2$ and varied the number of latent states $Q$ from $2$ to $4$ in the RESSM. For each replicate, we computed the cDIC values and the $\mbox{DIC}_k$ values ($k=1,2,3$) for the RESSMs with $Q=2,3,4$. The results are presented in Web Figure \ref{img:DIC}.

From Web Figure \ref{img:DIC}, we observed that all four methods favor the true model over the underfitting model with $Q=2$. However, only $\mbox{DIC}_3$ and cDIC show a preference for the true model over the overfitting model with $Q=4$, whereas $\mbox{DIC}_1$ and $\mbox{DIC}_2$ exhibit nearly identical DIC values for the two models with $Q=3$ and $Q=4$, suggesting a tendency to select overfitting models. This suggests a preference for using cDIC or $\mbox{DIC}_3$.
On the other hand, $\mbox{DIC}_3$ displays a considerably larger variation compared to the other three methods. For stable evaluation of the model performance, cDIC emerges as the preferred choice among the four methods.

\subsection*{C.6. Model comparison}

\begin{figure}[h]
    \begin{subfigure}{0.9\textwidth}
        \begin{flushright}
        \includegraphics[width=0.9\linewidth]{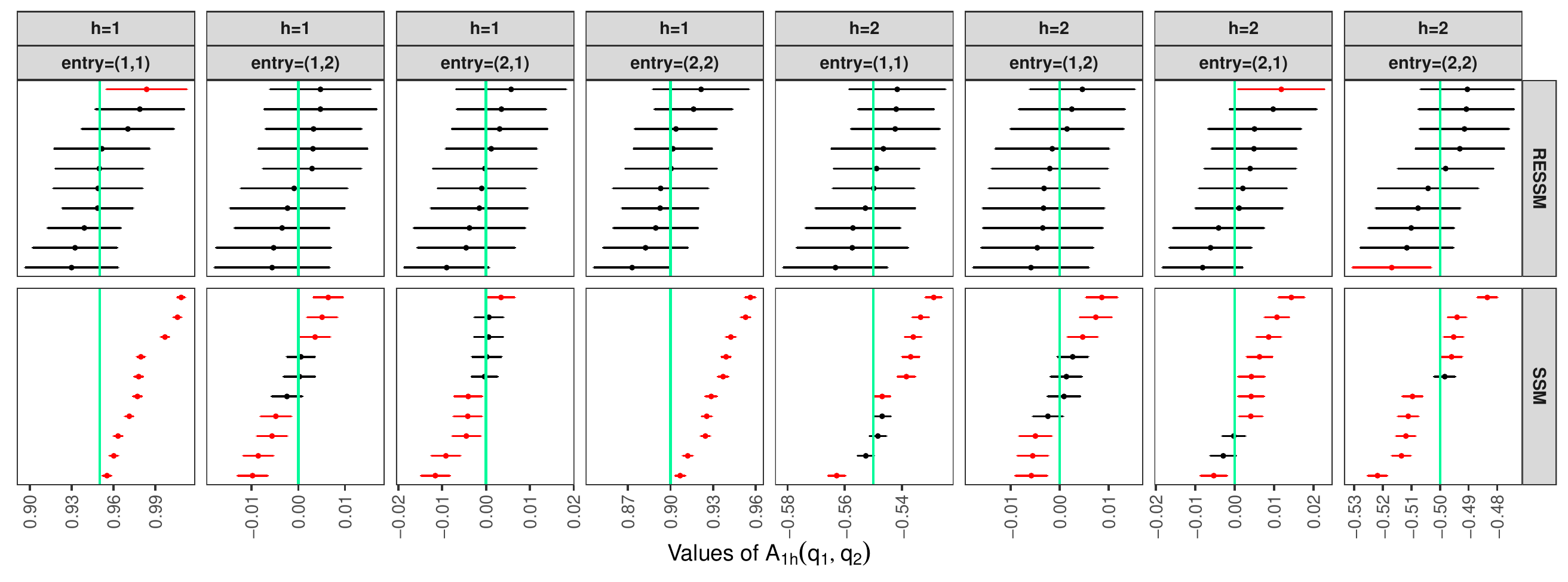}
        \end{flushright}
    \end{subfigure}
    \begin{subfigure}{0.9\textwidth}
        \begin{flushright}
        \includegraphics[width=0.9\linewidth]{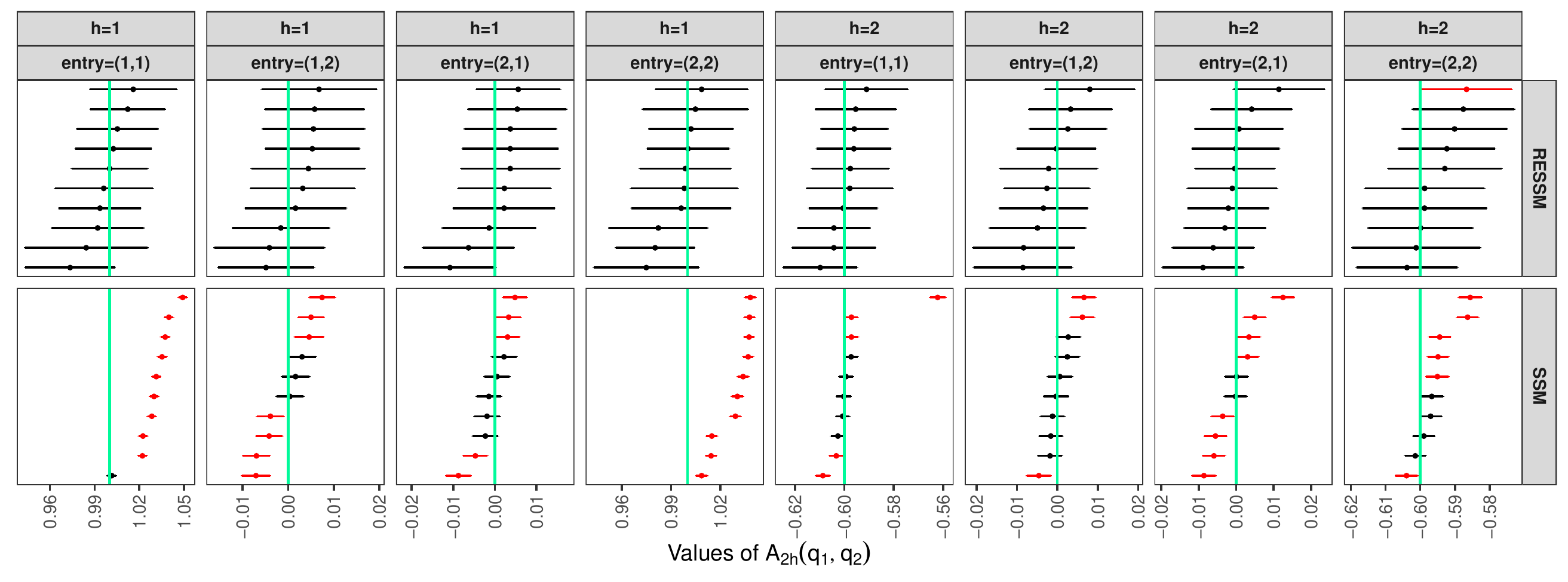}
        \end{flushright}
    \end{subfigure}
    \begin{subfigure}{0.9\textwidth}
        \begin{flushright}
        \includegraphics[width=0.9\linewidth]{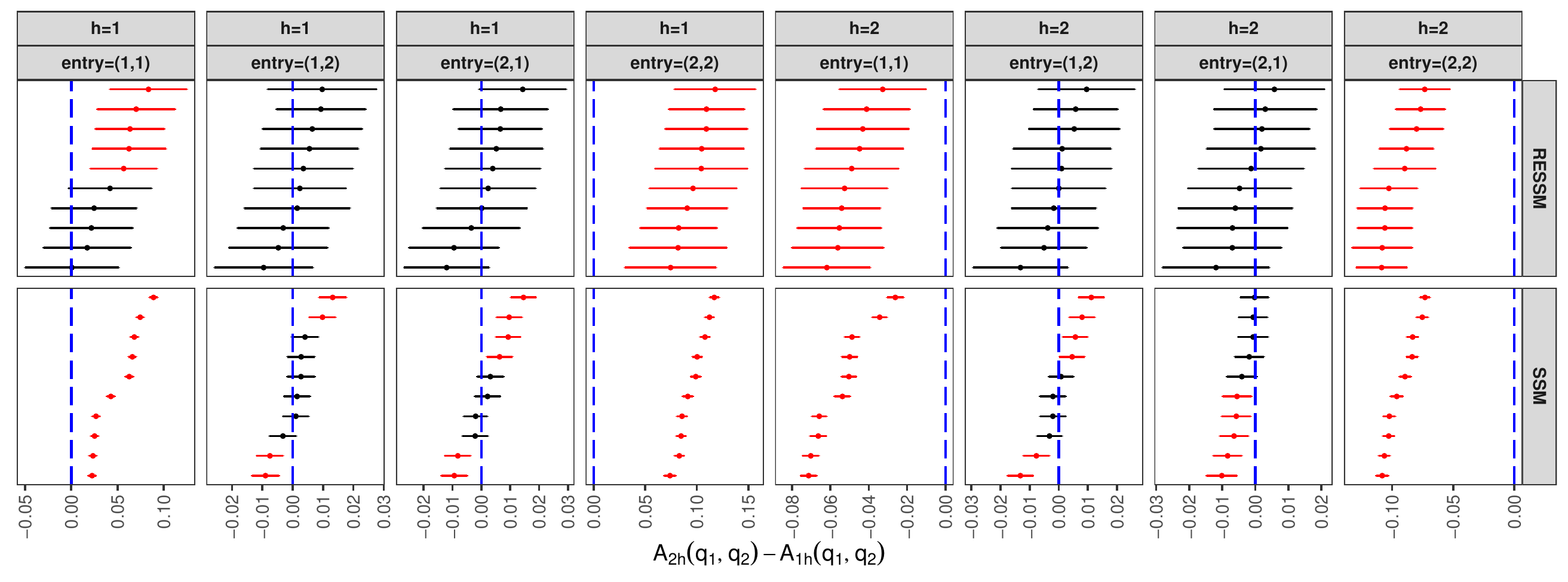}
        \end{flushright}
    \end{subfigure}
    \caption{The posterior means (dots) and $95\%$ credible intervals (bars) of the group-level $\mathbf{A}_{1h}$ for the Case group (Top Panel), the group-level $\mathbf{A}_{2h}$ for the Control group (Middle Panel), and group-level difference $\mathbf{A}_{2 h} - \mathbf{A}_{1 h}$ (Bottom Panel) using the proposed RESSM and the SSM by \cite{wang2022latent}. The green solid lines in the upper two panels are the true parameter values, the blue dashed lines in the bottom panel correspond to $x = 0$, and the red error bars are $95\%$ credible intervals that fail to cover the lines. }
  	\label{img:model_comparison}
\end{figure}

This simulation study compares the model performance between our proposed RESSM method and the latent state-space model (SSM) proposed by \cite{wang2022latent}.
We use two group (the MDD group and control group) with ($n_1=n_2=50$, $J=30$, $P=20$, $Q=2$, $m=2$).
In contrast to the RESSM, the SSM in \cite{wang2022latent} exhibits certain limitations, including: (i) a restriction to stationary data; (ii) a common spatial mapping matrix, where $\mathbf{\Theta}_{ri} \equiv \mathbf{\Theta}_{r}$; and (iii) restricted subject-level heterogeneity in temporal dynamical matrices, $\mathbf{A}_{ri} = \exp(\boldsymbol{\alpha}_r^{\top} \mathbf{X}_i) \mathbf{A}_{r}$, where $\mathbf{X}_i$ is a vector of observed time-invariant covariates. When there is no information about which covariates should be used, then $\mathbf{A}_{ri} \equiv \mathbf{A}_{r}$.
To ensure a fair comparison,  we only compared the SSM \citep{wang2022latent} with our proposed RESSM in terms of the estimation and inference performance for the group-level parameters. The results for group-level temporal dynamical matrices are presented in Web Figure \ref{img:model_comparison}.

The upper two panels of Web Figure \ref{img:model_comparison} illustrate the posterior means (depicted as dots) and $95\%$ credible intervals (represented by bars) for the group-level $\mathbf{A}_{1h}$ and $\mathbf{A}_{2h}$ ($h=1,2$) using the two methods. A notable contrast emerges when comparing the SSM to the proposed RESSM: the SSM exhibits considerably narrower credible intervals, leading to poor coverage rates when the EEG data exhibits between subject heterogeneity. 
Moving to the bottom panel of Web Figure \ref{img:model_comparison}, we observe the posterior means and $95\%$ credible intervals for the group-level difference $\mathbf{A}_{2h} - \mathbf{A}_{1h}$ using the two methods. In comparison to the RESSM, the SSM demonstrates a significantly larger type-I error, resulting in unreliable false positive results.
This discrepancy is attributed to the fact that the SSM ignores subject- and segment-level heterogeneity, thereby underestimating the variance. In contrast, the inference results for RESSM are more reasonable.
The inference for the group-level spatial mapping matrices exhibits similar patterns, and we omit the results here.

\section*{Web Appendix D}

This section presents additional details for the real data analysis discussed in Section 5 of the paper.

\subsection*{D.1. EEG data preprocessing}

It is well known that the quality of EEG preprocessing can significantly impact downstream analysis. A fundamental problem with EEG data analysis is its susceptibility to noise and artifacts. 
In the EMBRAC study, we follow the EEG data preprocessing framework outlined in \cite{yang2023learning}. The general preprocessing procedure is described as follows:
\textbf{(1)} To save memory and computational resources, the EEG data are resampled to $250$ Hz. 
\textbf{(2)} A notch filter is applied at $60$ Hz to remove line noise due to interference from the AC power line. 
\textbf{(3)} To obtain the desired frequency range of the data, we further apply a high-pass filter at $1$ Hz and a low-pass filter at $50$ Hz. 
\textbf{(4)} To address the noise in the channels, we adopt four automatic measures of detection of noisy channels and further interpolate bad channels \citep{bigdely2015prep}. Specifically, the four measures of automatic detection of noisy channels we use are extreme amplitudes (deviation criterion), lack of correlation with any other channel (correlation criterion), lack of predictability by other channels (predictability criterion) and unusual high frequency noise (noisiness criterion). 
\textbf{(5)} Eye-blink-related artifacts are removed using independent component analysis (ICA). 
\textbf{(6)} We segment the EEG data into two-second time segments. 
\textbf{(7)} To remove noisy high-amplitude artifacts contaminated by noisy time segments, we use a sensor-specific threshold learned from cross-validation and perform an automated rejection procedure \citep{jas2016automated}. 
\textbf{(8)} The filtered EEG data are re-referenced to the common average.

Note that in step (6), we chose the segment length of 2 seconds following existing literature in \cite{miraglia2016eeg} and \cite{yang2023learning}. In principle, it is possible to select an optimal segment length, but it is out of the scope of this paper.

\subsection*{D.2. Additional figures for EMBARC study}

\begin{figure}[h]
    \begin{subfigure}{0.92\textwidth}
        \begin{flushright}
        \includegraphics[width=0.93\linewidth]{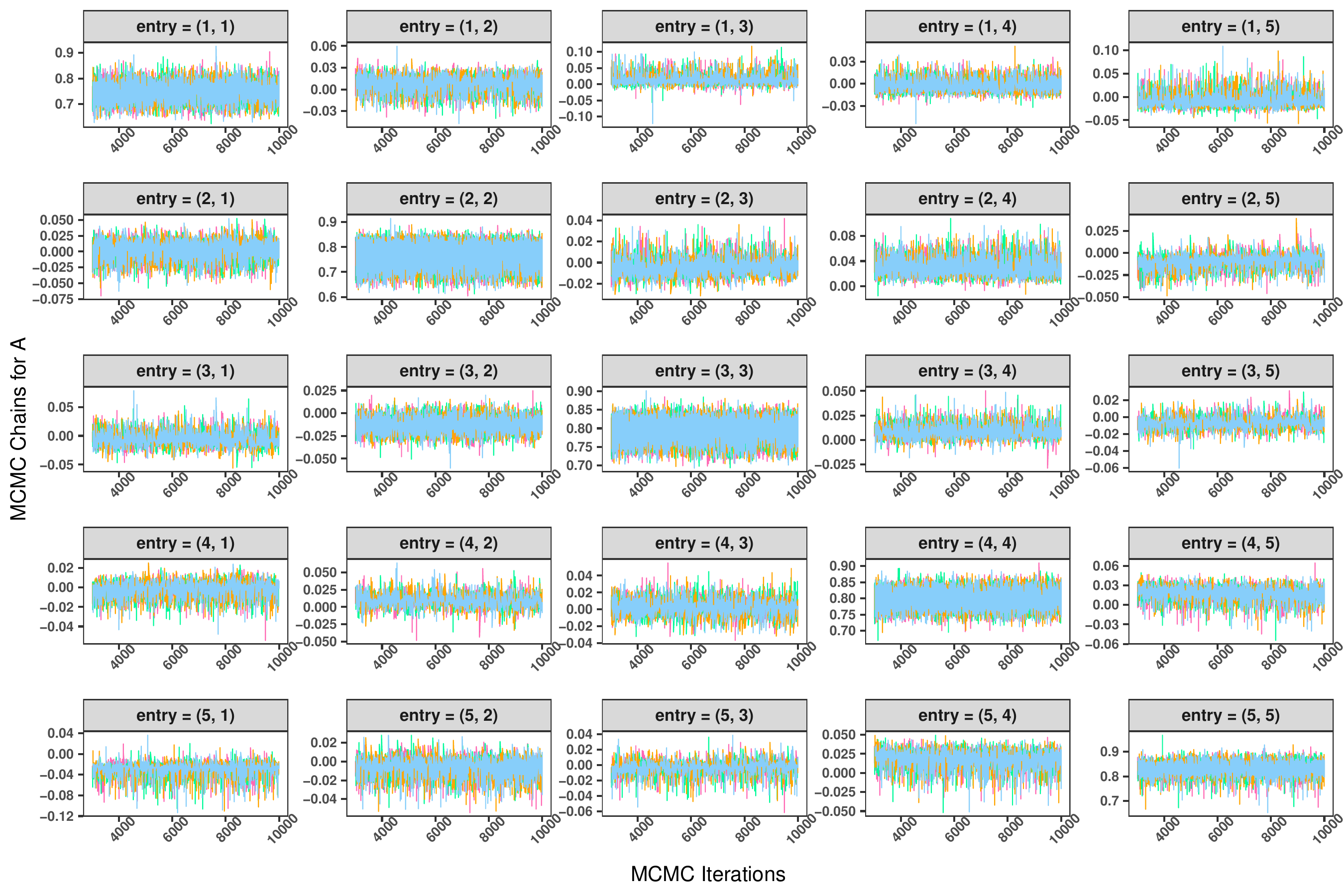}
        \end{flushright}
    \end{subfigure}
    \begin{subfigure}{0.92\textwidth}
        \begin{flushright}
        \includegraphics[width=0.93\linewidth]{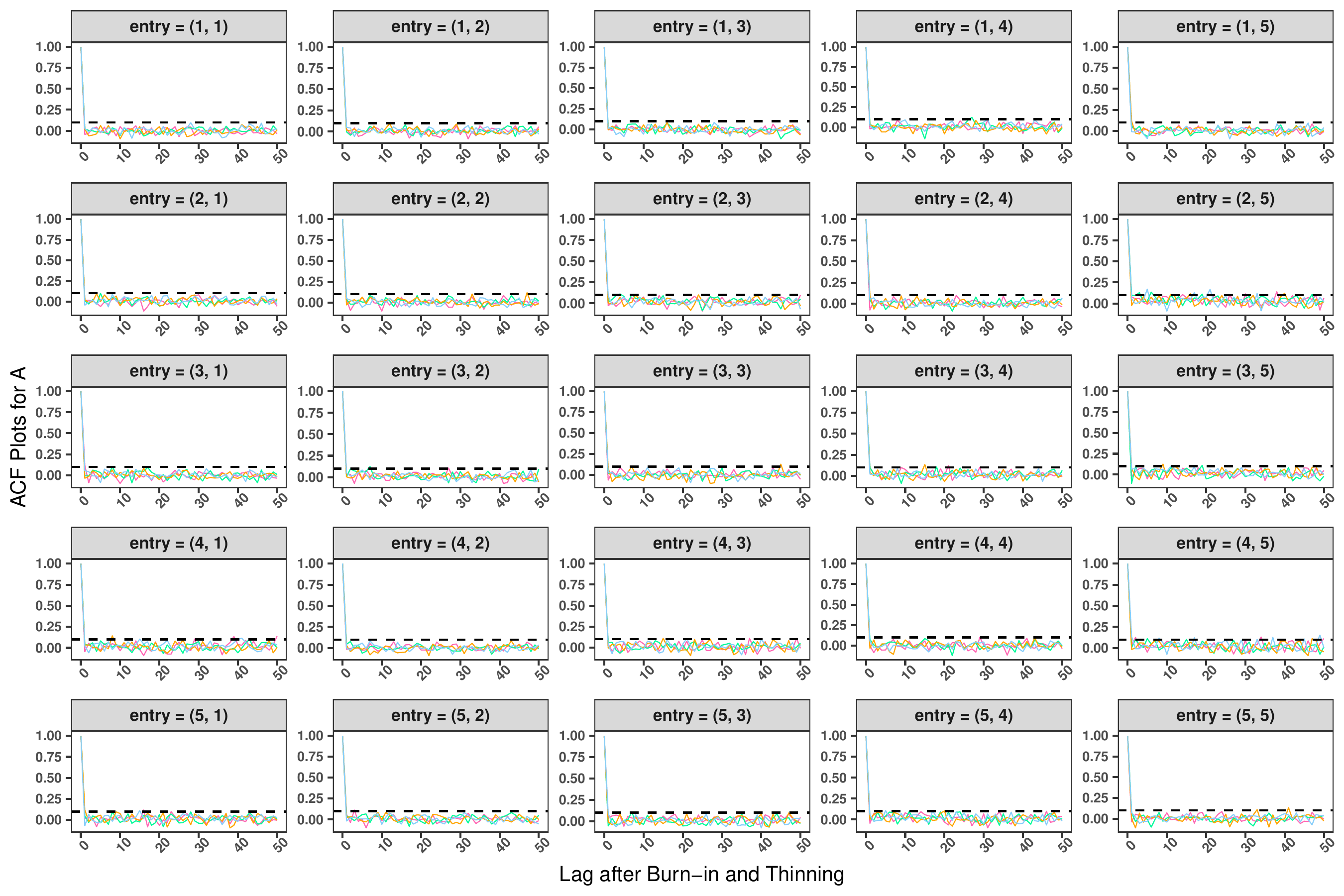}
        \end{flushright}
    \end{subfigure}
    \caption{Trace plots (top) and autocorrelation function (ACF) plots (bottom) for $\mathbf{A}_r$ $(r=1,2,3,4)$ to assess MCMC mixing and convergence. Each color represents one group. The black dashed lines in the ACF plots are $y=0.1$.}
  	\label{img:real_trace_acf}
\end{figure}

\begin{figure}[h]
 \centerline{\includegraphics[width=0.97\linewidth]{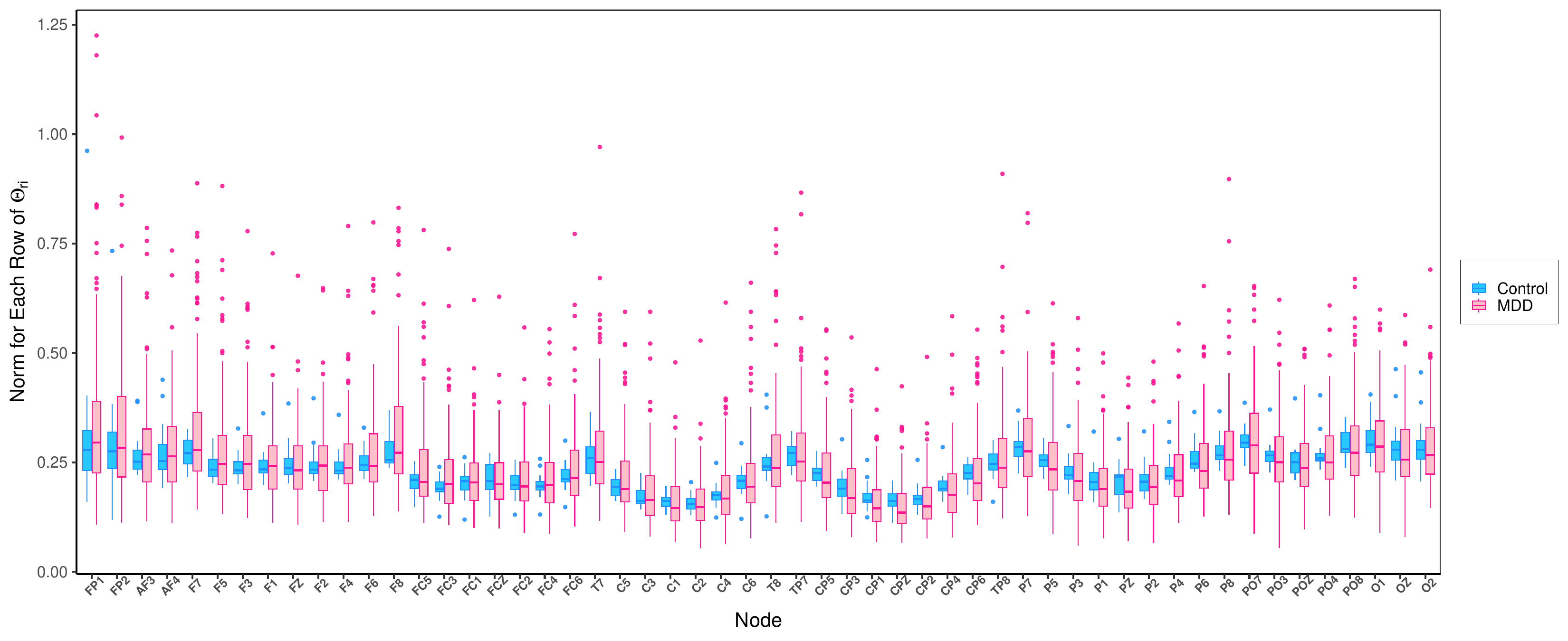}}
\caption{Boxplots depicting the norms of the $54$ rows within the subject-level spatial mapping matrices. Results for the MDD and control groups are colored in pink and blue, respectively.   }
\label{img:real_theta_subject}
\end{figure}

The MCMC diagnostics for $\mathbf{A}_r$ are presented in Web Figure \ref{img:real_trace_acf}.
The boxplot depicting the norms of the $54$ rows within the subject-level spatial mapping matrices is presented in Web Figure \ref{img:real_theta_subject}.

\begin{figure}[h]
    \begin{subfigure}{0.49\textwidth}
        \includegraphics[width=0.9\linewidth]{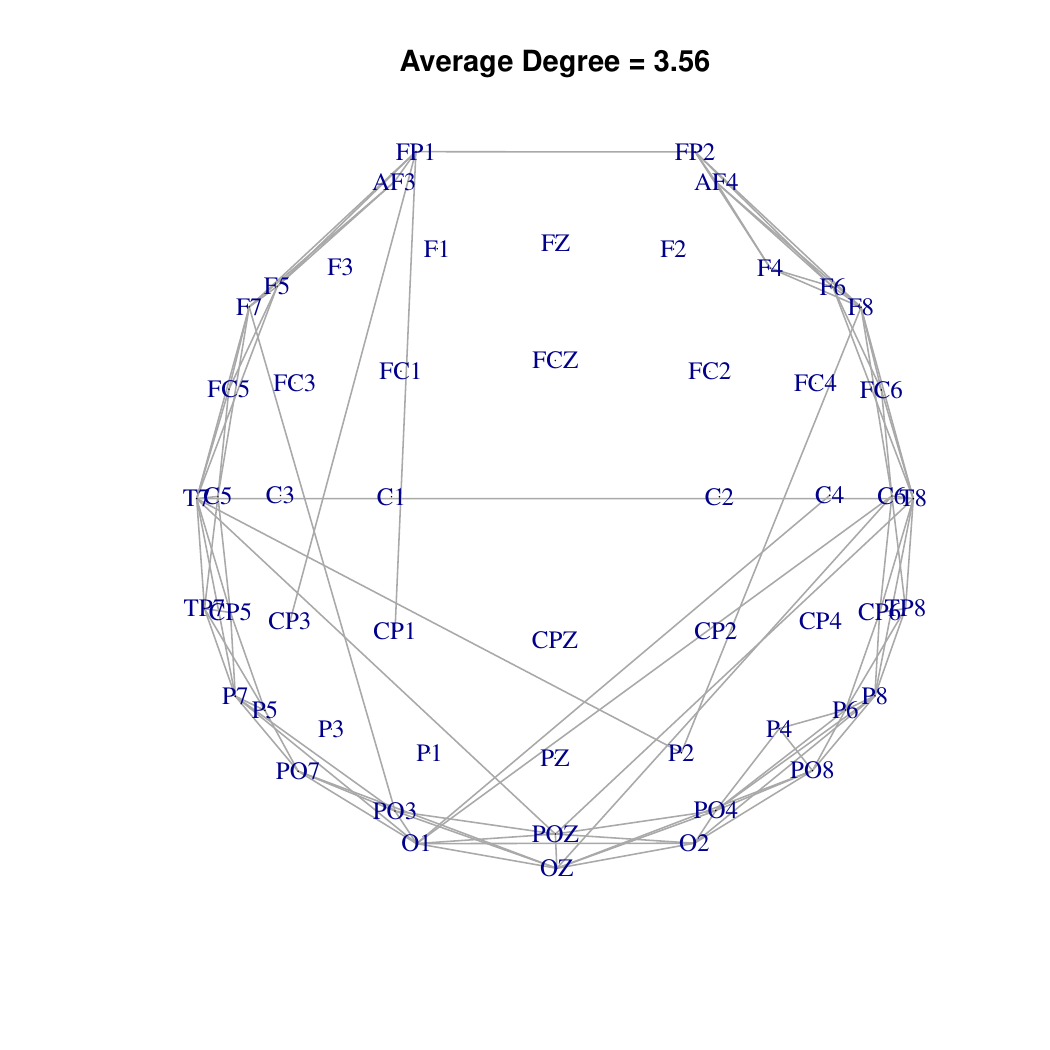}
        \vspace{-2em}
        \caption{MDD (TX)}
    \end{subfigure}
    \begin{subfigure}{0.49\textwidth}
        \includegraphics[width=0.9\linewidth]{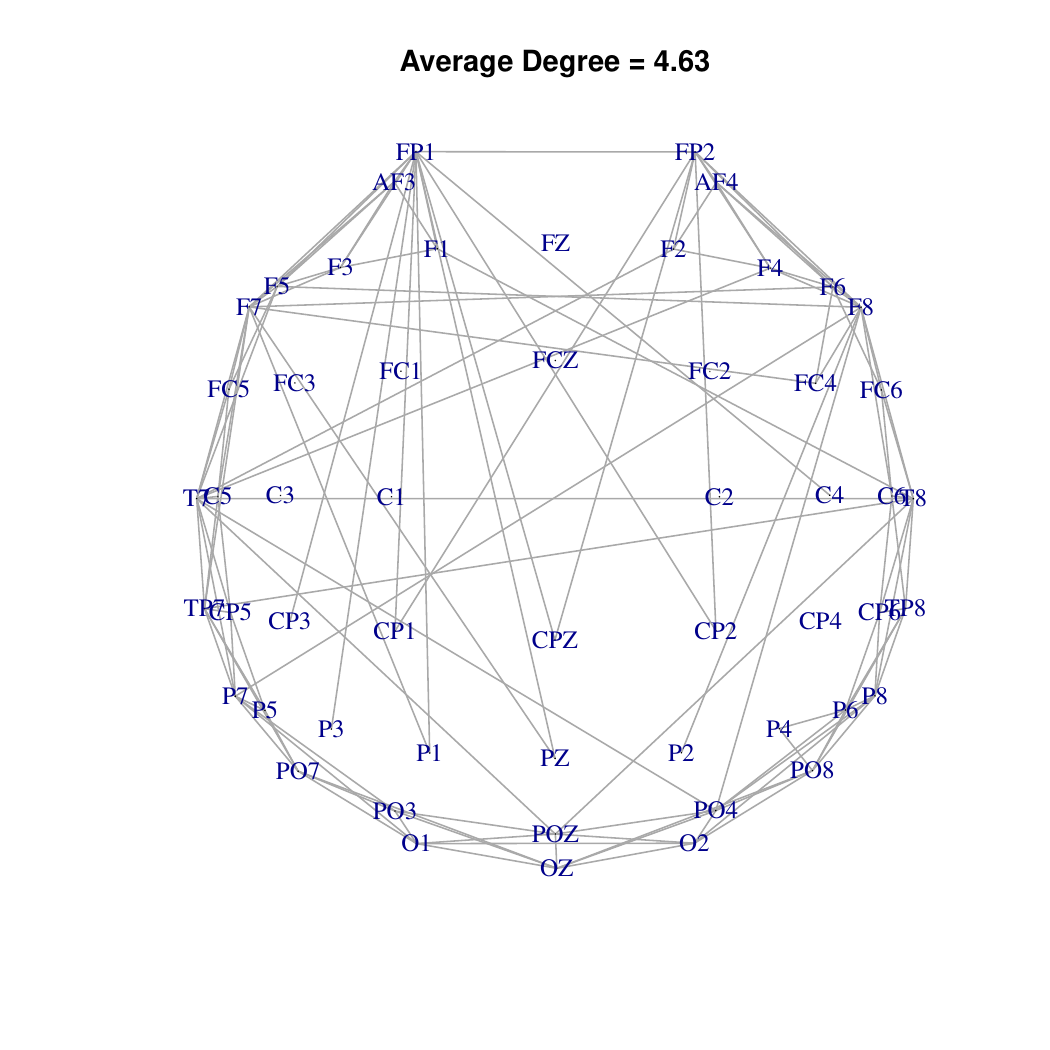}
        \vspace{-2em}
        \caption{Control (TX)}
    \end{subfigure}
    \begin{subfigure}{0.49\textwidth}
        \includegraphics[width=0.9\linewidth]{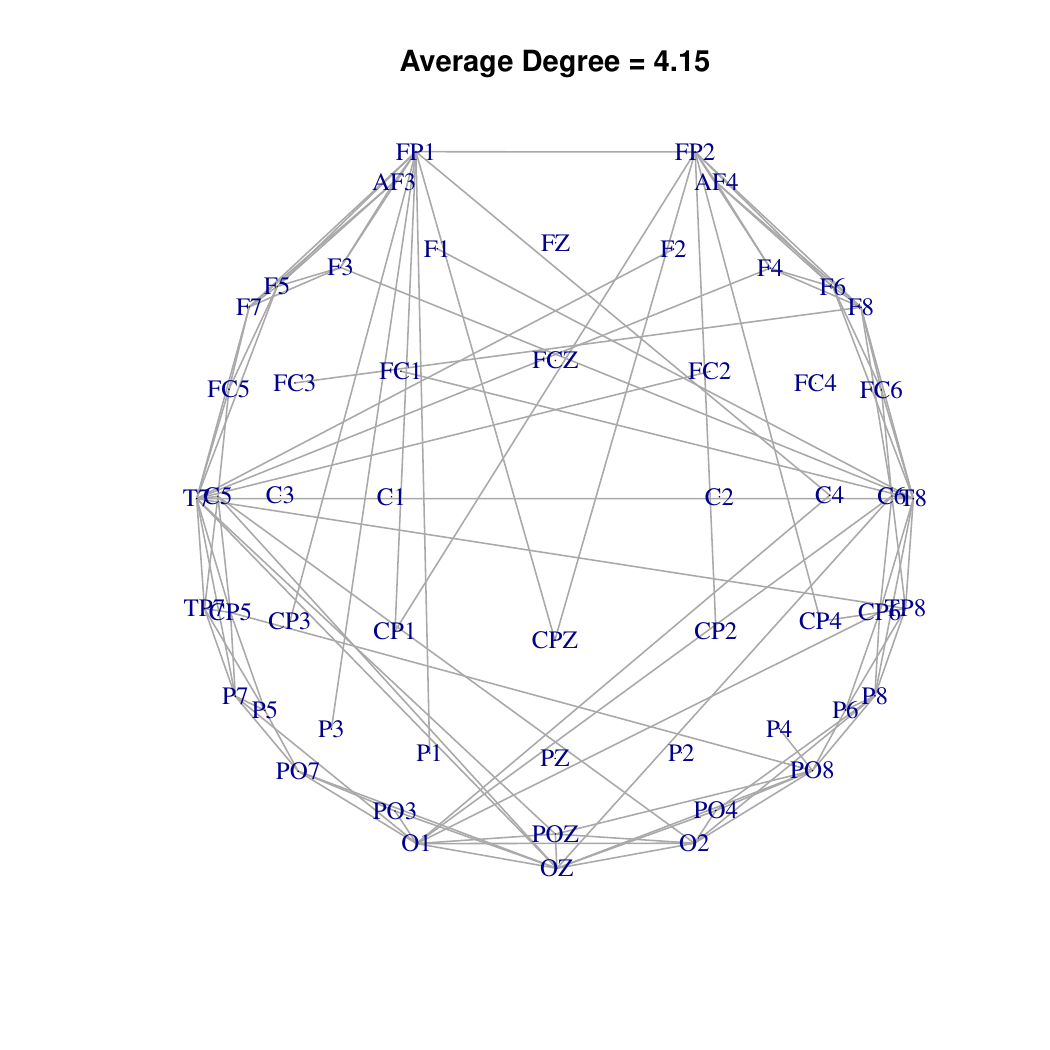}
        \vspace{-2em}
        \caption{MDD (CU)}
    \end{subfigure}
        \begin{subfigure}{0.49\textwidth}
        \includegraphics[width=0.9\linewidth]{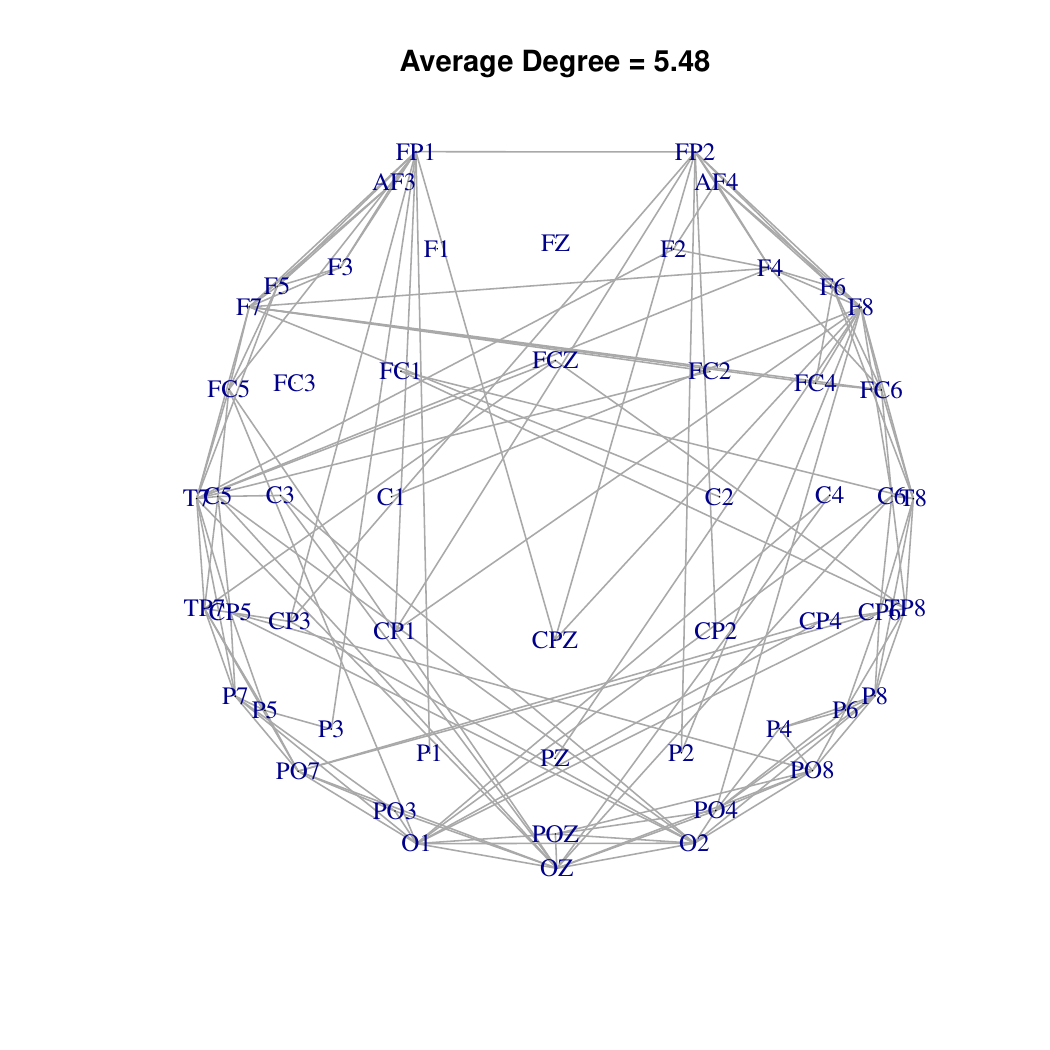}
        \vspace{-2em}
        \caption{Control (CU)}
    \end{subfigure}
    \caption{Visualization of group-level directional connectivity matrices for the MDD and control groups at sites TX and CU. Directionality is disregarded, employing undirected graphs. A threshold of $0.05$ is applied to eliminate connections with absolute values smaller than the threshold. Average degree is defined as the total number of connections for each node.}
  	\label{img:connectivity_graph}
\end{figure}

\begin{figure}[h]
    \begin{subfigure}{0.32\textwidth}
        \includegraphics[width=0.945\linewidth]{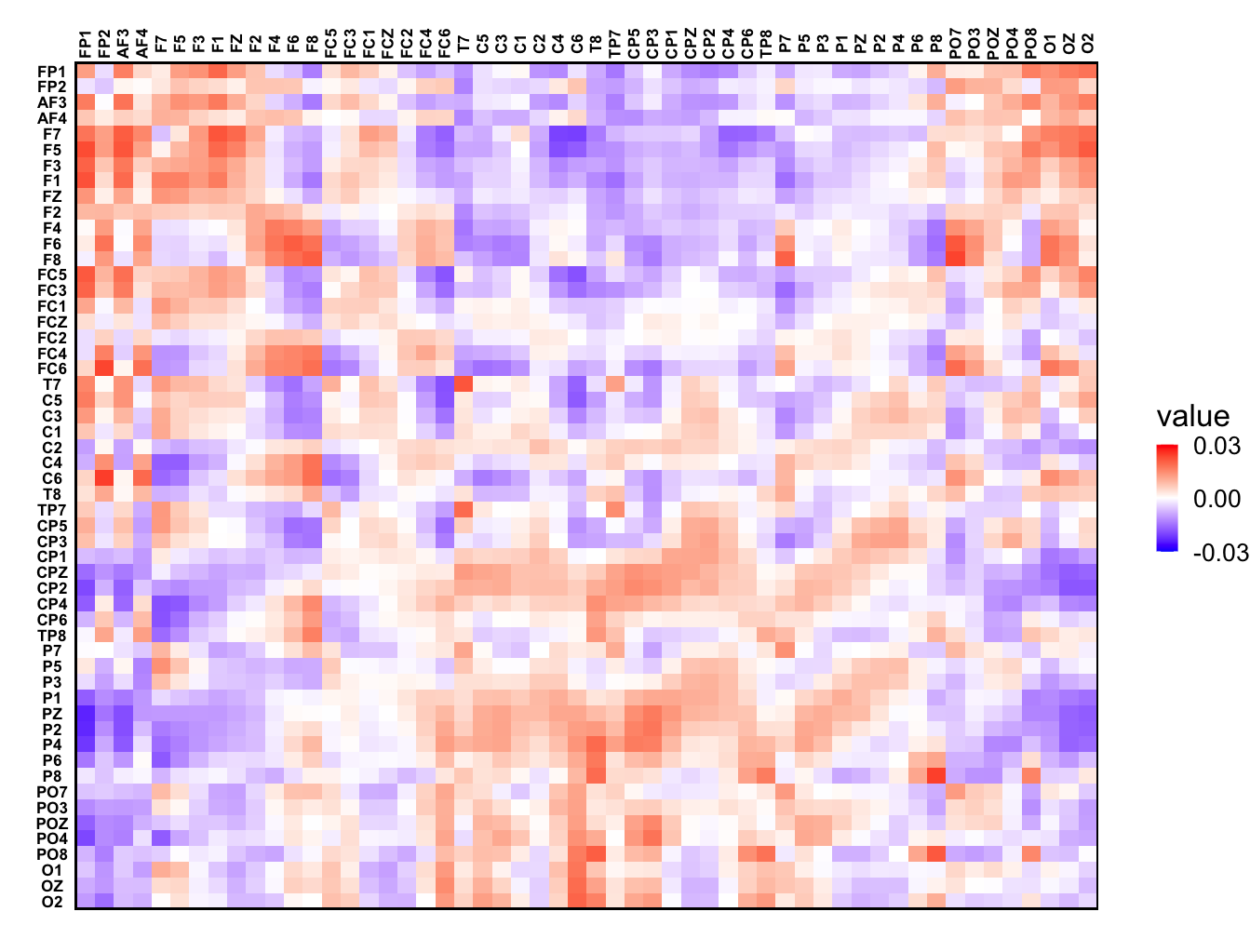}
        \caption{TX}
    \end{subfigure}
    \begin{subfigure}{0.32\textwidth}
        \includegraphics[width=0.95\linewidth]{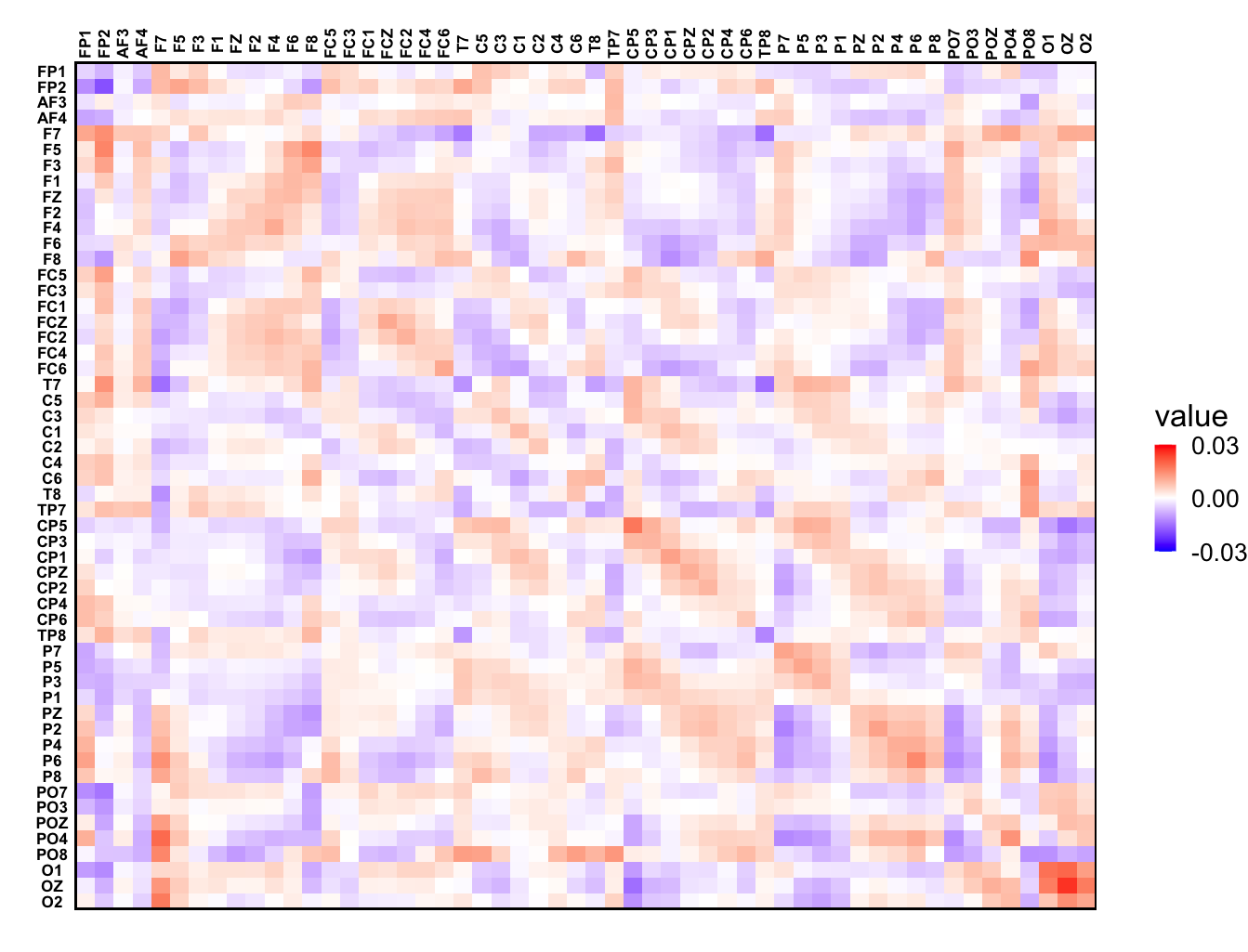}
        \caption{CU}
    \end{subfigure}
    \begin{subfigure}{0.32\textwidth}
        \includegraphics[width=0.95\linewidth]{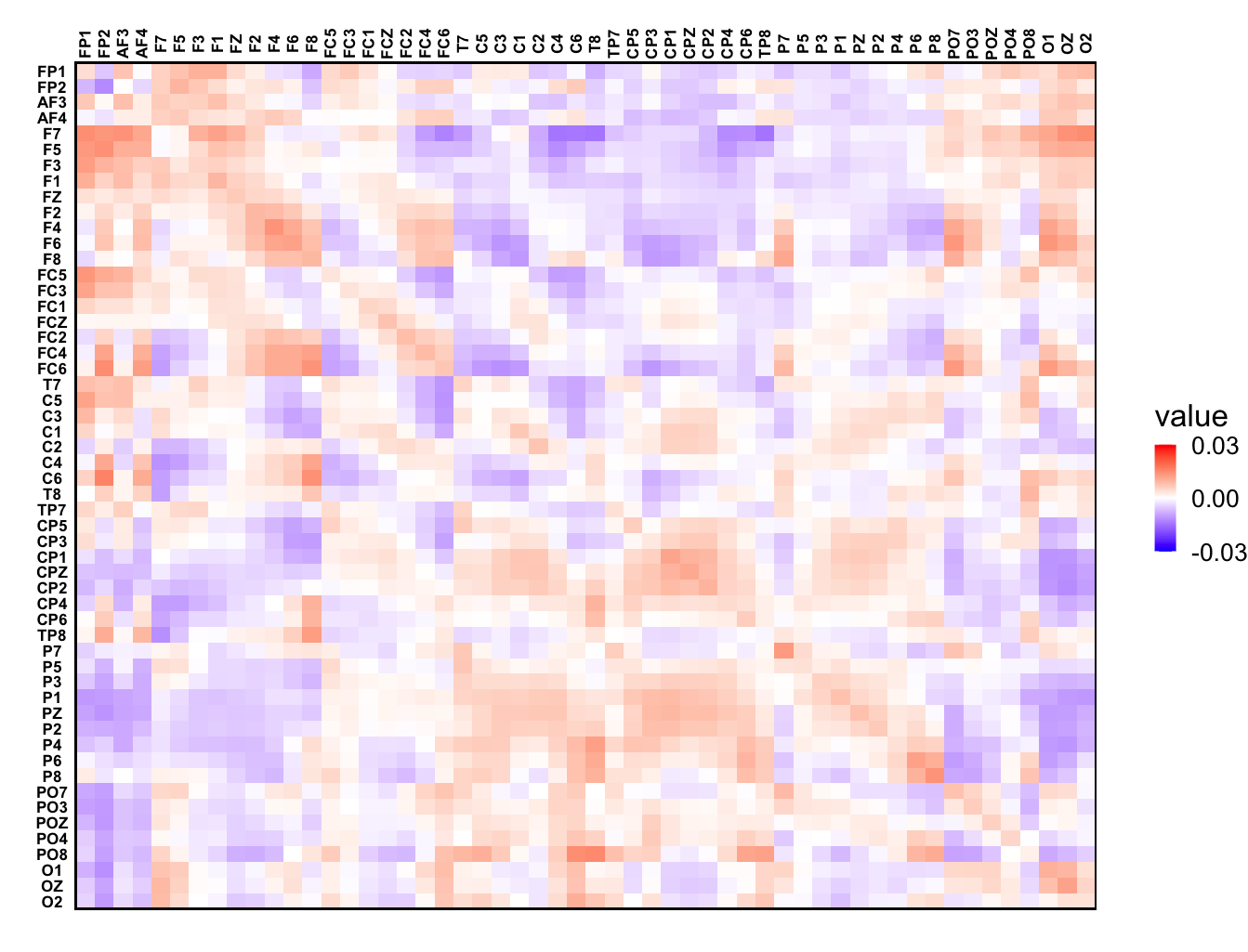}
        \caption{Both sites}
    \end{subfigure}
    \caption{The posterior mean differences in directional connectivities between the MDD and control groups for site TX, site CU, and the combined data from both sides.}
  	\label{img:DMC_new}
\end{figure}

For brain connectivity, a visualization of group-level directional matrices for the MDD and control groups at sites TX and CU is shown in Web Figure \ref{img:connectivity_graph}. Directionality is not shown and the networks are visualized as undirected graphs. A threshold of $0.05$ is applied to eliminate connections with absolute values smaller than the threshold. Define node degree as the total number of (edge) connections for one node, and average degree as the average of all node degrees for one network.
In Web Figure \ref{img:connectivity_graph}, we observed that patients with MDD exhibit reduced connectivity compared to the control groups, indicated by smaller average degrees (i.e. MDD groups have average degrees of $3.56$ and $4.15$ for sites TX and CU,  Control groups have average degrees of $4.63$ and $5.48$ for sites TX and CU). This observation aligns with the findings presented in \cite{veer2010whole}.
Meanwhile, the posterior mean differences in directional connectivities between the MDD and control groups for site TX, site CU, and the combined data from both sides are shown in Web Figure \ref{img:DMC_new}. We observed a more significant difference in connectivities between the MDD and control groups at site TX than at site CU. This difference may be due to variations in the experimental conditions at different sites and warrants further investigation.

\begin{figure}[h]
 \centerline{\includegraphics[width=0.97\linewidth]{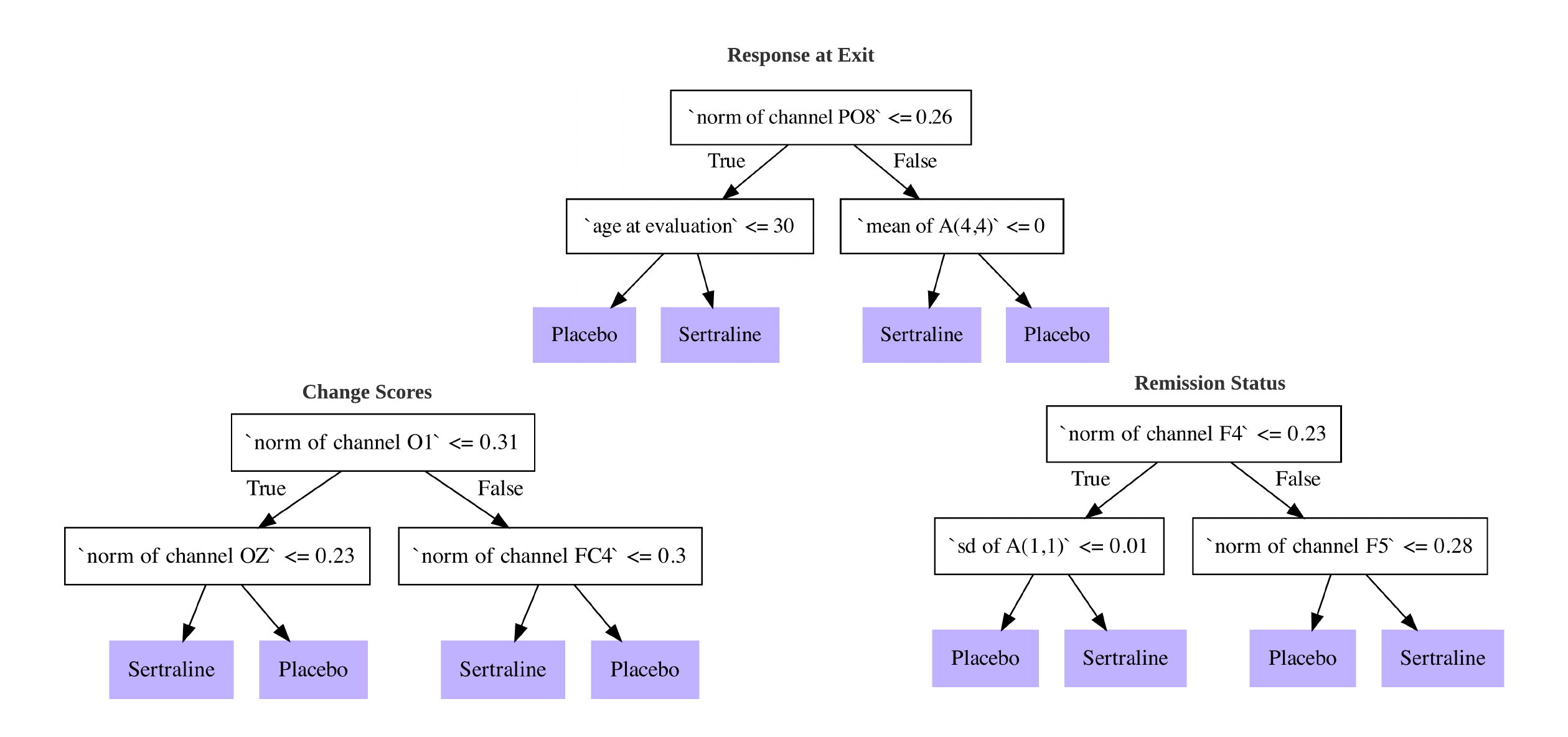}}
\caption{Optimal depth-2 policy tree learned by optimizing the augmented inverse propensity weighting loss function}
\label{f:policy_tree}
\end{figure}

For predicting heterogeneous treatment effects (HTE) and constructing optimal treatment decision rules using the features extracted by RESSM, the following figures are presented.
Web Figure \ref{f:var_importance} shows the top ten most important features from causal forests trained with clinical, demographic, asnd RESSM features using response at exit, change of the HAMD17 scores and remission status as outcome respectively. Web Figure \ref{f:policy_tree} shows the optimal depth-2 policy tree fitted using the algorithms by \cite{athey2021policy} using response at exit, change of the HAMD17 scores and remission status as outcome respectively.

To evaluate the predictive value of the subject-level EEG features for the CATE using causal forest, we compute the root mean squared error (RMSE) of the transformed outcome via ten repeats of 5-fold cross-validation. The transformed outcome for the $i$-th subject is defined as $Z_i^{*} = Z_i \cdot (W_i - p)/(p \cdot (1 - p))$ \citep{tian2014simple}, 
where $Z_i$ is one of the three treatment outcomes discussed in Section 5 in the paper, $W_i = 0, 1$ is an indicator variable for treatment and $p = P(W_i=1)$.

\begin{figure}[h]
 \centerline{\includegraphics[width=0.97\linewidth]{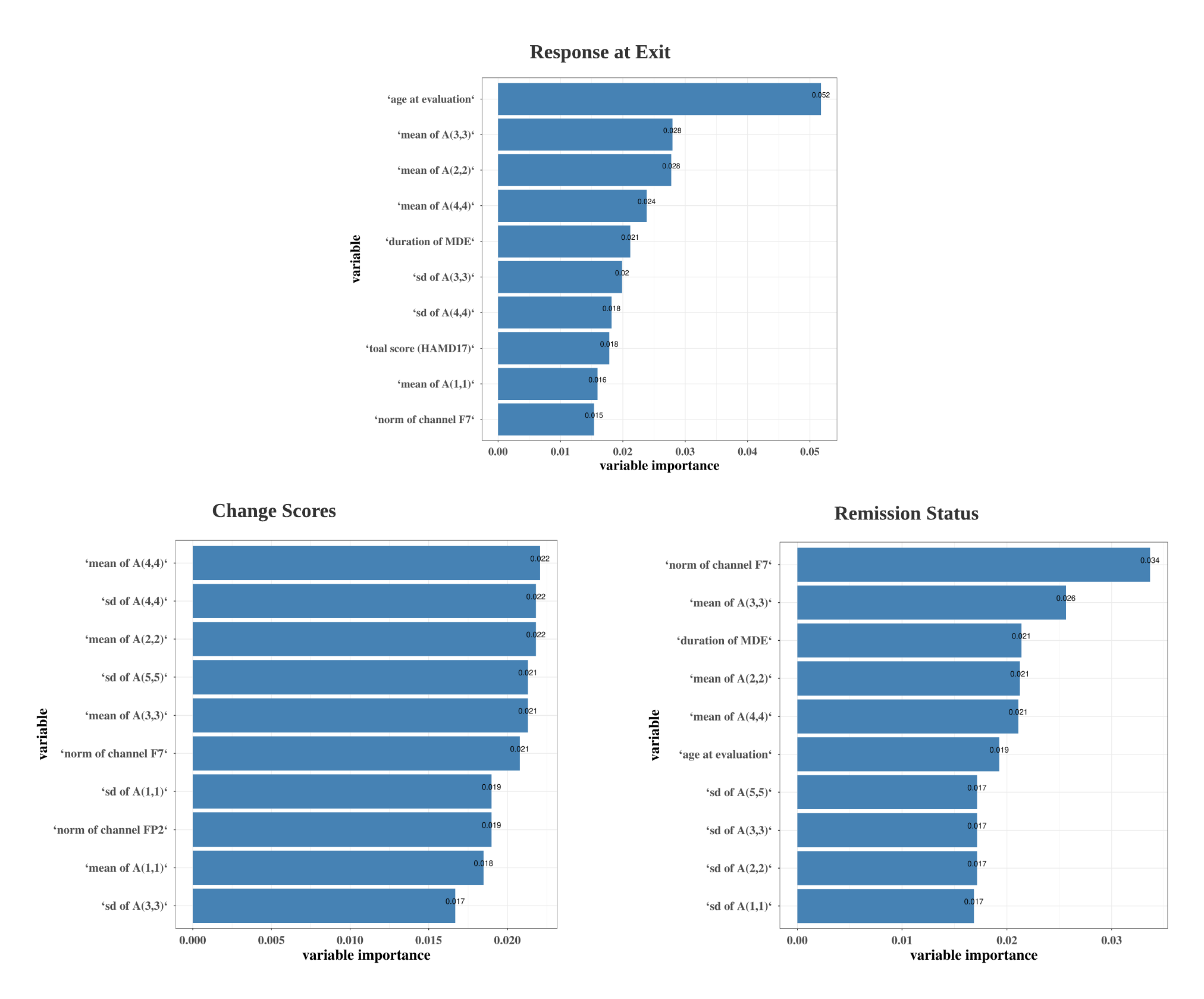}}
\caption{Variable importance measure of the top ten most important features}
\label{f:var_importance}
\end{figure}

\end{document}